%% file: main.tex
\documentclass{amsart}
\usepackage[a4paper, margin=2cm]{geometry}
\usepackage[T1]{fontenc}
\usepackage[utf8]{inputenc}

\usepackage{graphicx}

\usepackage{amsmath}

\usepackage{amsthm}
\usepackage{natbib}
\usepackage{amsxtra}

\usepackage[colorlinks]{hyperref}

\usepackage[dvipsnames,svgnames]{xcolor}

\usepackage{macro}    

\usepackage{nicematrix}

\usepackage{float}

\usepackage{amssymb,amscd}

\usepackage{subcaption}

\usepackage{mathtools}

\usepackage{txfonts}

\usepackage{makecell}

\usepackage{tabularx}

\usepackage{booktabs} 

\usepackage{dirtytalk}

\usepackage{pgfplots}
\pgfplotsset{compat=1.17}

\usepackage{tikz}
\usetikzlibrary{calc}
\usetikzlibrary{math}
\usetikzlibrary{shapes.arrows}
\usetikzlibrary{decorations,decorations.markings}
\usetikzlibrary{decorations.pathreplacing}

\usepackage[normalem]{ulem} 
\theoremstyle{remark}

\usepackage[backgroundcolor=blue!10]{todonotes}

\newcommand{\todoC}[1]{{\color{magenta}#1}}
\newcommand{\todoR}[1]{{\color{black}#1}}

\date{}

\begin{document}


\title{Symmetries of (quasi)periodic materials: \\
Superposability vs. Indistinguishability}

\author{Markus Husert}
\address{Sorbonne Université, CNRS, Institut Jean Le Rond d\lq Alembert, UMR 7190, 75005 Paris, France}
\email{first.author@email.com}

\author{Christelle Combescure}
\address{IRDL, UMR CNRS 6027, Univ. Bretagne Sud, Bretagne INP, ENSTA, Institut Polytechnique de Paris, Univ. Brest, Lorient, France}
\email{first.author@email.com}

\author{Renald Brenner}
\address{Sorbonne Université, CNRS, Institut Jean Le Rond d\lq Alembert, UMR 7190, 75005 Paris, France}
\email{first.author@email.com}

\author{Nicolas Auffray}
\address{Sorbonne Université, CNRS, Institut Jean Le Rond d\lq Alembert, UMR 7190, 75005 Paris, France}
\email{first.author@email.com}

\date{\today}

\subjclass[2020]{74Q15, 74A40} 
\keywords{keyword1, keyword2, keyword3}

\begin{abstract}
This work is devoted to the study of the symmetries of (quasi)periodic architectured materials.  
The main goal is to extend the symmetry criterion of superposability for strictly periodic to quasiperiodic materials.  
For this purpose, the weaker symmetry criterion of indistinguishability is used. It relies on a statistical description of the mesostructure and is defined in terms of the spatial autocorrelation functions of the material under consideration.  
By using the representation of these autocorrelation functions in Fourier space, the space groups of both periodic and quasiperiodic materials can be obtained.
In this context, an image-processing methodology is proposed to identify the key characteristics of a material’s space group (i.e its point group and its symmorphism) directly from the Fourier transform of the mesostructure.  
The method is validated on synthetic two-dimensional images of (quasi)periodic architectured materials and it is pointed out, as an illustrative example, that the rotational symmetry of the classical Penrose tiling is of order ten.  
\end{abstract}

\maketitle

 \tableofcontents

 \include{intro.tex}
\input{periodic.tex}

 \input{quasiperiodic.tex}
\input{method.tex}

 \input{results.tex}

 \input{conclusion.tex}

 \section*{Acknowledgements}
The authors would like to thank Denis Gratias for drawing their attention to the notion of indistinguishability at the core of this paper, as well as for his valuable advice.

\bibliographystyle{elsarticle-harv}
\bibliography{quasisymmetry}

\appendix

\include{appendix}

\end{document}

%% file: intro.tex
\section*{Introduction}
\begin{figure}[H]
    \centering
    \includegraphics[width=0.7\linewidth]{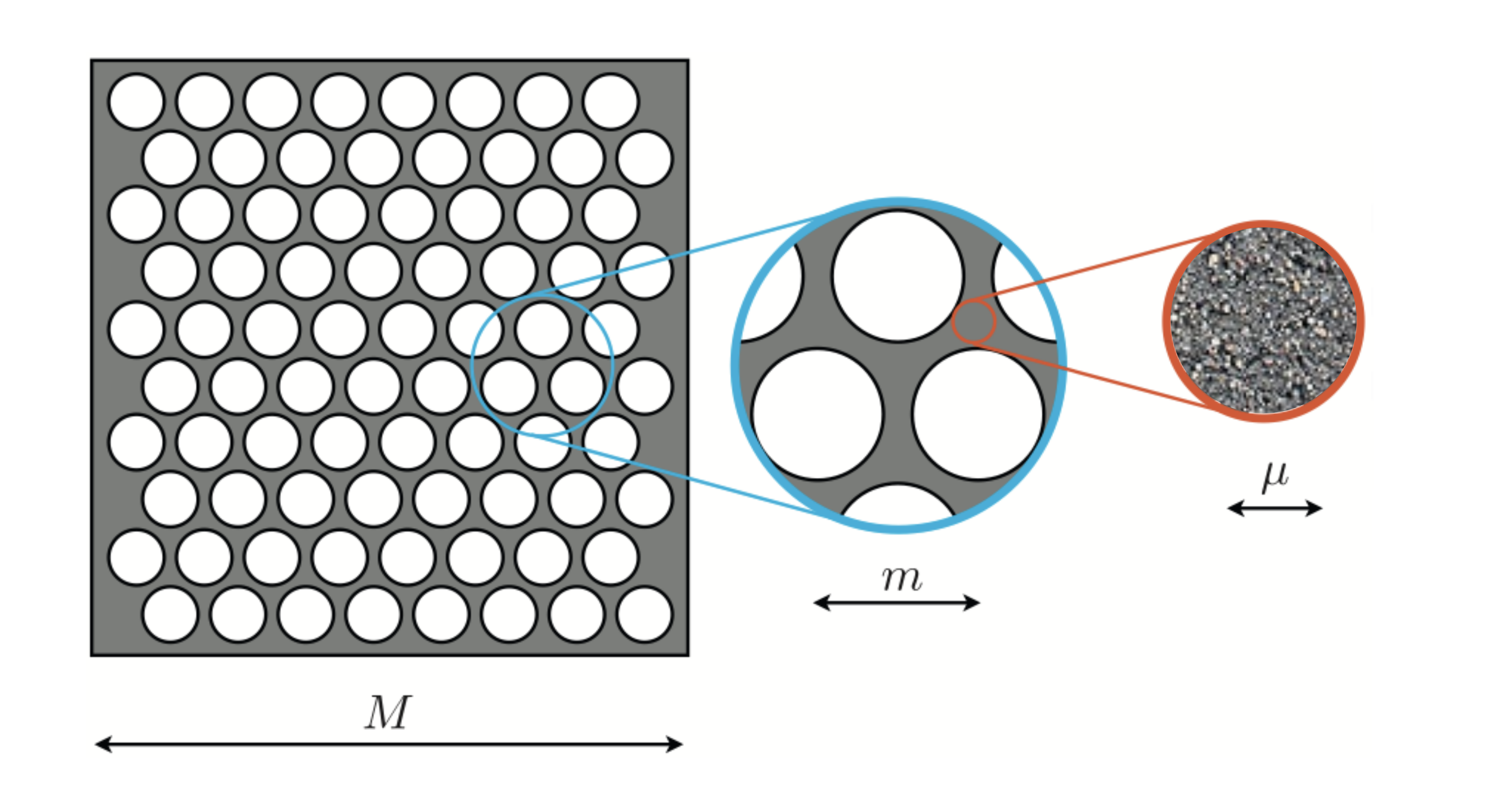}
    \caption{\todoR{Characteristic scales of an architectured material}}
    \label{fig:archi}
\end{figure}

With the rise of additive manufacturing, recent years have seen the emergence of architectured materials in engineering. These materials differ from conventional bulk materials such as steel or polymers in that they feature one or more additional levels of material organisation between their microstructural scale -- the physico-chemical scale -- and their macrostructural scale where is defined the geometry of the structure they compose \citep{brechet_2013,poncelet_2018}. This organisation is schematically represented in \autoref{fig:archi}. Beyond the immediate advantages of material savings and structural lightening, the ability to control internal architecture enables architectured materials to exhibit new functionalities such as tunable band gaps for elastic or acoustic wave-propagations \citep{rosi_2024, Bayat2018} or designed instabilities \citep{Kochmann2017,Azulay2024}. When these functionalities are unconventional, such materials are referred to as metamaterials \citep{craster_2023}.\\

In this development, determining the internal architecture becomes a new critical parameter in material design. A fundamental question arises: how should this architecture be organised? Traditionally and, as we will see, somewhat simplistically one contrasts random organisation on the one hand with periodic organisation on the other. While not entirely inaccurate, this classification is nonetheless overly rigid. It is worth noting that this dichotomy is likely rooted in the theoretical modelling tools available: stochastic approaches on the one hand and classical crystallographic frameworks on the other.\\

\begin{figure}[H]
    \centering
    \includegraphics[width=0.3\linewidth]{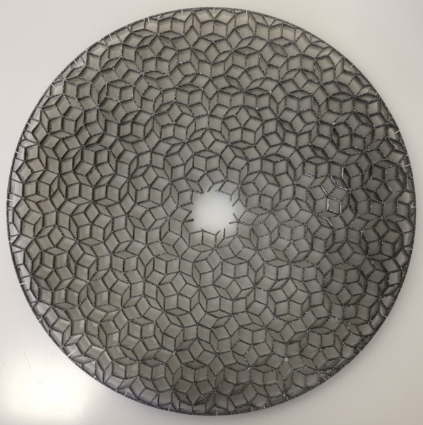}
    \caption{Laser cut disk of Penrose honeycomb used for Brazilian test \citep{somera2022effective}.}
    \label{fig:Penrose_Honey_Comb}
\end{figure}

Let us set aside modelling tools for now. 
In fact, undertaking a comparison merely based on the stochastic or deterministic property of a material results in the forfeiture of a fundamental characteristic intrinsic to periodic media: the notion of symmetry. Indeed, within periodic systems, symmetry is articulated through the stringent notion of invariance under a set of affine isometries constituted by combinations of proper and improper rotations and translations \citep{mcweeny_1963,armstrong_1988}. Moreover, the combination of rotational and translational invariance is not automatic and the set of rotation orders that are compatible with translational symmetry is governed by the crystallographic restriction theorem. This theorem states that periodic structures in $\mathbb{R}^2$ and $\mathbb{R}^3$ cannot be invariant under rotations of order 5 or of any order $n \geq 7$\footnote{A rotation of order $n$ is a rotation by an angle $2\pi / n$.}.
The theoretical framework leading to the crystallographic restriction theorem is directly inherited from the tools developed by crystallographers starting in the 19th century \citep{michel_2001}. Physics and chemistry at the end of the 20th century experienced a Copernican revolution with the accidental discovery of quasicrystals by \cite{shechtman_1984}. These crystals are atomic structures whose organisation is incompatible with translational symmetry (for instance with rotational symmetry of order 5), yet they display well-defined diffraction patterns. Such aperiodic crystals appear locally disordered (due to the absence of a translational lattice), but exhibit long-range order (with an essentially discrete diffraction spectrum). As a result, the notion of long-range order proves to be a more appropriate concept for describing modes of matter organisation than periodicity, which underlies the classical dichotomy between random and periodic structures \citep{lifshitz_2011}. Such deterministic structures, with the Penrose tiling serving as perhaps the most renowned example, can be efficiently produced at millimetre scales via conventional laser cutting techniques (\autoref{fig:Penrose_Honey_Comb}) or through additive manufacturing processes, where they are referred to as \textit{quasiperiodic architectured materials}.
In recent years, quasiperiodic architectured materials have attracted increasing interest in solid mechanics \citep{Badiche2000, Beli2020, Moat2022,Imediegwu2023, rosa_stiff_2024}. The ability to combine local disorder with a high degree of rotational symmetry opens the door to novel properties. Besides the obvious higher isotropy degree of these structures, the apparent local disorder enhances the fracture toughness of the structure, as cracks encounter difficulty in propagating \citep{glacet_2018}. These structures are also of interest due to their potential to generate complete band gaps \citep{steurer_2007, Bouchitte2010, Beli2020}. They have also been shown to display 
oriented localised deformation bands under tension loading \citep{Badiche2000}.

\begin{figure}[H]
    \centering
    \begin{subfigure}[b]{0.3\linewidth}
        \centering
        \includegraphics[width=\linewidth]{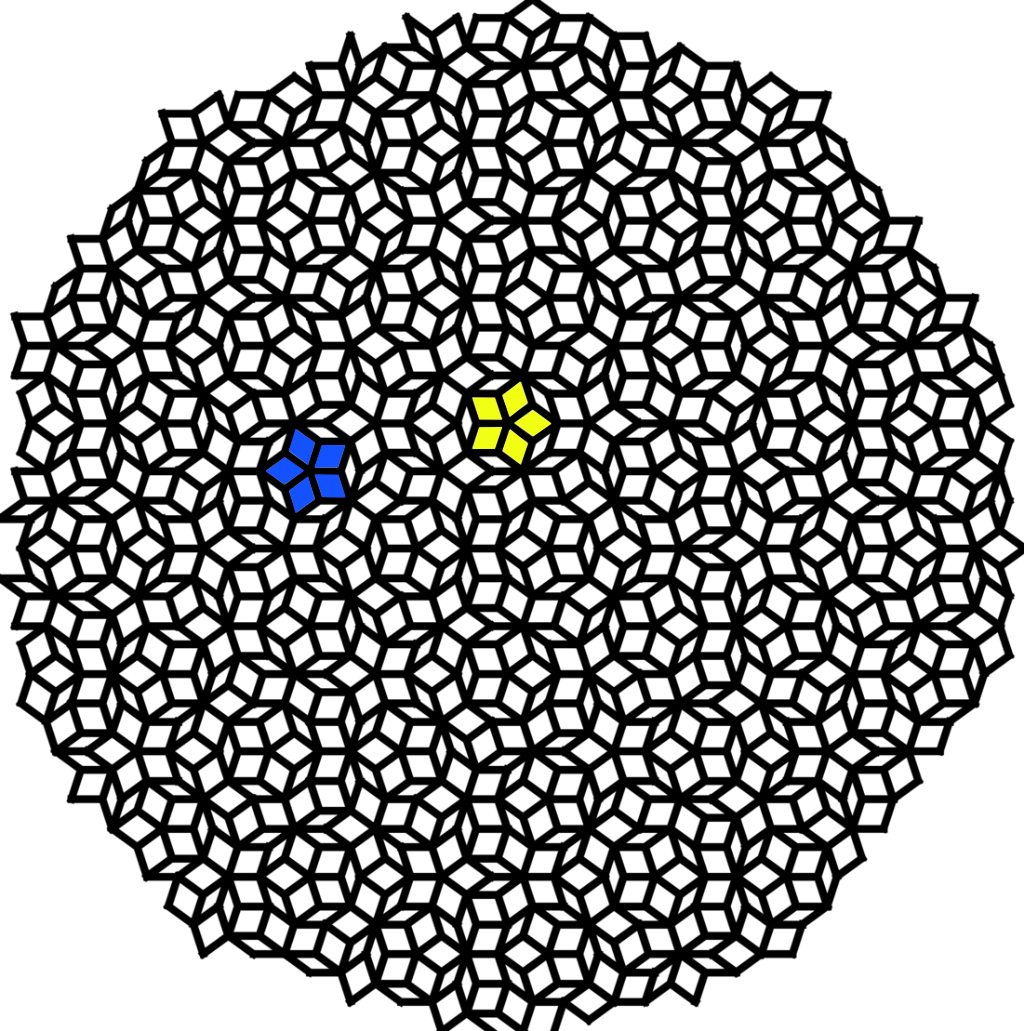}
        \caption{Original Penrose tiling}
        \label{fig:Penrose}
    \end{subfigure}
    \hspace{1.5cm}
    \begin{subfigure}[b]{0.3\linewidth}
        \centering
        \includegraphics[width=\linewidth]{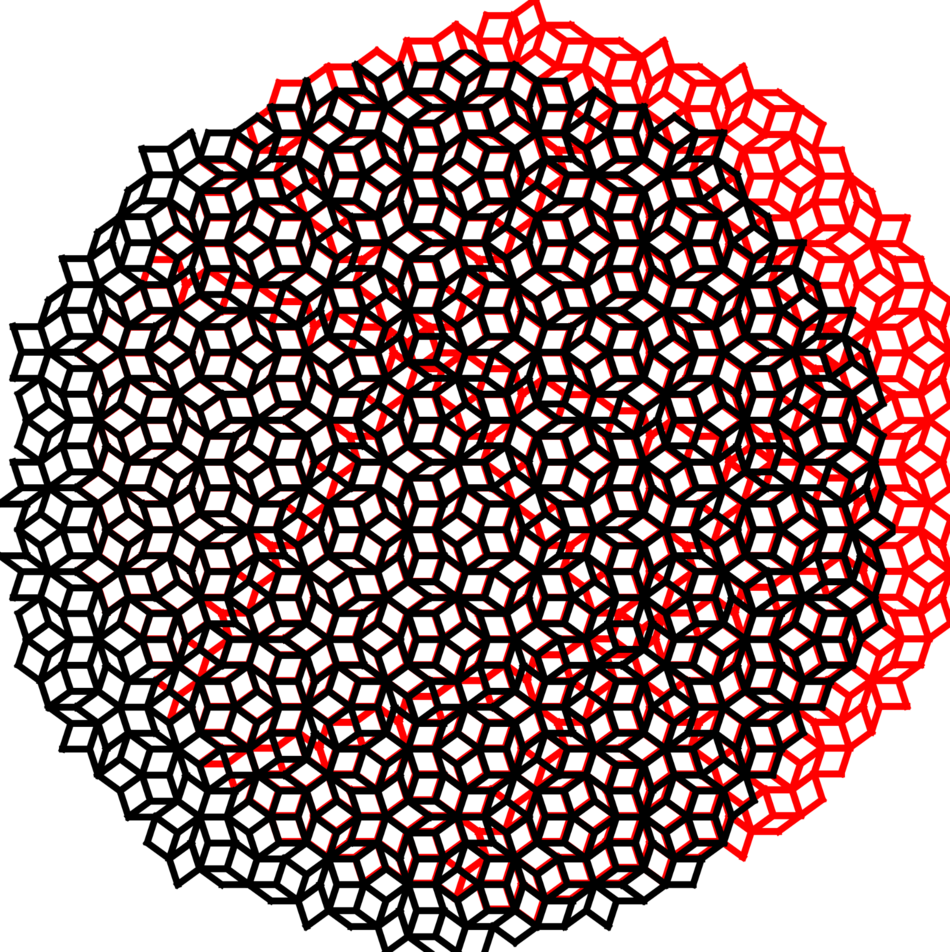}
        \caption{Original (black) and rotated (red) tilings}
        \label{fig:Penrose_overlay}
    \end{subfigure}
    \caption{Illustration of the mismatch of the Penrose tiling for a $\frac{2\pi}{10}$ rotation combined with a translation. On figure (a)\todoR{ two five-pointed stars with different orientations are highlighted in yellow and blue. On figure (b) defect lines (\textit{worms}) arise from the lack of translational invariance in the tiling.}}
    \label{fig:Imperfect_Invariance}
\end{figure}

From a modelling perspective of the effective properties emerging from these architectured materials, the order of rotational invariance of the medium is an important parameter. Indeed, according to Curie's principle \citep{curie_1894}, it is related to the anisotropy properties of the effective constitutive tensors. Furthermore, Hermann’s theorem~\citep{Hermann_1934,gluge_systematic_2021} specifies that in 2D (respectively in 3D), if the order of rotational symmetry exceeds the order of the tensor modelling the effective property, the latter will be hemitropic\footnote{ hemitropic means invariant with respect to every rotation but no mirror invariance.} (respectively transversely hemitropic). Then, quasiperiodic materials lifting the crystallographic restriction on the rotational order, they allow for a broader family of isotropic deterministic materials. One could then conclude that most of the emerging properties of quasiperiodic architectured materials rely on the high-order rotational symmetry of the architecture. Stated as such, this assertion-frequently encountered in the literature is nevertheless problematic. Let us consider a generic fragment of the Penrose tiling \autoref{fig:Penrose}. Despite its apparent symmetry, there exists no isometry that maps it onto itself up to a translation. \autoref{fig:Penrose_overlay} represents in red a rotated Penrose tiling by a rotation of $\frac{2\pi}{10}$ superposed to the original Penrose tiling after a translation. One observes that the invariance by superposition is incomplete. \todoR{There is no global translation that can eliminate all the defects (a.k.a. worms) observed in \autoref{fig:Penrose_overlay}. Only finite regions can be superimposed}. The usual notion of symmetry as geometric invariance under superposition is thus ill-suited to this type of organisation.

If one observes the repeating patterns appearing on \autoref{fig:Penrose}, one notices \emph{stars} that are invariant under rotations of order 5 and possess mirror symmetries. Such symmetry corresponds to the  dihedral group $\DD{5}$, and it is often attributed to the Penrose tiling in the solid mechanics literature \citep{IMEDIEGWU2023111922,rosa_stiff_2024}. However, upon closer inspection, these stars appear in two variants that differ by an angular shift of $\frac{2\pi}{10}$\todoR{, highlighted in yellow and blue in~\autoref{fig:Penrose}}, and both variants occur with equal statistical frequency. Consequently, one might argue that the actual symmetry is $\DD{10}$. In the absence of a precise and operational definition, it is impossible to resolve this ambiguity.


Thus, the question addressed in this article is: in what precise sense can a quasiperiodic pattern be considered symmetric? As we shall see, this question has been resolved in condensed matter physics via the generalisation of superposability to the broader notion of \textit{indistinguishability}. 
Although in its current form this formulation appears to be little known within the solid mechanics community, it is in fact intimately connected to the study of the effective properties of heterogeneous materials \citep{milton_2002}.\\

It is well established  that the effective physical properties admit an infinite-series representation involving mesostructure correlation functions of successive orders \citep{brown_1955, torquato_2002}.
Indistinguishability refers precisely to the invariance of correlation functions under a set of transformations, rather than to the invariance of the structure itself. Then the symmetries of a medium can be characterised through its correlation functions by identifying the set of isometries that leave these functions invariant. The medium transformed by these operations is not necessarily superposable with the original medium, yet it remains indistinguishable from the original one at large scale.  This set of transformations defines the indistinguishability group, which generalises the concept of the superposability group of a medium. From a homogenisation standpoint, two heterogeneous media are indistinguishable if they share identical effective physical responses.\par
The notion of indistinguishability 
is naturally characterised in reciprocal space, or Fourier space, as presented by \citet{mermin_1991} and \citet{lifshitz_1996}. It is important to note that this characterisation is constructive, in the sense that it can be implemented algorithmically to identify the indistinguishability group of a function or an image. This is precisely what is proposed in this article. After introducing the concept of indistinguishability, we present a procedure that allows, from synthetic images of mesostructures, the determination of their symmetry properties in the sense of indistinguishability. In the case of periodic media, this notion coincides, as expected, with the classical notion of symmetry by superposition. The proposed algorithm works on the pixel values of an image and can therefore be applied to real mesostructure images obtained in an experimental context.\par
Continuing along a similar line of inquiry, in a recent study \cite{poncelet_2023} proposed a methodology that uses the diffraction diagram of either experimental or synthetic images to identify symmetry-breaking bifurcations within periodic media. However, their methodology is restricted to periodic media and focuses solely on the analysis of the diffraction diagram, i.e. the amplitudes of the Fourier coefficients, which provides partial information on symmetries. In contrast, the procedure proposed in this article offers a more comprehensive approach, since it relies on a thorough analysis of both the amplitude as well as the phase of the Fourier coefficients.
When studying the symmetries of a (quasi)periodic material, it is important not to limit oneself to its point group, but also to take into account the affine symmetries of the geometry, particularly their symmorphic or non-symmorphic character. A space group is symmorphic if its associated point group operations can be fully realised at a single spatial location, as opposed to being distributed across inequivalent positions within the unit cell. As highlighted in the literature, the symmorphic characteristics of a material have been found to significantly influence the formation of bandgaps in periodic materials \citep{Zhang2019} and the possibility for the creation of soft modes \citep{liu_2021,zheng_2023}. Additionally, materials that experience symmetry-breaking bifurcations may exhibit deformation patterns that differ solely in their symmorphic character \citep{Azulay2024}. Consequently, it is essential to develop a procedure capable of identifying the space group of a material, including its symmorphy.\\

The article is organised as follows. In the first section, we recall the fundamental concepts and definitions related to periodic media. This construction allows us to introduce the classical superposability-based symmetry group that characterises geometric invariance in direct space, commonly referred to as the \textit{space group}. Moreover, the concept of symmorphism is introduced in the context of periodic materials. The second section deals with the notion of quasiperiodic functions and examines their associate representation in Fourier space. The invariance properties of correlation functions are then presented, along with their implications in Fourier space. This allows us to extend the notion of superposability-based symmetry -- also referred to as \textit{strong symmetry} -- to indistinguishability-based symmetry -- also referred to as \textit{weak symmetry}. The third section is devoted to the algorithmic implementation of these concepts and the final section focuses on the application of this methodology to various periodic and quasiperiodic mesostructures.

%% file: periodic.tex
\section{Symmetry of periodic heterogeneous materials}\label{s:Period_Het_Mat}
As a first step, it is necessary to define the notion of a multidimensional periodic function. This will lead us to introduce two fundamental structures: the direct and the reciprocal lattice groups. This is the object of~\autoref{ss:notion_period}.
These concepts will then allow us to define, in~\autoref{ss:period_mat}, the geometry of a periodic material through a periodic function $\rho$. This function, which is not necessarily continuous, is referred to as the density function.
For periodic structures, a symmetry is defined as an affine transformation that leaves the density function invariant. This invariance can be interpreted geometrically as bringing the structure into coincidence with itself; it is an operation that is expressed directly in physical space. This notion of symmetry can be seen as describing short-range order. The long-range order of periodic structures is a consequence of this. By construction, the periodicity translations are symmetries of the geometry. However, the full symmetry group typically includes additional operations, such as rotations, reflections, and their combinations. The full set of these operations constitutes the \emph{space group} of the material. This is the object of~\autoref{ss:Sym_Inv}.

Before proceeding, we begin by introducing some notations and definitions related to the considered geometric framework. In this document, we denote by $\EE^d$ the $d$-dimensional Euclidean affine space, and by $\mathbb{V}^d$ its associated vector space. In the present context, we consider $d=2, 3$. Up to the choice of an origin $\mathrm{O}$, we may identify $\EE^d$ with $\mathbb{V}^d$, and similarly, a choice of basis yields a non-canonical identification $\mathbb{V}^d \simeq \mathbb{R}^d$. Throughout this document, Einstein's summation convention over repeated indices will be used.

\subsection{On the notion of periodicity}
\label{ss:notion_period}

To introduce the notion of periodic functions of several variables, let us begin by recalling some elementary facts from the one-dimensional case.
A real-valued function $f$ is said to be periodic if there exist non-zero numbers $v$ such that
\begin{equation*}
\forall x \in \mathbb{R},\quad f(x + v) = f(x).
\end{equation*}
$T$, called the fundamental period of $f$, denotes the minimal positive number that satisfies this property.
One can then define the following set $\latt := \{t=nT,\ n \in \mathbb{Z}\}$.
This set is isomorphic to $(\mathbb{Z}, +)$, the Abelian group of integers, with the trivial property:
\begin{equation*}
\forall x \in \mathbb{R},\ \forall t \in \latt,\quad f(x + t) = f(x).
\end{equation*}
\todoR{This property expresses that the set $\latt$ is composed of all the translations $t$ that leave the function $f$ invariant. These translations belong to the symmetry group\footnote{It is only a subgroup \todoR{of the full symmetry group of $f$ since the latter} may contain other elements for instance the centrosymmetry $x\mapsto-x$ if the function $f$ is even.} of $f$.}
\todoR{The same idea can be expressed by introducing} the translation operator $\mathcal{T}_{t}$ on functions\footnote{$\mathcal{T}_{t}$ is a representation of $(\mathbb{Z}, +)$ on the space of functions $f$.} such that
\ben
\forall x\in\RR,\quad \mathcal{T}_{t}[f(x)]:=f(x-t).
\een
\todoR{Within this formalism, the symmetry of $f$ with respect to $\latt$ can be expressed as} 
\ben
\forall t\in\latt,\quad \mathcal{T}_{t}[f(x)]=f(x)
\een

Let us now move to the multidimensional case for which periodicity exists along different directions. To formalise this idea, consider a family of $d$ linearly independent vectors $\{\vT_i\}$. And let $\latt$ denote again the set of their linear integer combinations
\beq
\latt:=\{\vt=n_i\vT_i\enspace,\enspace n_i \in \ZZ\}=\mathrm{Span}_{\mathbb{Z}}\left(\vT_i\right), i\in \{1,..,d\} .
\label{eq:latt}
\eeq
This set is a discrete non-compact Abelian group called the \emph{lattice group}\footnote{This structure is often referred to as a $\mathbb{Z}$-module. An R-module is a generalisation of the notion of vector space, in which the scalars are taken from a ring rather than a field. In particular, a $\mathbb{Z}$-module may be viewed as a "vector space" over the ring of integers.}.

The linearly independent vectors $\{\vT_i\}$ that define the lattice group are called the basis vectors of $\latt$. They correspond to directional fundamental periods. Their choice is not unique, since two bases linked by an element of the general linear group $\GLG(d,\ZZ)$ yield the same lattice group \citep{pitteri_2002}. Associated with the lattice basis vectors, a parallelepiped domain $\UC$ of $\EE^d$, known as \emph{the primitive unit cell}, can be defined (see \autoref{fig:latticepoint}):
\ben
\UC
=\{y_i\vT_i\enspace|\enspace 0\leq y_i <1\}.
\een

\begin{figure}[h!]
	\centering
	\begin{subcaptionblock}{0.47\linewidth}
        \centering
		\begin{tikzpicture}[scale=1]
			\foreach \i in {-2, ...,2}{
				\foreach \j in {-2, ...,2}{
					\draw[fill=black] (\i,\j) circle (0.08);
				}
			}
			\draw[line width=1pt,ForestGreen,-{stealth}](0,0)--(1,0)node[at end,anchor=north]{$\ve{T}_1$};
			\draw[line width=1pt,ForestGreen,-{stealth}](0,0)--(0,1)node[at end,anchor=east]{$\ve{T}_2$};
			\draw[line width=1pt,red] (-0.5,-0.5)rectangle (0.5,0.5);
			\node[red,anchor=north]at (0,-0.5){$\UC$};
		\end{tikzpicture}
		\caption{Square lattice group}
	\end{subcaptionblock}
	\begin{subcaptionblock}{0.47\linewidth}
        \centering
		\begin{tikzpicture}[scale=1.1]
			\tikzmath{\xone=1;\yone=0;\xtwo=0.5;\ytwo=sqrt(3)/2;};
			\def\idces{{{-2, 1}, {-2, 2}, {-1, -1}, {-1, 0}, {-1, 1}, {-1, 2}, {0, -2}, {0, -1}, {0, 0}, {0, 1}, {0, 2}, {1, -2}, {1, -1}, {1, 0}, {1, 1}, {2, -2}, {2, -1},{2,0},{-2,0}}}
			\foreach \i in {0, ..., 18}{
				\pgfmathsetmacro{\x}{\idces[\i][0]}
 				\pgfmathsetmacro{\y}{\idces[\i][1]}
				\fill[black] (\xone*\x+\xtwo*\y,\yone*\x+\ytwo*\y) circle (0.08);
			}
			\draw[line width=1pt,ForestGreen,-{stealth}](0,0)--(1,0)node[at end,anchor=north]{$\ve{T}_1$};
			\draw[line width=1pt,ForestGreen,-{stealth}](0,0)--(\xtwo,\ytwo)node[at end,anchor=east]{$\ve{T}_2$};

			\draw[line width=1pt,red] (-0.5*\xone-0.5*\xtwo,-0.5*\yone-0.5*\ytwo)--(-0.5*\xone+0.5*\xtwo,-0.5*\yone+0.5*\ytwo)--(+0.5*\xone+0.5*\xtwo,0.5*\yone+0.5*\ytwo)--(+0.5*\xone-0.5*\xtwo,0.5*\yone-0.5*\ytwo)--cycle;
			\node[red,anchor=north]at (0,-0.5){$\UC$};
		\end{tikzpicture}
		\caption{Hexagonal lattice group}
	\end{subcaptionblock}
	\caption{Examples of point sets of periodic lattices in $\RR^2$ together with their basis vectors ${\color{ForestGreen}\ve{T}_1}$, ${\color{ForestGreen}\ve{T}_2}$ and their associated unit cell $\UC$.
    }
    \label{fig:latticepoint}
\end{figure}
Similarly to the unidimensional case, a function $f$ is $\latt$-periodic if
\begin{equation*}
\forall \vt\in\latt,\quad \mathcal{T}_{\vt}[f(\vx)]:=f(\vx -\vt)=f(\vx),
\end{equation*}
with $\vx$ the vector position of a point $\mathrm{P}\in\EE^d$. 
$\latt$ is again a subgroup of the symmetry group of $f$. 

The lattice group $\latt$ induces the following equivalence relation between the points of $\EE^d$:
\ben
\vx \sim \vx'\qquad \Leftrightarrow \qquad\vx - \vx' \in \latt
\een
\todoR{which means that two points are said to be equivalent under the action of the lattice group $\latt$ if they are separated by a vector belonging to $\latt$.}
The equivalence class of point $\mathrm{P}$ with respect to the relation $\sim$ is called the $\latt$-orbit of $\mathrm{P}$. \todoR{It corresponds to the set of all the points equivalent to $\mathrm{P}$ under the action of the lattice group $\latt$.}
The primitive unit cell $\UC$, defined by the basis vectors $\{\vT_i\}$, is a fundamental domain for $\latt$ since it contains exactly one point per $\latt$-orbit. Moreover, the primitive unit cell contains all the information about the periodic function, which means that the whole function can be constructed from it.\par


This formalism has to be extended \todoR{from functions} to measures in order to generate \emph{point sets} of the Euclidean affine space. To that aim consider $\delta(\vx)$, the Dirac measure satisfying
\ben
\delta(\vx)=
\begin{cases}
1 \quad\text{if}\ \vx=\ve{0}\\
0 \quad\text{otherwise}
\end{cases}
.
\een
This measure indicates a point $\mathrm{P}$ localised at the origin $\vx_{0}$. The Dirac measure $\delta_{\ve{y}}$ associated with a point \todoR{located at $\ve{y}$ is defined by translation: $\delta_{\ve{y}}(\vx)=\mathcal{T}_{\ve{y}}[\delta(\vx)]=\delta(\vx-\ve{y})$.
Let us denote by $\Lambda_{P}$ the $\latt$-orbit of the point $P$ defined by
\ben
\Lambda_{P}=\{ \vx\in\EE^d, \vx=\vx_{0}+\vt \quad (\vx_{0},\vt) \in \UC\times\latt\}.
\een
}
This defines what is called a \emph{lattice point set}\footnote{It is important to distinguish the \textit{lattice group} from the \textit{lattice point set}, the latter being the geometrical expression of the former.}. 
The associated measure $\Pi_{\Lambda_{P}}$ is the Dirac comb
\ben
\Pi_{\Lambda_{P}}\defeq \sum_{y\in\Lambda_{P}}\delta_{\ve{y}}.
\een
Two examples of lattice point sets are presented in \autoref{fig:latticepoint} along with basis vectors of the parent lattice group and their associated primitive unit cells.\par




The structure of $\latt$-periodic functions is described by introducing the reciprocal lattice group, denoted by $\latt^{\star}$. To that end, let us consider the following elementary function
\beq
\label{eq:elemnt}
f_{\ve{k}}(\vx) = \e^{2 \iu \pi \ve{k}\cdot\vx}.
\eeq
\todoR{This function is an eigenfunction for the translation operator $\mathcal{T}_{\vt}$ since \beq\label{eq:egein}
    \mathcal{T}_{\vt}[f_{\ve{k}}(\vx)]:=f_{\ve{k}}(\vx -\vt)=\e^{-2 \iu \pi \ve{k}\cdot\vt}f_{\ve{k}}(\vx).
\eeq
}
\todoR{where $\e^{-2 \iu \pi \ve{k}\cdot\vt}$ acts as an eigenvalue for the translation operator. Requiring the $\latt$-periodicity of the elementary function $f_{\ve{k}}$, $\mathcal{T}_{\vt}[f_{\ve{k}}(\vx)]=f_{\ve{k}}(\vx)$, imposes the following condition on the reciprocal vector $\ve{k}$ }
\ben
\forall \vt \in \latt,\quad \ve{k} \cdot \vt \in \ZZ.
\een
This condition is satisfied for all $\ve{k}$ of the form
\ben
\ve{k} = \nu_{i} \vT^{\star}_{i},\quad \nu_{i} \in \ZZ,\qquad \vT^{\star}_{i} \cdot \vT_{j} = \delta_{ij}
\een
Hence, from the basis $\vT_{j}$ of $\latt$, a basis $\vT^{\star}_{i}$ of the reciprocal lattice group $\latt^{\star}$ is defined so that
\ben
\latt^{\star}=\{\nu_{i} \vT^{\star}_{i}\enspace|\enspace \alpha_i \in \ZZ\}=\mathrm{Span}_{\mathbb{Z}}\left(\vT^{\star}_{i}\right), i\in \{1,..,d\}.
\een

\noindent It results that, by $\mathbb{C}$-linearity, the general shape of a $\latt$-periodic function is constructed from the eigenfunctions introduced in~\autoref{eq:elemnt} as
\beq\label{eq:perfunc}
f(\vx) = \sum_{\vk\in \dual{\latt}} c_{\vk}\e^{2\iu \pi \vk\cdot \vx},
\eeq
in which $c_{\vk}\in\mathbb{C}$.
Conversely, starting from a periodic function $f$, the coefficients $c_{\vk}$ in \autoref{eq:perfunc} are determined by the Fourier transform  of $f$.
Formally\footnote{The Fourier transform is introduced and used here in a formal manner. Strictly speaking, one should distinguish between functions and distributions and address the associated convergence issues. These subtleties are deliberately set aside, and we refer the reader to \citet{baake_2013} for a more rigorous presentation of this tool.}, the Fourier transform $\hat{f}$ of a function $f$ is defined as
\ben
    \mathcal{F}\left[f\right](\vk) = \widehat{f}(\vk)\defeq \int_{\RR^{d}} f(\vx) \e^{-2 \iu \pi  \vk\cdot\vx}\mathrm{d} \vx,
    \label{eq:fourierdef}
\een
with inverse transform
\ben
    \mathcal{F}^{-1}\left[{\widehat{f}}\right](\vx) =
    f(\vx)
    \defeq\int_{\RR^{d}}
\widehat{f}(\vk) \e^{2 \iu \pi  \vk.\vx} \mathrm{d} \vk.
\een
In the case where $f$ is a real-valued function its Fourier transform is hermitian, that is,
\ben
    \widehat{f}(-\vk)=\widehat{f}^*(\vk),
\een
with $\widehat{f}^*$ the complex conjugate of $\widehat{f}$.



\subsection{Heterogenous periodic material}\label{ss:period_mat}
\todoR{Using the previously defined notions of multidimensional periodic functions, it is now possible to describe the structure of a heterogeneous periodic material using its density function. For the sake of simplicity, this article is restricted to the case of a heterogeneous two-phase material but the previously introduced formalism can be applied to any periodic heterogeneous material.}
Let us consider an infinite two-phase material occupying a domain denoted by $\Omega$. Each constituent (or phase) $r$ ($r=1,2$) has a domain $\Omega_r$ satisfying
\beq\label{eq:bondedphase}
\begin{cases}
 \Omega_{1}\cup\Omega_{2}=\Omega\\
 \Omega_{1}\cap\Omega_{2}=\varnothing
\end{cases}.
\eeq
It results that $\Omega_{2}$ can simply be defined as $\Omega/ \Omega_{1}$ the complement of $\Omega_{1}$ in $\Omega$.
In the sequel, the density function $\rho$ is defined as the characteristic, or indicator, function of the domain $\Omega_1$, that is
\beq
\rho(\vx):=\mathrm{I}_{\Omega_{1}}(\vx)=
\begin{cases}
	1 &\text{ if } \vx\in\Omega_1\\
    0 &\text{ otherwise} \\
    \end{cases}.
    \label{eq:charfun}
\eeq
The density function $\rho$ allows to describe the whole domain $\Omega$ since, with the assumptions listed in \autoref{eq:bondedphase}, $\mathrm{I}_{\Omega_{1}}(\vx)+\mathrm{I}_{\Omega_{2}}(\vx)=1,\,\forall \vx\in \Omega$.
Within this framework, a heterogeneous material is said to be $\latt$-periodic if its density function $\rho$ is itself $\latt$-periodic:
\beq\label{eq:period}
    \forall{\vx} \in {\RR^{d}},\  \forall{\vt} \in {\latt} \quad \rho(\vx+\vt)=\rho(\vx).
\eeq
\todoR{Moreover, as stated in the previous subsection,} since the density function $\rho$ is $\latt$-periodic, it is entirely determined by its values within a primitive unit cell of $\latt$. \todoR{Let $\varrho$ denote the restriction of the density function to a primitive unit cell $\UC$, that is $\varrho(\vx)=\rho(\vx),\;\forall \vx\in\UC$ and 0 elsewhere. $\varrho$ describes the \emph{elementary periodic pattern} of the heterogeneous material}.
From this restriction, the density function \todoR{can be} expressed as a convolution between the elementary periodic pattern and the Dirac comb associated to the lattice point set $\Lambda$:
\ben
\rho(\vx)=\left(\varrho\ast\Pi_{\Lambda}\right)(\vx).
\een
The interest of this latter formulation is that it allows a direct formal construction and interpretation of the Fourier transform $\hat\rho(\ve{k})$ of the density. Indeed, applying the convolution theorem \citep{baake_2013} gives
\beq
    \hat\rho(\ve{k}) = \hat{\varrho}(\ve{k})\  \hat{\Pi}_{\Lambda}(\ve{k}),
    \label{eq:fourcoeff-periodic}
\eeq
with $\hat{\varrho}$ the Fourier transform of $\varrho$ and $\hat{\Pi}_{\Lambda}$ the Fourier transform of the Dirac comb defined by
\ben
\hat{\Pi}_{\Lambda} = \Pi_{\Lambda^{\star}}
\een
where $\Lambda^{\star}$ is the lattice point set associated with the reciprocal lattice group $\latt^{\star}$.

\subsection{Superposability-based symmetry}\label{ss:Sym_Inv}

\todoR{In the previous subsections, periodicity, i.e. invariance under translations, was the only symmetry operation considered. The lattice group $\latt$ was thus identified as a subgroup of the symmetry group of $f$. We now introduce the notion of superposability-based symmetry to define the full symmetry of a heterogeneous material. }\todoR{This is the classical notion of symmetry.}

Let $\E{d}$ denote the group of Euclidean motions in dimension $d$. This group can be expressed as the semi-direct product:
\ben
\E{d} = \mathrm{O}(d) \ltimes \mathbb{R}^d
\een
where:
\begin{itemize}
\item $\mathrm{O}(d)$ is the orthogonal group, consisting of all linear isometries $\dT{Q}$ of $\mathbb{R}^d$. It includes:
\begin{itemize}
\item proper orthogonal transformations with $\det(\dT{Q}) = 1$, corresponding to rotations;
\item improper orthogonal transformations with $\det(\dT{Q}) = -1$, corresponding to reflections and compositions of reflections with rotations.
\end{itemize}
\item $\mathbb{R}^d$ represents the group of translations.
\end{itemize}
\todoR{An affine transformation $\gelem \in \E{d}$ is classically} \citep{mcweeny_1963} denoted by the pair $\gelem = (\dT{Q} \mid \ve{v})$, where $\dT{Q} \in \mathrm{O}(d)$ and $\ve{v} \in \mathbb{R}^d$ represents the translational part. 
Its action $\star$ on a point of $\EE^d$ with coordinates $\vx$ is given by
\ben
\gelem\star\vx=\dT{Q}\cdot\vx+\ve{v}.
\een
Denoting by $\gelem^{-1}$ the inverse of $\gelem$, \todoR{the following} action can be defined on functions as\footnote{\todoR{Since $\vx$ and $f(\vx)$ belong to different vector spaces, the group action of $\gelem \in  \E{d}$ on each of them are necessarily different. We chose to keep the same notation for both, however, for the sake of readability, as is customary in the mathematical community.}}
\ben
\gelem\star f(\vx) := f(\gelem^{-1}\star \vx).
\een

\todoR{For a heterogeneous material, the invariance of the density function is referred to as \textit{superposability}, and the associated transformations are called \textit{strong symmetries}.  Invariance under superposition means that affine transformations exist that maps the material exactly onto itself.} The affine symmetry group of a material $\mathcal{S}$ is therefore defined as the set of affine isometries leaving the density function invariant:
\ben
\mathcal{S}:=\{\gelem\in\E{d}|\ \gelem\star \rho(\vx) = \rho(\vx) \}.
\een
This group is called the \emph{space group}; it is a discrete subgroup of $\E{d}$.
\\ \par




Space groups are used in crystallography to describe the geometric arrangement of perfect crystals. They are obtained as the combination of two groups:
\begin{enumerate}
    \item the point group $\mathcal{P}\subset\text{O}(d)$, which contains proper and improper rotations, \todoR{termed point symmetries, since they leave at least one point of the space invariant;}
    \item the lattice group $\latt\subset\RR^d$, which characterises the discrete translations that leave the material invariant.
\end{enumerate}

\begin{figure}[h!]
    \centering
    \begin{subfigure}[c]{0.48\linewidth}
    \centering
    \includegraphics[width=.7\textwidth]{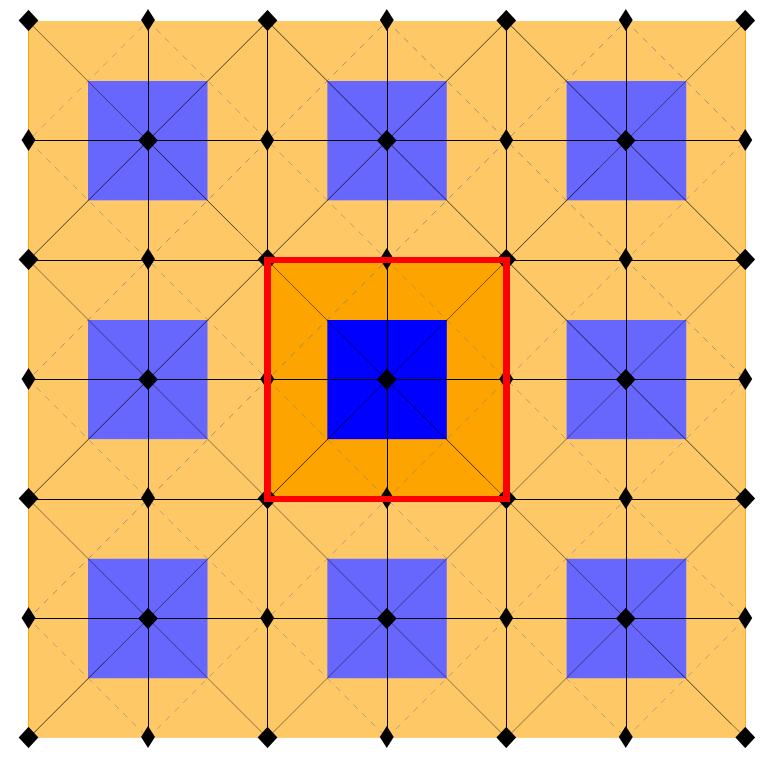}\caption{p4mm}
     \label{fig:p4m_s}
    \end{subfigure}
    \begin{subfigure}[c]{0.48\linewidth}
    \centering
     \includegraphics[width=.7\textwidth]{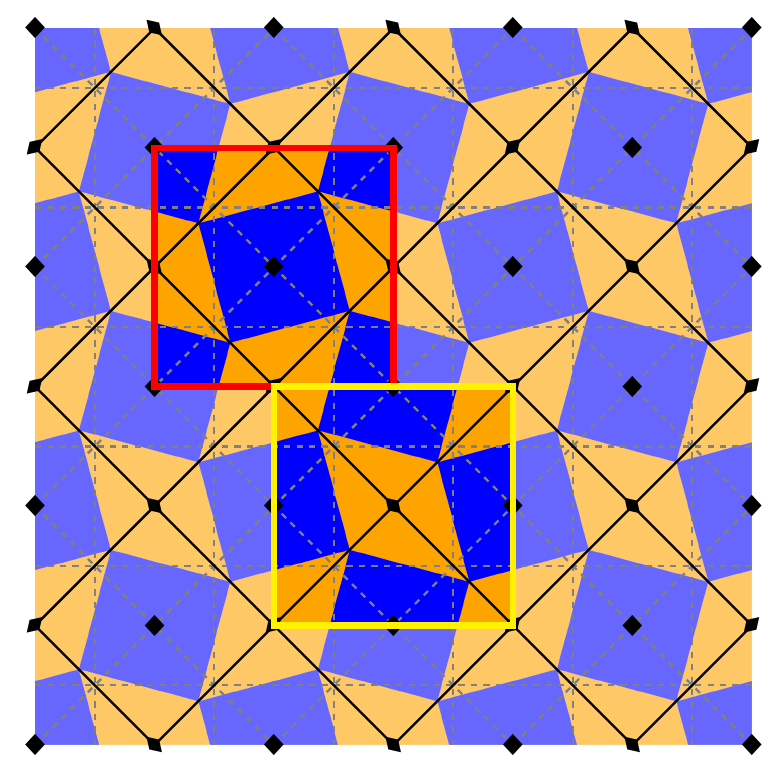}\caption{p4g}
     \label{fig:p4g_s}
    \end{subfigure}
    \caption{Two tilings with different space groups but identical point group $\DD{4}$. $\Diamondblack$ denotes centres of rotation of order 4, $\blacklozenge$ centres of rotation of order 2, solid black and dashed gray lines represent mirror and glide (mirror combined with translation) lines, respectively. \todoR{Possible primitive unit cells are highlighted in red (p4mm) or red and yellow (p4g).}}
    \label{fig:p4m-p4g-comp}
\end{figure}

The lattice group $\latt$ has a quite simple definition as it is the subgroup of $\mathcal{S}$ that contains only pure translations represented by elements $(\te{I}~|~\vt)$\footnote{A distinction is introduced here between arbitrary translations $\ve{v}\in \mathbb{R}^d$ and translations that are elements of the lattice group $\vt\in\latt$ as defined in \autoref{eq:latt}.}.
The point group is less direct to define. Formally, it is isomorphic to the quotient of the space group $\mathcal{S}$ by the lattice group $\latt$, i.e. $\mathcal{P}\cong\mathcal{S}/\latt$.
Practically, this means that the point group can be defined by setting to zero the translational part $\ve{v}$ of the space group elements $\left(\te{Q}~|~\ve{v}\right)$. Note that there is no requirement for the point group $\mathcal{P}$ to be a subgroup of $\mathcal{S}$ \citep{hiller_1986}. \todoR{Based on this criterion, space groups fall into two categories:}
\begin{itemize}
    \item \emph{symmorphic}: the point group is a subgroup of the space group. Then, there exists a point in the primitive unit cell where all the elements of the point group are realised (\autoref{fig:p4m_s}). 
    \todoR{This corresponds to the simplest situation, in which the point groups of the tiling and of the primitive unit cell coincide.}
    \item \emph{non-symmorphic}: the point group is not a subgroup of the space group. 
    Non-symmorphic space groups contain elements $\gelem=\left(\te{Q}~|~\ve{v}\right)$ with $\ve{v}\notin \latt$. \todoR{These elements correspond to glide mirrors or screw axes (in 3D).}
   In such a case, the elements of the point group are realised at distinct points of the primitive unit cell. For instance, for a non-symmorphic space group with mirror planes and axis of rotational symmetries, these elements of symmetry (mirror planes and rotation axis) will not all intersect at the same point of the primitive unit cell as would be the case for a symmorphic space group (\autoref{fig:p4g_s}).
\end{itemize}


\todoR{To illustrate the concepts introduced above, consider the two materials whose images are displayed in \autoref{fig:p4m-p4g-comp}. Although they share the same lattice group $\latt$ and point group $\mathcal{P}$,  they belong to different space groups $\mathcal{S}$: the first (\autoref{fig:p4m_s}) is p4mm, which is symmorphic, while the second (\autoref{fig:p4g_s}) is p4g, which is non-symmorphic.}


\todoR{
These mesostructure share the same lattice group $\latt$, one primitive cell of which is delimited by a red or yellow border. On theses figures, black lines represent mirror axes while a black diamond, resp. rhombus, symbol represents a centre of rotation of order 4, resp. order 2.
}
\todoR{An analysis of the primitive unit cells highlighted in red suggests that the point group of the first tiling  is $\DD{4}$, whereas it is $\ZZZ_4$ for the second.}
\todoR{However, in this latter case, the primitive cell highlighted in yellow would instead suggest $\DD{2}$ as the point group. This discrepancy arises from the fact that, in the p4g tiling, the mirror axes intersect at a twofold rotational center represented by a rhombus symbol rather than at the fourfold center. Moreover, in this case, the horizontal and vertical mirrors only exist as glide mirrors combining mirror and translation symmetries and represented with dashed lines. All these peculiarities come from the non-symmorphic nature of the tiling.}
\todoR{
To be fully explicit, let us detail the set of generators of each space group. To this end, 
$\vp{h}$ denotes the horizontal mirror, $\vp{r}_{n}$ the rotation of angle $\frac{2\pi}{n}$ and $\vT_{i}$ the lattice basis vectors. We then have:
 \begin{enumerate}
     \item $\{\left(\dT{I}~|~\vT_{1}\right),\left(\dT{I}~|~\vT_2\right),\left(\dT{r}_4~|~0\right),\left(\vp{h}~|~0\right)\}$. One can recognise in these generators the generators of the lattice group $\latt=\{\left(\dT{I}~|~\vT_{1}\right),\left(\dT{I}~|~\vT_{2}\right)\}$ and of the point group $\mathcal{P}=\DD{4} =\{\left(\dT{r}_4~|~0\right),\left(\vp{h}~|~0\right)\}$. Both these groups are then subgroups of the space group of the material which is the symmorphic space group p4mm.
     \item $\{\left(\dT{I}~|~\vT_{1}\right),\left(\dT{I}~|~\vT_{2}\right),\left(\dT{r}_4~|~0\right),\left(\vp{h}~|~\frac{(\vT_{1}+\vT_{2})}{2}\right)\}$. One can recognise in these generators the generators of the lattice group $\latt=\{\left(\dT{I}~|~\vT_{1}\right),\left(\dT{I}~|~\vT_{2}\right)\}$. However, the generators of the point group $\mathcal{P}=\DD{4}$ 
    do not appear as generators of the space group. Consequently, only the lattice group is a subgroup of the space group of this material. This space group is the non-symmorphic space group p4g. Remember that it has been said that the point group can be defined by setting to zero the translational part $\ve{v}$ of the space group elements $\left(\te{Q}~|~\ve{v}\right)$. 
    If the translational part of these glide elements is set to zero, then the point group appears to be  $\DD{4}$.
 \end{enumerate}}

%% file: quasiperiodic.tex
\section{Symmetry of quasiperiodic heterogeneous materials}

Based on the classical notion of symmetry for periodic media, the case of quasiperiodic materials is now addressed. \todoR{In particular, we highlight that a natural framework for describing their effective physical properties is provided by $n$-point correlation functions. Although traditionally used for random materials, this tool is here applied to the characterization of deterministic mesostructures.}

\subsection{Heterogeneous quasiperiodic material}

\begin{figure}[H]
    \centering
        \includegraphics[width=0.5\textwidth]{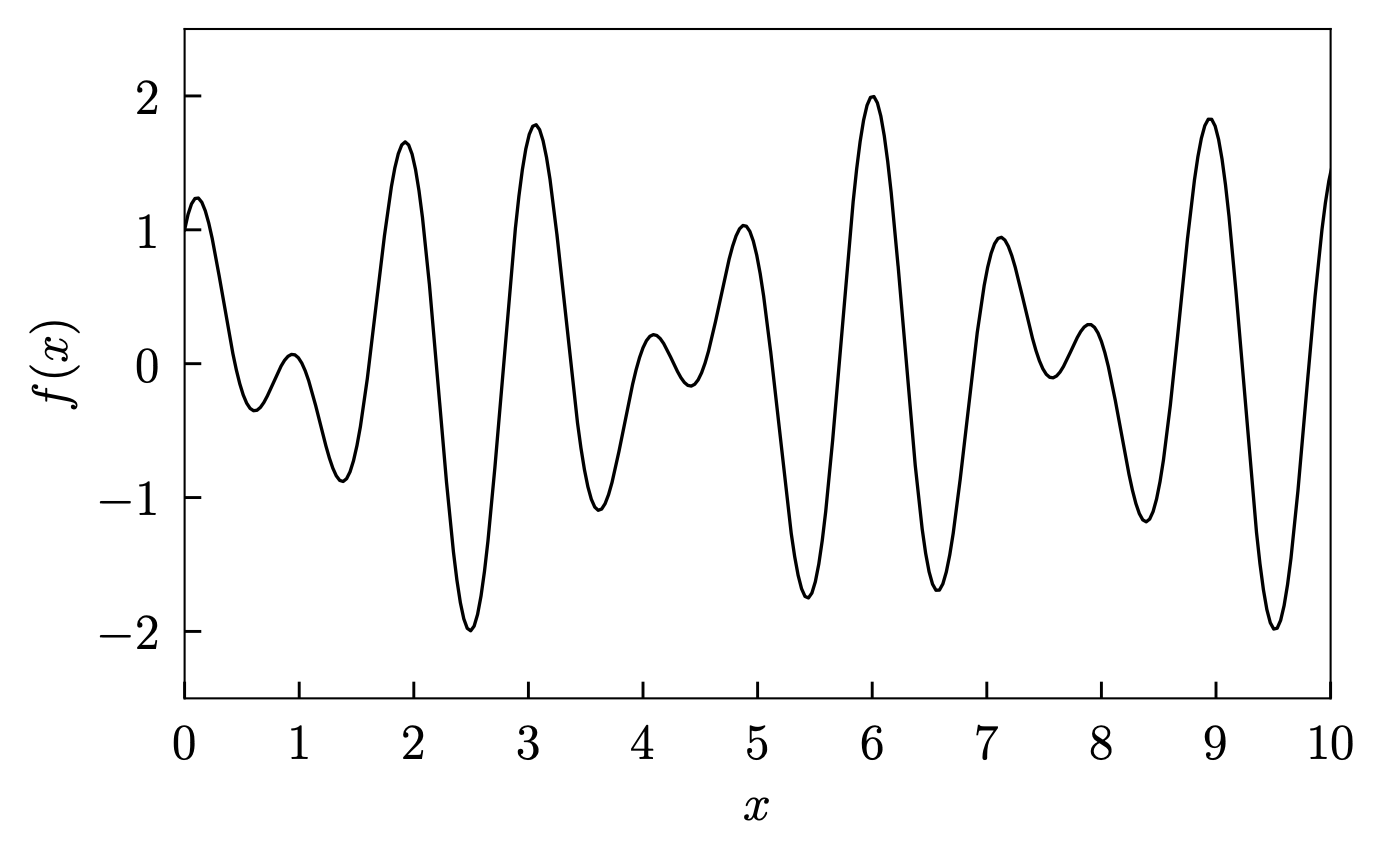}
        \caption{Illustrative example of a quasiperiodic function $f_\text{qp}(x)=\cos{2\,\pi x}+\sin{\sqrt{2}\,\pi x}$. The two incommensurate periods are ($T_1=1,\,T_2=\sqrt{2}$) and the resulting function is aperiodic. }
        \label{fig:quasifunc}
    \end{figure}

As with periodic functions, we begin by introducing the notion of quasiperiodic functions in the one-dimensional setting.
To do so, let us consider a periodic function $f_\text{p}$ with period $T$. Its Fourier series reads
\beq\label{eq:1Dperfun}
f_\text{p}(x) = \sum_{m=-\infty}^{+\infty} c_m \e^{2\iu \pi k\,x}\quad \text{with}\quad c_m = \frac{1}{T}\int_{0}^{T} f_\text{p}(x)\e^{-2\iu \pi k\,x}\mathrm{d} x,\quad k=\frac{m}{T},
\eeq
and it satisfies, by definition, $f(x+nT)=f(x),\:\forall x\in \RR,\:\forall n\in \ZZ$.
A quasiperiodic function $f_\text{qp}$
is a generalization of \autoref{eq:1Dperfun} in which the wave numbers $k$ are no longer multiples of a fundamental frequency, but integer linear combinations of $m_i$ incommensurate\footnote{\todoR{Incommensurate here means that the ratio between the frequencies is irrational.}} frequencies $\frac{1}{T_i}$ \citep{bohr_1926}, that is
\beq
f_\text{qp}(x) = \sum_{m_1=-\infty}^{+\infty}\ldots\sum_{m_\mu=-\infty}^{+\infty} c_{m_1\ldots m_\mu} \e^{2\iu \pi (k_1+\ldots+k_\mu)x},\quad k_i=\frac{m_i}{T_i},\quad
m_i \in \ZZ,
\eeq
with $c_{m_1\ldots m_\mu}$ the Fourier-Bohr coefficients. \todoR{An example of a quasiperiodic function defined by two Fourier-Bohr coefficients is given in~\autoref{fig:quasifunc}.}
\noindent Within this framework, the density function $\rho(\vx)$ of a \todoR{$d$-dimensional quasiperiodic heterogeneous material takes the form}
\beq
     \label{eq:qpdef}
     \rho(\vx)= \sum_{\vk \in \mathcal{M}} \hat{\rho}(\vk) \e^{2 \iu\pi \vk\cdot \vx}. 
 \eeq
 \todoR{
 Contrary to the periodic case, there exists no lattice group corresponding to translations in real space. However, a quasiperiodic material can be defined via a lattice group in the reciprocal space. Accordingly, the reciprocal lattice group
 \ben
    \mathcal{M}=\mathrm{Span}_{\mathbb{Z}}\left(\vk_i\right), i\in \{1,..,\mu\}.
 \een
 is defined directly using the $\mu$ fundamental frequencies $\vk_i$. For a quasi-periodic function, the (reciprocal) lattice group forms a $\mathbb{Z}$-module of rank $\mu$, strictly larger than the physical dimension $d$.} The symmetry group of $\mathcal{M}$ will be referred to as the  \emph{holohedry} in the following. \todoR{It corresponds to the symmetry group of the reciprocal lattice point set\footnote{In crystallography, the holohedry is defined as the symmetry group of the lattice point set. In the quasiperiodic case, however, no such lattice exists, and the definition is therefore formulated in the reciprocal space.}.}\\

The complex Fourier(-Bohr) coefficients of the quasiperiodic density $\rho(\vx)$ are defined by the complex valued function
\beq\label{eq:fourcoeff}
    \hat{\rho}(\vk) =\displaystyle \lim_{\Omega \rightarrow \infty} \frac{1}{|\Omega|} \int_\Omega \rho(\vx) \e^{-2 \iu \pi \vk\cdot\vx} \mathrm{d} \vx,\quad \forall \vk\in\mathcal{M}.
\eeq
\todoR{The above equation generalises the conventional notion of the Fourier coefficient of a periodic function to the Fourier-Bohr coefficient of a quasiperiodic function by integrating over a domain $\Omega$ going to infinity instead of the finite sized primitive unit cell used in~\autoref{eq:fourcoeff-periodic}. For the sake of conciseness, the Fourier-Bohr coefficients are simply referred to as Fourier coefficients in the following.}

\todoR{\subsection{On the notion of indistinguishability}}
\label{sec:indisting}
In contrast to periodic materials, there is no physical translation that leaves invariant the density $\rho$ of a quasiperiodic material.  The notion of symmetry must therefore be addressed differently.
Following developments in condensed matter physics \citep[see, for instance][]{mermin_1992}, \todoR{when interested by the physical properties of a (quasi)periodic material,} the invariance of \todoR{this material} is more appropriately characterised in terms of its macroscopic tensorial 
properties rather than its mesoscopic geometry.


\todoR{The effective physical properties of a heterogeneous medium can be expressed as infinite series involving mesostructural correlation functions of successive orders \citep{brown_1955, torquato_2002, milton_2002}. The quantities relevant for characterising the macroscopic tensorial properties are therefore the set of $n$-point density autocorrelation functions, defined as \citep{torquato_2002}
\ben
	C_n(\ve{x}_1,\ldots,\,\ve{x}_n)=\lim_{\Omega \rightarrow \infty} \frac{1}{|\Omega|}\int_\Omega \rho(\ve{x}_1-\vx) \,\ldots\, \rho(\ve{x}_n-\vx) \mathrm{d} \vx.
\een
Within the homogenisation framework, the macroscopic continuum is assumed to be homogeneous, meaning that its effective properties are independent of the macroscopic point considered. A direct consequence of this translational invariance is that the functions  $C_n$ only depend on the relative positions $\ve{x}_i-\ve{x}_j$.
From a physical perspective, the correlation functions admit a natural statistical interpretation. The one-point correlation function  $C_1$  represents the volume fraction of phase 1 ($C_1=|\Omega_1|/|\Omega|$), whereas  $C_n(\ve{x}_1,\ldots,\,\ve{x}_n)$ gives the probability that $n$ points located at $\ve{x}_1,\ldots,\,\ve{x}_n$ belong to phase 1.}

\todoR{A new notion of symmetry, known as \emph{indistinguishability}, based on these autocorrelation functions and relative to their invariance with respect to orthogonal transformations, is now introduced.}
Two media with density functions $\rho$ and $\rho'$ are said to be \textit{indistinguishable} if their autocorrelation functions $C_n$ coincide for all orders $n$ \citep{rokhsar_1988,lifshitz_1996}. As already mentioned, this implies that the corresponding materials display identical effective properties \citep{milton_2002}.
This criterion is less restrictive than the invariance of the density functions themselves. As a consequence, two media which are superposable by an isometry are indistinguishable but the converse is not true.\par
\todoR{From a symmetry perspective, correlation functions are invariant under translations but not, in general, under rotations. In other words, while a translated density is indistinguishable from the original, this property does not generally hold for arbitrary rotations.
The limiting case where correlation functions are rotationally invariant at all orders $n$ has been considered in the literature \citep{kroner_1977} and is termed \textit{statistical isotropy of infinite order}. This notion plays a central role in deriving bounds on the effective elastic properties of isotropic random media and provides a classical example of indistinguishability in solid continuum mechanics.}

In the following, we will refer to indistinguishability as \emph{weak symmetry} and the relationship between indistinguishable densities is noted $\rho\sim\rho^\prime$. \todoR{As shown in~\citep{mermin_1992}, the coincidence condition for the correlation functions of two densities $\rho$ and $\rho'$ at arbitrary order can be expressed as a set of relations between their Fourier coefficients}\footnote{The vectors $\vk^{(i)}$ denote arbitrary elements of $\mathcal{M}$ and should not be confused with the fundamental frequencies $\vk_i$.}
\begin{equation}\label{eq.CondFour}
\forall n\in \mathbb{N},~
\hat{\rho}'(\vk^{(1)})\hat{\rho}'(\vk^{(2)})\,\ldots\,\hat{\rho}'(\vk^{(n)})=\hat{\rho}(\vk^{(1)})\hat{\rho}(\vk^{(2)})\,\ldots\,\hat{\rho}(\vk^{(n)}),\quad \forall (\vk^{(1)},\ldots,\vk^{(n)})\in \mathcal{M}
\end{equation}
under the condition that $\sum_{i=1}^n \vk^{(i)}=\mb{0}$. \todoR{This condition is the Fourier-space expression of the translational invariance of the correlation functions, which reflects the homogeneity of the macroscopic properties.
\ben
\forall \ve{v}\in\RR^d,\quad
	C_n(\ve{x}_1+\ve{v},\ldots,\,\ve{x}_n+\ve{v})=C_n(\ve{x}_1,\ldots,\,\ve{x}_n)
\een
Thus, in Fourier space, a translation introduces a phase factor $e^{2\iu \pi (\vk^{(1)}+\dots+\vk^{(n)})\cdot \ve{v}}$. For the correlation functions $C_{n}$
 to remain invariant, this phase must be equal to 1 for any translation vector $\ve{v}$, which leads to the stated condition.
Despite their simpler form, the Fourier indistinguishability relations \autoref{eq.CondFour} must hold at every order $n$, rendering this criterion impractical to verify. However,
it has been shown~\citep{mermin_1992} that, generically, the indistinguishability conditions in Fourier space can be reduced to the following conditions}
\beq\label{eq:gauge}
\hat{\rho}'(\vk)=\e^{2 \iu \pi  \chi(\vk)}\hat{\rho}(\vk),
\eeq
with $\chi$ a linear gauge function (to within an integer), meaning that
\beq\label{eq:linearcond}
\chi(\vk^{(1)})+\chi(\vk^{(2)})\equiv\chi(\vk^{(1)}+\vk^{(2)}),\quad\forall (\vk^{(1)},\vk^{(2)})\in \mathcal{M},
\eeq
where $\equiv$ denotes equality modulo 1.
By linearity, a gauge function $\chi$ acting on a (quasi)periodic density ${\rho}$ with fundamental frequencies $\vk_1,..,\vk_{\mu}$ is completely determined by its values on the fundamental frequencies, since
\beq
\chi(\vk=m_1\vk_1+..+m_{\mu}\vk_{\mu})\equiv
m_1\chi(\vk_1)+..+m_{\mu}\chi(\vk_{\mu}), \quad
\forall
\vk \in \mathcal{M}.
\label{eq:gaugefunextrapolation}
\eeq
\todoR{Rigorously, these conditions must be tested for every frequency $\vk \in \mathcal{M}$ in order to be satisfied. In practice, however, they will only be checked for a finite number of frequencies as discussed in details in \autoref{subsubfreq}.}
\todoR{It should be noted that the condition given by \autoref{eq:gauge} may be satisfied even if \autoref{eq:linearcond} is not. In particular, the first condition implies $|\hat{\rho}'|=|\hat{\rho}|$, that is, the agreement of the diffraction patterns. Therefore, the agreement of the diffraction patterns is a necessary but not sufficient condition for indistinguishability.}

The gauge \autoref{eq:gauge} should be compared with \autoref{eq:egein}, revealing that a gauge transformation can be interpreted as a generalised translation.\footnote{\todoR{This result links the Fourier-based and hyperspace descriptions of quasicrystals \citep{duneau_1985}, which are not discussed here. }}
\todoR{More precisely, any gauge function $\chi(\vk)$
can be decomposed into a part related to a $d$-dimensional rigid body displacement $\ve{u}$ and, if $\mu>d$, a remaining component $\ve{\psi}$ called \textit{phason}~\citep{lifshitz_2011} as $\chi(\vk)=\frac{\ve{u}\cdot \vk}{2 \pi}+\ve{\psi}(\vk)$.
In the periodic case with $\mu=d$, the phason component $\ve{\psi}$ vanishes and any gauge functions is determined by only the rigid body displacement $\ve{u}$.
\todoR{It follows by~\autoref{eq:perfunc},} that for any two indistinguishable periodic densities $\rho$ and $\rho^{\prime}$, there exists a displacement vector $\ve{u}$ such that $\rho^{\prime}(\ve{x})=\rho(\ve{x}+\ve{u})$, i.e $\rho$ and $\rho^\prime$ are superposable in real space. This shows that the criterion of indistinguishability reduces to superposability for periodic densities.}\\


\todoR{\subsection{Indistinguishability-based symmetry}}
\label{sec:genspacegroup}
The preceding definitions of indistinguishability and gauge functions allow to generalise the notion of space groups to the quasiperiodic case.  The gauge \autoref{eq:gauge} allows to define the full set of symmetry operations of the correlation functions which corresponds to the set of mesoscopic transformations that leave the macroscopic properties invariant.

An orthogonal transformation $\mb{Q}$ is said to be a \textit{weak symmetry} of a density function $\rho$ if
\beq
\label{eq:phasefundef}
\hat{\rho}(\mb{Q}\vk)=\e^{2 \iu \pi  \Phi_{\mb{Q}}(\vk)}\hat{\rho}(\vk),
\eeq
with phase functions $\Phi_{\mb{Q}}$ satisfying the gauge-linearity condition introduced in \autoref{eq:linearcond}\footnote{For periodic materials, the phase function reads $\Phi_\mb{Q}(\mb{k})=-(\mb{Q}\mb{k})\cdot\mb{v}$ for each element $\left(\mb{Q}~|~\ve{v}\right)$ of the space group $\mathcal{S}$.}.
Accordingly, \todoR{weak} symmetry transformations leave the Fourier density invariant up to a gauge function. \todoR{Thus, in Fourier space, phase functions replace translations, which are encoded by the lattice group in a direct-space description, whenever such a description is possible.}\par

 \todoR{A phase function is a linear mapping from $\mathcal{M}$ to $\mathbb{R}/\mathbb{Z}$, i.e., the real numbers modulo $1$, satisfying
\ben
\Phi(\vk^{(1)}+\vk^{(2)})\equiv\Phi(\vk^{(1)})+\Phi(\vk^{(2)}),\quad\forall (\vk^{(1)},\vk^{(2)})\in \mathcal{M},
\een
Such a mapping can be regarded as a group homomorphism from $(\mathcal{M}, +)$ to $(\mathbb{R}/\mathbb{Z}, +)$. We denote by $\mathcal{F}^\star = \mathrm{Hom}(\mathcal{M}, \mathbb{R}/\mathbb{Z})$ the set of these homomorphisms, which forms a abelian group, the group of gauge functions.}

The generalised space group $\mathcal{S}^\star$\todoC{\footnote{The $\star$ notation has been chosen since the weak symmetry operations are defined with respect to values in the reciprocal space.}} is the set of elements $\gelem=\lbrace\mb{Q}~|~\Phi_\mb{Q}\rbrace$ with group operation
\beq
\lbrace\mb{Q}~|~\Phi_\mb{Q}\rbrace\lbrace{\mb{Q}'}~|~\Phi_{\mb{Q}'}\rbrace\equiv \lbrace\mb{Q}{\mb{Q}'}~|~\Phi_{\mb{Q}\mb{Q}'}\rbrace
\label{eq:groupop}
\eeq
where
\ben
\Phi_{\mb{Q}\mb{Q}'}(\mb{k})\equiv\Phi_{\mb{Q}}(\mb{Q}'\mb{k})+\Phi_{\mb{Q}'}(\mb{k}).
\een

\todoR{Similarly to the periodic case, this generalised space group is defined by combination of two groups: $\mathcal{P}^\star$ and $\mathcal{H}^\star$, in which $\mathcal{H}^\star$, the group of \textit{phase functions}, is a discrete subgroup of $\mathcal{F}^\star$.} 
The point group $\mathcal{P}^{\star}$ of a quasiperiodic material can be defined by setting to zero the phase shift $\Phi_\mb{Q}$ of the generalised space group elements $\lbrace\mb{Q}~|~\Phi_\mb{Q}\rbrace$ 



\todoR{The concept of symmorphism, as introduced in the context of direct space, can be naturally formulated in reciprocal space. In doing so, it allows this notion -- originally associated with periodic media -- to be extended to quasiperiodic media. In the direct space,} a space group is said to be symmorphic if there exists a point in at which the entire point group is realised. In other words, invariance is achieved without requiring any translation. In Fourier space, this condition reads
\ben
\hat{\rho}(\mb{Q}\ve{k})=\hat{\rho}(\ve{k}),\quad \forall \mb{Q}\in\mathcal{P}^\star,
\een
meaning that the Fourier coefficients are left invariant under the action of any point group element. \todoR{In other words, indistinguishability is achieved without requiring any phase shift.}
It results, that a generalised space group is said to be \textit{symmorphic} provided there exists a gauge function $\chi$ since they \beq\label{eq:ConSym}
\Phi_{\mb{Q}}(\mb{k}) +\chi(\mb{Q}\mb{k}-\mb{k}) \equiv 0,\quad \forall \mb{Q}\in\mathcal{P}^\star.
\eeq
\todoR{In this case, the Fourier-space group is obtained as
 \begin{equation}
 \mathcal{S}^\star = \mathcal{P}^\star \ltimes \mathcal{H}^\star.
 \end{equation}}
To express this condition, let us now examine how a set of phase functions transforms under the action of a gauge transformation. This situation is described by the following commutative diagram:
\begin{figure}[H]
    \centering
       \begin{equation}
\begin{CD}
\hat\rho(k) @>\quad e^{2i\pi\chi(k)}\quad>> \hat\rho^\prime(k)\\
@V e^{2i\pi\Phi_{\mb{Q}}(k)}VV @VVe^{2i\pi\Phi^\prime_{\mb{Q}}(k)}V\\
\hat\rho(\mb{Q}k) @>\quad e^{2i\pi\chi(\mb{Q}k)}\quad >> \hat\rho^\prime(\mb{Q}k)
\end{CD}
\end{equation}
    \caption{Gauge equivalence relationships}
    \label{fig:eqvgauge}
\end{figure}
It states that two sets of phase functions $\Phi_{\mb{Q}}$ and $\Phi'_{\mb{Q}}$, describing indistinguishable densities $\rho$ and $\rho'$, related by a gauge function $\chi$, belong to the same symmetry class \citep{rokhsar_1988}.
The gauge transformation relating $\Phi_{\mb{Q}}$ and $\Phi'_{\mb{Q}}$ reads
\beq\label{eq:gaugtrans}
\Phi'_{\mb{Q}}(\mb{k}) \equiv \Phi_{\mb{Q}}(\mb{k}) +\chi(\mb{Q}\mb{k}-\mb{k}).
\eeq
Hence a space group is symmorphic if there exists a gauge function $\chi$, independent of $\mb{Q}$, such that the transformed phase functions $\Phi'_{\mb{Q}}(\mb{k})$ are equivalent to zero for every $\mb{Q}\in\mathcal{P}^\star$ and $\mb{k}\in\mathcal{M}$.\newline{}

%% file: method.tex
\section{Numerical determination of space group properties}
\label{sec:method}

\begin{figure}[H]
	\begin{tikzpicture}[block/.style={
      rectangle,
      draw=blue,
      thick,
      fill=blue!20,
      text width=10em,
      align=center,
      rounded corners,
      minimum height=2em
    }]
		\node[anchor=center,inner sep=0] (image1) at (0,0) {\includegraphics[width=3.7cm]{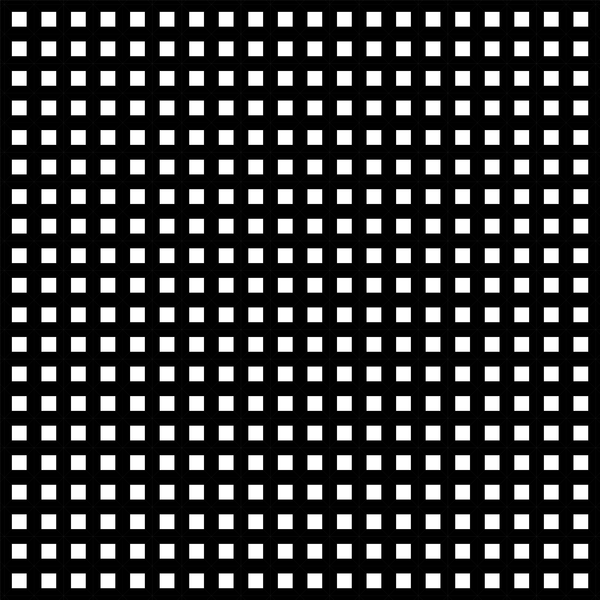}};

		\node[anchor=west,inner sep=0] (image2) at ($(image1.east)+(1.4,0)$) {\includegraphics[width=5.5cm]{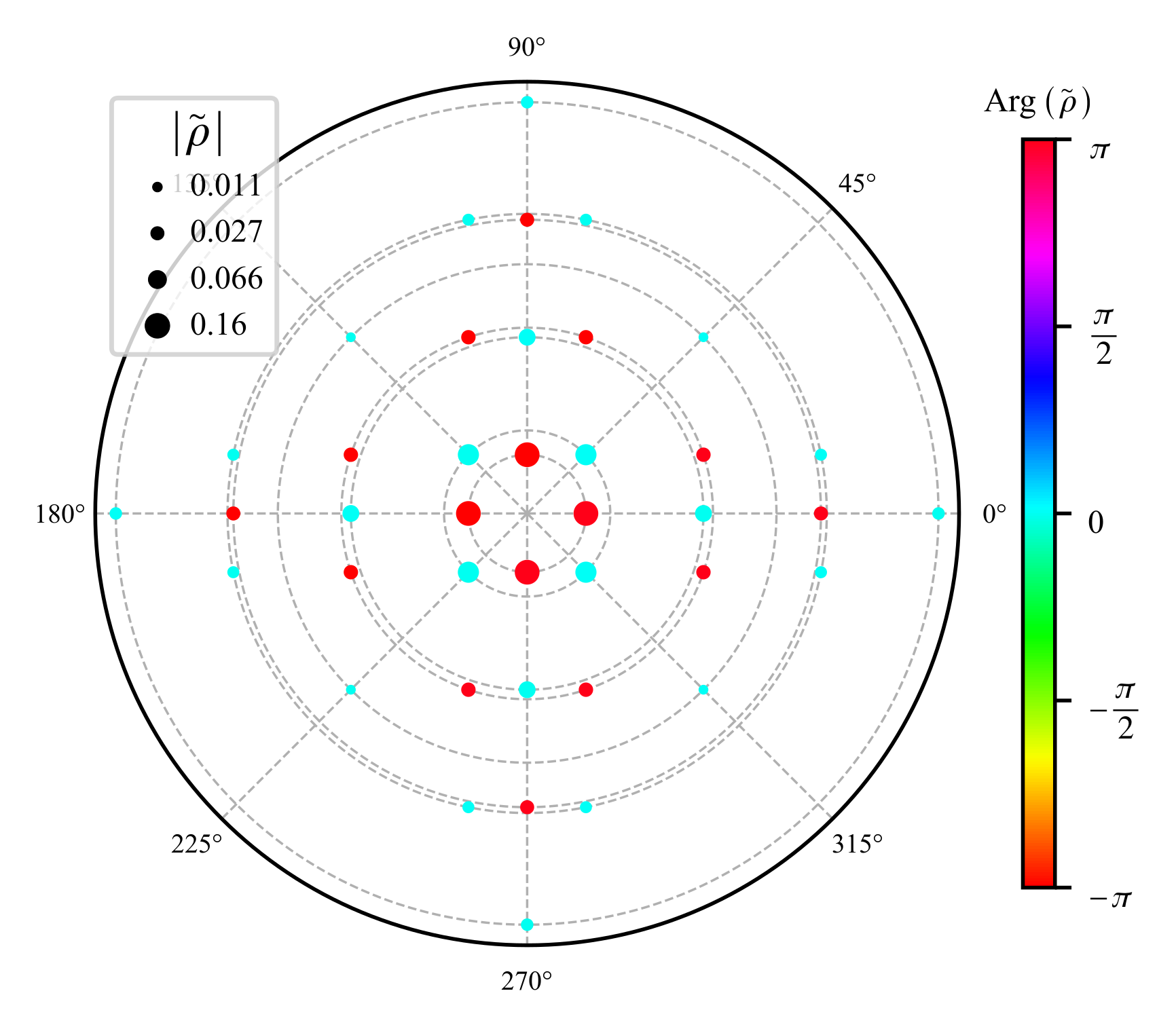}};
		\node[anchor=north] (label2) at (image2.south) {\textbf{Fourier coefficients}};
		\draw[-{stealth}] (image1.east)--(image2.west) node[midway,anchor=south]{Step 1};
		\node[block,anchor=west] (block) at ($(image2.east)+(1.4,0)$){
		\makecell[l]{
			\textbullet Point group\\
			\textbullet Symmorphic character
		}
	};
	\draw[-{stealth}] (image2.east)--(block.west) node[midway,anchor=south]{Step 2};

    \node[anchor=north] at (image1.south|-label2.north) {\textbf{Studied image}};
    \node[anchor=north] at (block.south|-label2.north) {\textbf{Space group properties}};

	\end{tikzpicture}
	\caption{The proposed two-step procedure that determines the point group and symmorphic character of an image of a mesostructure.}
	\label{fig:procedure}
\end{figure}

This section presents the developed methodology that allows to determine symmetry properties of a two-dimensional heterogeneous mesostructure. The input data is a grayscale image of the real or synthetic mesostructure.
As illustrated in \autoref{fig:procedure}, the procedure involves two steps: extracting the Fourier coefficients from the input image, followed by post-processing to determine its point group and symmorphic character.
\subsection{Calculation of Fourier coefficients}
In order to obtain a set of Fourier coefficients from the studied image, two issues must to be adressed:
\begin{enumerate}
    \item the definition of the Fourier coefficient $\hat{\rho}(\vk)$ (see \autoref{eq:fourcoeff}) must be adapted to account for the discrete pixel grid of the image; 
    \item a set of fundamental spatial frequencies $\mathbf{k}_i$, along with a finite subset of their integer combinations $\mathcal{M}^{\select} \subset \mathcal{M}$, must be chosen to enable the subsequent analysis of the corresponding Fourier coefficients $\hat{\rho}(\mathcal{M}^{\select})$.
\end{enumerate}
\subsubsection{Fourier transform of an image}
The pixel values of the image are referred to as ${\rho}_{ij}$ with $1\leq i,j \leq n$, where $i$ and $j$ index the rows and columns of the grid respectively and $n$ denotes the numbers of pixels in a row or a column. The values ${\rho}_{ij}$ themselves lie between zero and one corresponding to black or white pixels respectively. \\
The complex-valued Fourier coefficient of an arbitrary frequency $\left[\vk\right]=
\begin{bmatrix}
	k_x\\
	k_y
\end{bmatrix}
$, with regard to the standard orthonormal base, 
is then given by\footnote{The factor $\frac{1}{n^2}$ allows to have a consistent definition of the Fourier coefficient regardless of the resolution of the studied image.}
\begin{equation}
	\hat{\rho}(\vk)\defeq\frac{\sum_{i,j=1}^{n}\rho_{ij} \e^{-2 \pi \iu (k_xj+k_yi)/n}}{n^2},
    \label{eq:fourierdef}
\end{equation}
which amounts to the approximation of the limit in \autoref{eq:fourcoeff} by a square region $\Omega$ of sufficient size. Using this definition, the Fast-Fourier Transform (FFT) $\hat{\rho}_{kl}\defeq\hat{\rho}\left(
\begin{bmatrix}
	k\\
	l
\end{bmatrix}
\right)$ is then introduced as the grid of Fourier coefficients of integer valued frequencies $\begin{bmatrix}
	k\\
	l
\end{bmatrix} \in \ZZ^2$.
\subsubsection{Selection of spatial frequencies to study}\label{subsubfreq}
In the present case, the number $\mu$ of fundamental frequencies is known a-priori for all studied materials. For periodic heterogeneous materials, it is recalled that it is equal to the dimension of the physical space (here $d=2$). For quasiperiodic ones, it is equal to the number of dimensions of their minimal embedding \citep{baake_guide_1999}. In general, the number $\mu$ is unknown and needs to be assumed. It is, however, limited to multiples of $\totient(N)$, where $\totient$ denotes Euler's totient function and $N$ the order of rotational symmetry of the holohedry~\citep{rabson_1991}.\\

\begin{figure}[H]
	\centering
	\begin{subcaptionblock}{0.4\linewidth}
		\centering
		\begin{tikzpicture}
			\node[anchor=south west,inner sep=0] (image) at (0,0){\includegraphics[width=\linewidth]{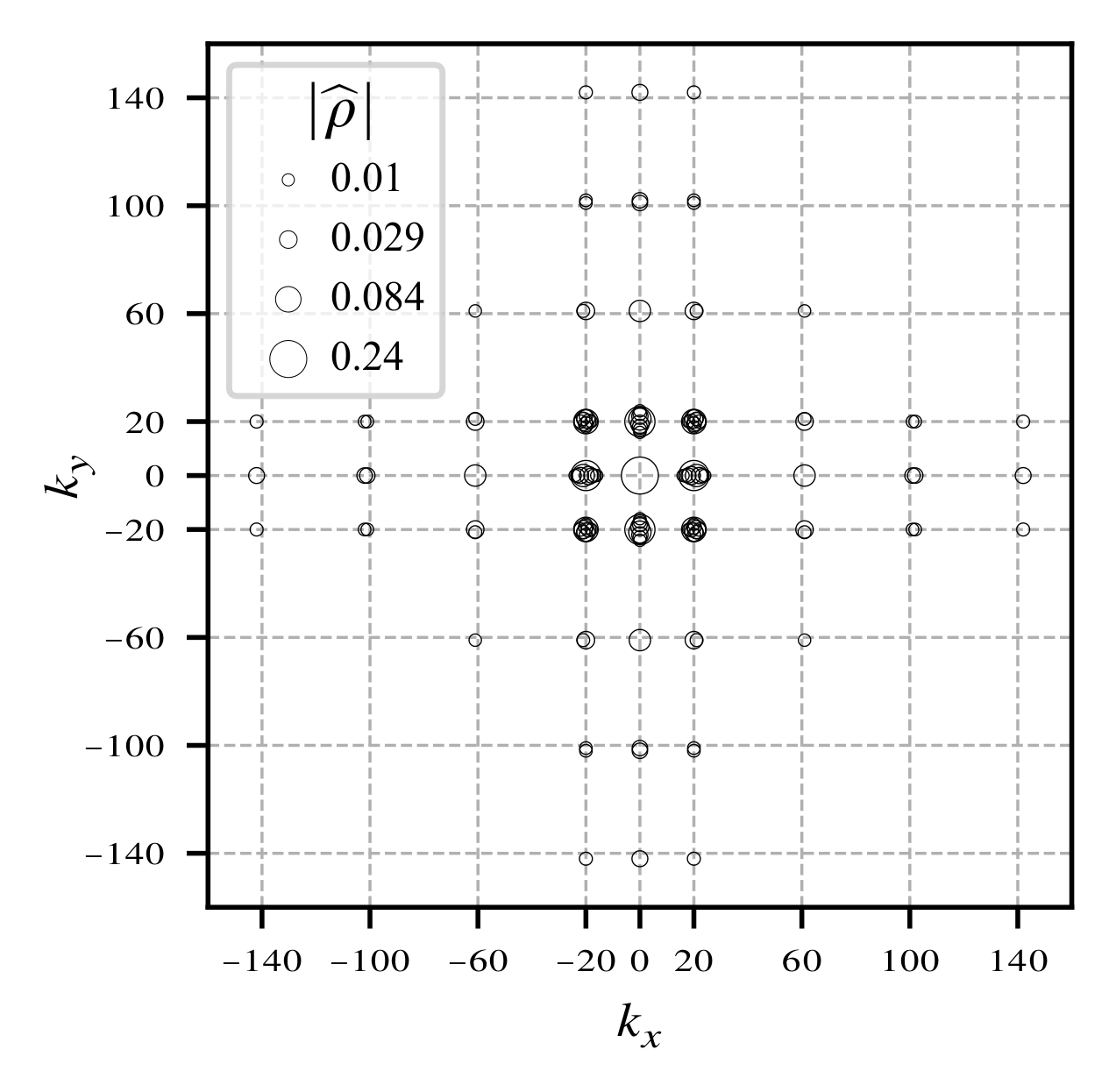}};

			\begin{scope}[x=(image.south east),y=(image.north west)]
				\coordinate (base) at (152/244.416,1-104.5/239.232);
				\draw[red] (base) circle (0.1cm);
				\node[red,anchor=west] at (base.east){$\vk_1^{\guess}$};
				\coordinate (base) at (140/244.416,1-93/239.232);
				\draw[red] (base) circle (0.1cm);
				\node[red,anchor=south] at (base.east){$\vk_2^{\guess}$};
			\end{scope}
		\end{tikzpicture}
		\caption{}
		\label{fig:identify-fundamental-frequencies-a}
	\end{subcaptionblock}
	\begin{subcaptionblock}{0.55\linewidth}
		\centering
		\includegraphics[width=\linewidth,trim={0 0.1cm 0 0},clip]{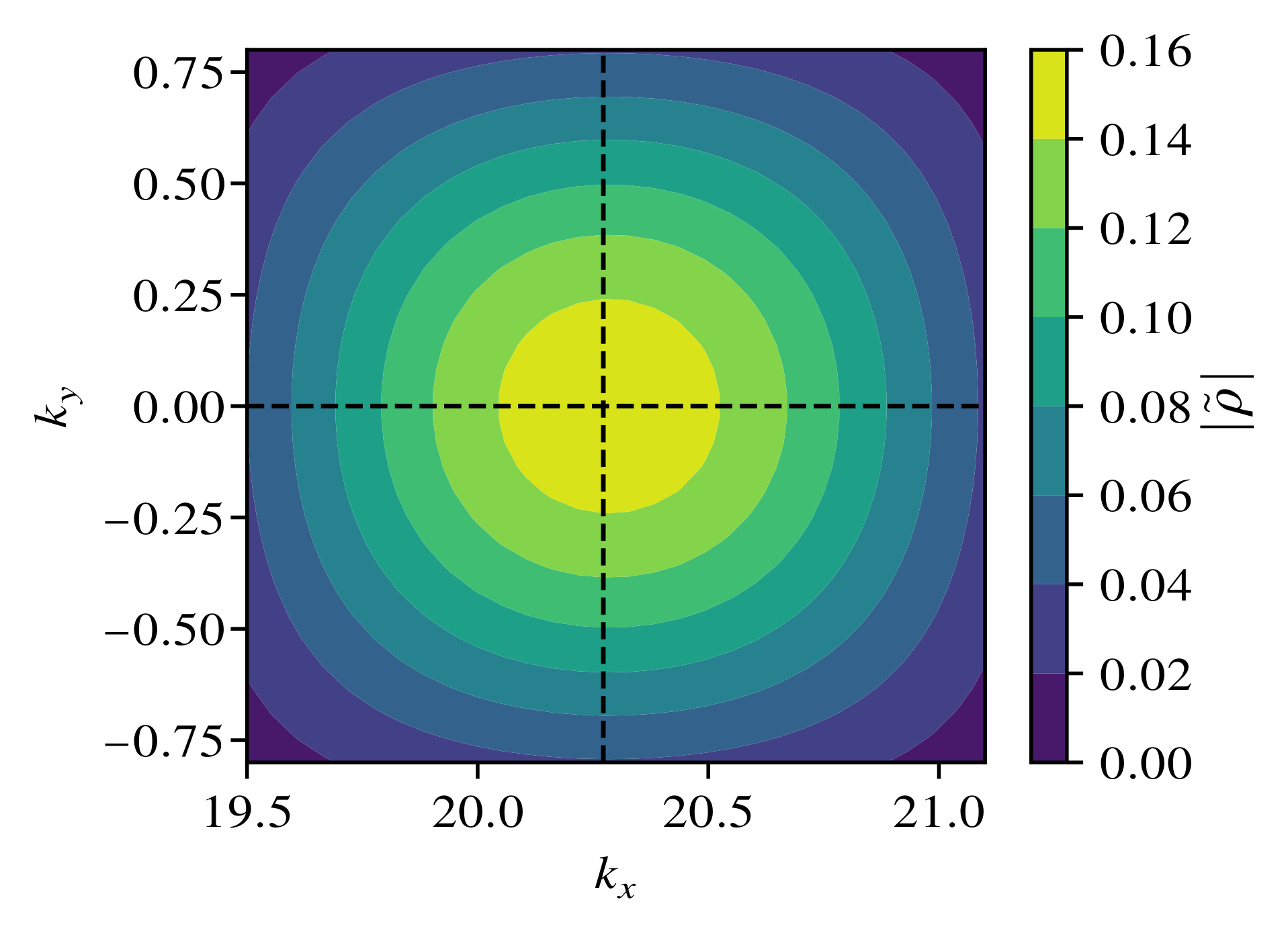}
		\caption{}
		\label{fig:identify-fundamental-frequencies-b}
	\end{subcaptionblock}
	\caption{Amplitudes of the Fourier transform of a periodic image: (a) Values of diffraction diagram $|\hat{\rho}_{kl}|$ above threshold of $\thd=0.01$. A derived first guess of the fundamental frequencies $\vk_1^{\guess}$ and $\vk_2^{\guess}$ is circled in {\color{red}red}. (b) Amplitudes $|\hat{\rho}|$ in the vicinity of $\vk_1^{\guess}$. The local maximum is indicated by the intersection of dashed lines.}
	\label{fig:identify-fundamental-frequencies}
\end{figure}

It may seem rather intuitive to select the fundamental frequencies from the frequencies contained in the FFT.
However, since those frequencies are arranged on a square grid, the compatible point groups would be restricted to subgroups of $\DD{4}$ and the coordinates of $\vk_i$ to integer values. Consequently, such a strategy is inappropriate in the generic case.
The amplitudes of the FFT do, however, reveal information about the reciprocal lattice $\mathcal{M}$. By considering only the values above an arbitrarily chosen threshold $\thd$, the approximative positions of some elements of the $\ZZ$-module $\mathcal{M}$ are revealed. These positions are referred to as peaks in the following.\\
The procedure is as follows. First, $\mu$ distinct peaks, denoted $\vk_i^{\guess}$ with $1\leq i\leq\mu$, are selected by inspection as first guesses  of the fundamental frequencies,  Second, the precise values of the fundamental frequencies $\vk_i$ are obtained by locating the local maxima of the amplitudes $|\hat{\rho}|$ in the vicinity of $\vk_i^{\guess}$.
These two steps are illustrated  in \autoref{fig:identify-fundamental-frequencies-a} and \autoref{fig:identify-fundamental-frequencies-b}, respectively. 
Finally, a finite set $\mathcal{M}^{\select}$ of integer multiples of the fundamental frequencies $\vk_i$ is chosen so as to account for the remaining peaks. In practice, this set is taken as a subset of all the frequencies
$$\mathcal{M}^{\select}\defeq\left\{\sum_{i=1}^{\mu}\alpha_i \vk_i~|~\alpha_i \in \ZZ \text{ and } \sum_{i=1}^{\mu}\alpha_i \leq \fl\right\}$$
that is, all linear combinations involving at most $\fl$ multiples of the fundamental frequencies\footnote{The limitation $\fl$ on the number of fundamental frequencies used to index the detected peaks corresponds to a truncation of the Fourier series of $\rho$. Lowering the amplitude threshold $\thd$ in the FFT reveals additional peaks that need to be indexed,  which in turn requires including more terms in the Fourier series.}. 


If the majority of the detected peaks can be indexed using $\mathcal{M}^{\select}$, the chosen $\vk_i$ and $\mathcal{M}^{\select}$ are confirmed. Otherwise, the procedure must be repeated with a new set of initial guesses $\vk_i^{\guess}$ until a consistent indexing is achieved.\\
The subsequent analyses are carried out using the Fourier-coefficients $\hat{\rho}(\mathcal{M}^{\select})$ determined from \autoref{eq:fourierdef}.
Their amplitudes $|\hat{\rho}(\mathcal{M}^{\select})|$ constitute the classical Fraunhofer diffraction diagram~\cite[Section 8.2]{lipson_optical_2010}.

\subsection{Determination of space group properties}
In this section, we detail the methodology used to detect the symmetry features of a (quasi)periodic pattern based on the Fourier coefficients extracted from its image.
\subsubsection{Detection of the point group}
\label{sec:pointgroup}
This subsection focuses on determining the point group of the pattern present in the image, while the question of symmorphism will be treated in the next subsection.
To this end, suitable deviation measures are introduced to quantify how elements of a candidate point group depart from the indistinguishability condition stated in \autoref{eq:phasefundef}.

The holohedry -- defined as the set of linear isometries of the diffraction diagram -- is considered as the initial point group candidate $\mathcal{P}^{\guess}$. This hypothesis is then checked by evaluating whether the indistinguishability condition for each of its generators is satisfied\footnote{The generators of a group are defined as a subset of its elements such that any other element can be expressed as a composition of a finite number of generators. While the generators of a dihedral group $\DD{N}$ are not uniquely determined, in this work they are always chosen to be the clockwise rotation of angle $\frac{2\pi}{N}$ together with one of its mirrors. (The tilings studied in this work allow to always choose the horizontal mirror.) If a generator $\mb{Q}$ is indeed a symmetry operation, then it possesses a phase function $\phi_{\mb{Q}}(\vk)$. If this is the case for each generator of the point group, then the phase function of any other point group element must be compatible with gauge-linearity by \autoref{eq:groupop}.}.\\
The compatibility of a generator $\mb{Q}$ 
 is assessed using an overall deviation measure  $\symerror$ constructed from two distinct components: an amplitude deviation
$\amplerror$ and a phase deviation 
$\gaugeerror$ defined as follows:
\begin{itemize}
    \item the amplitude deviation $\amplerror$ checks  whether $\mb{Q}$ leaves invariant the amplitudes of the Fourier coefficients: $|\hat{\rho}(\vk)|=|\hat{\rho}(\mb{Q}\vk)|$. Consequently, it is defined as the relative difference $\amplerror\left(\te{\Q},\vk\right)\defeq 
    \reldiff{\abs{\tilde{\rho}{}(\vk)}}{\abs{\tilde{\rho}{}(\te{\Q}\vk)}}$ with $\reldiff{x}{y}\defeq\frac{\abs{x-y}}{\mathrm{max}(x,y)}$;
    \item the phase deviation $\gaugeerror_{\mb{Q}}$ evaluates the discrepancy of the phase. 
    If $\Phi_{\mb{Q}}$ is indeed a gauge function, then its values on $\mathcal{M}^{\select}$ can be extrapolated from the fundamental frequencies using \autoref{eq:gaugefunextrapolation}.
    Therefore, the deviation $\gaugeerror_{\mb{Q}}$ can be defined as the difference between the actual and extrapolated values of $\Phi_{\mb{Q}}(\vk)$
    \beq
    \gaugeerror_{\mb{Q}}(\vk=n_1\vk_1+..+n_{\mu}\vk_{\mu})
    \defeq d^{\mathbb{Z}}\left(
    		{\Phi_{\mb{Q}}(\vk)},
    		{n_1\Phi_{\mb{Q}}(\vk_1)+..+n_{\mu}\Phi_{\mb{Q}}(\vk_{\mu})}
    	\right)
    	\label{eq:def-gauge-deviation-lin}
    \eeq
    $d^{\mathbb{Z}}$ denotes the difference modulo 1 of two real numbers, which is defined as
    \ben
    d^{\ZZ}(x,y)\defeq \mathrm{Min}\left(\mathrm{frac}(x-y),1-\mathrm{frac}(x-y)\right)
    \een
    with $\mathrm{frac}$ the fractional part of a number.
\end{itemize}
In addition, the overall error $\symerror$ must fulfil the following requirements:
\begin{itemize}
	\item if both $\amplerror$ and $\gaugeerror_{\mb{Q}}$ are $0$, $\symerror$ is equal to $0$; 
	\item if $\amplerror$ or $\gaugeerror_{\mb{Q}}$ reach their maximum value ($1$ and $\frac{1}{2}$, respectively), $\symerror$ is equal to $1$.
\end{itemize}
A bilinear interpolation is thus proposed to define $\symerror$
 \beq
 \symerror(\mb{Q},\vk)=1-\left(1-\amplerror(\mb{Q},\vk)\right)\left(1-2\gaugeerror_{\mb{Q}}(\vk)\right).
 \label{eq:def-symerror}
 \eeq
The discrepancy must be evaluated for each generator of the candidate group $\mathcal{P}^{\text{guess}}$ to determine whether it qualifies as a point group of weak symmetry. If this is not the case, the analysis must be repeated for all subgroups of $\mathcal{P}^{\text{guess}}$, and so on.

\subsubsection{Detection of symmorphism}
\label{sec:detect-symm}
This subsection now explains how the symmorphism of an image can be inferred from its Fourier coefficients.

As a preliminary remark, it is worth noting that non-symmorphism can occur only in space groups whose point group is \todoR{conjugate to} the dihedral group $\mathrm{D}_N$ with even $N$
~\citep{rabson_1991}. All other space groups are necessarily symmorphic.

As detailed in \autoref{sec:indisting}, a space group is symmorphic when there exists a gauge function $\chi$ that cancels the phase functions of all point-group generators via the transformation in \autoref{eq:gaugtrans}. We demonstrate below that, for any generalized space group defined by the phase functions of its point-group generators, only finitely many gauge functions are admissible. This ensures that the symmorphic nature of the space group can be established with certainty.
The generators of $\DD{N}$ are chosen to be the horizontal mirror $\vp{h}$ and the rotation $\vp{r}$ of angle $\frac{2\pi}{N}$, with their respective phase functions ${\Phi}_{\vp{h}}$ and ${\Phi}_{\dT{r}}$.
~\cite{rabson_1991} showed that any phase function ${\Phi}_{\dT{r}}$ is cancelled by the gauge function
\beq
	{\chi}_{\dT{r}}(\vk)\equiv \frac{1}{4}\left({\Phi}_{\dT{r}}(\vk)+{\Phi}_{\dT{r}}(\dT{r}\vk)+..+{\Phi}_{\dT{r}}(\dT{r}^{(N/2-1)}\vk)-{\Phi}_{\dT{r}}(\dT{r}^{N/2}\vk)-..-{\Phi}_{\dT{r}}(\dT{r}^{(N-1)}\vk)\right).
	\label{eq:chi_r_def}
\eeq
As all of the remaining rotations in $\DD{N}$ are obtained by repeated application of $\vp{r}$, their phase functions are equally cancelled by ${\chi}_{\dT{r}}$.
Accordingly, the problem of finding the gauge function $\chi\equiv{\chi}_{\dT{r}}+{\chi}_{\vp{h}}$ that cancels the phase functions of the entire point group is reduced to finding a second gauge function ${\chi}_{\vp{h}}$ that cancels ${\Phi}_{\vp{h}}$ without changing the already nullified phase functions of the rotational point group elements.\\
The rotation $\dT{r}_2$ of angle $\pi$ belongs to any dihedral group $\DD{N}$ with $N=2^m$. Its phase function ${\Phi}_{\dT{r}_2}$ transforms under application of ${\chi}_{\vp{h}}$ as
\beq
	{\Phi}_{\dT{r}_2}'(\vk)\equiv {\Phi}_{\dT{r}_2}(\vk)+\chi_{\vp{h}}(\dT{r}_2\mb{k}-\mb{k})\equiv {\Phi}_{\dT{r}_2}(\vk)+\chi_{\vp{h}}(-2\mb{k})\equiv {\Phi}_{\dT{r}_2}(\vk)-2\chi_{\vp{h}}(\mb{k}).
	\label{eq:chi_h_def}
\eeq
Consequently, in order to keep ${\Phi}_{\dT{r}_2}(\vk)$ nullified modulo 1, $\chi_{\vp{h}}$ may only take values of either $0$ or $\frac{1}{2}$ (modulo 1). Furthermore, as $\chi_{\vp{h}}$ is completely determined by its values on the fundamental frequencies $\vk_1,..,\vk_{\mu}$ by virtue of \autoref{eq:gaugefunextrapolation}, there are only $2^{\mu}$ distinct choices of $\chi_{\vp{h}}$ available.\\
If any of the $2^{\mu}$ resulting options of $\chi\equiv{\chi}_{\dT{r}}+{\chi}_{\vp{h}}$ cancels both $\Phi_{\vp{h}}$ and $\Phi_{\vp{r}}$, the space group is symmorphic. 

%% file: results.tex
\section{Results}
This section presents the results obtained by applying the proposed method to images of periodic and quasiperiodic heterogeneous materials. The first subsection is devoted to periodic mesostructures. In this specific case, as previously discussed, the notions of superposability and indistinguishability coincide. However, employing the Fourier transform provides a systematic means of identifying not only the point group of such heterogeneous materials but also their symmorphic nature. 
\todoR{While determining the point group of a periodic heterogeneous material may appear straightforward, the identification of non-symmorphic point groups can be challenging when relying solely on visual inspection. In such cases, the proposed Fourier-based algorithm provides a clear and unambiguous determination.}

In the second subsection, the method is applied to quasiperiodic heterogeneous materials, for which identifying the weak point group directly from the image becomes significantly more challenging, if not impossible. The selected examples highlight the usefulness of the proposed algorithm, particularly in cases where superposability and indistinguishability do not coincide. 

\subsection{Periodic heterogeneous materials}

\todoR{Throughout this section, all symmetries are understood in their classical strong sense.}

Although centred images are typically assumed when working with periodic heterogeneous materials, practical considerations -- such as image misalignment during acquisition or shifted numerical samples -- may lead to a off-centred base image. The consequences that such a shift can have on the symmetry analysis is first investigated. Accounting for this effect is essential, since this specific centring choice is no longer possible for quasiperiodic media.

In the second part, three tilings sharing the same square lattice are examined. These examples serve as introductory demonstrations of the method and illustrate how the distinction between symmorphic and non-symmorphic space groups manifests in the Fourier domain.

\subsubsection{Influence of shift}
\label{sec:shift}

\begin{figure}[H]
\centering
\begin{subfigure}[c]{0.3\linewidth}
\includegraphics[width=\linewidth]{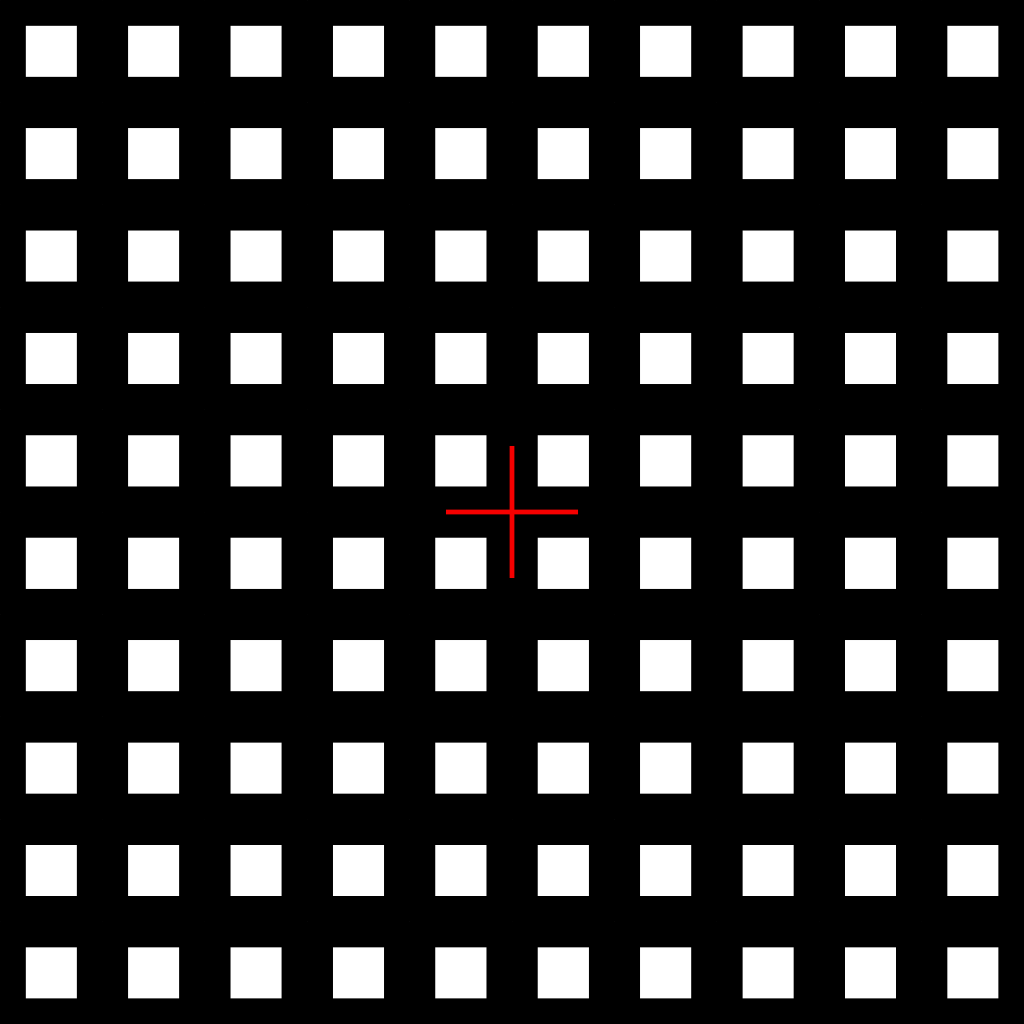}
\caption{}
\label{fig:p4mtilingcent}
\end{subfigure}
\hspace{2em}
\begin{subfigure}[c]{0.3\linewidth}
\includegraphics[width=\linewidth]{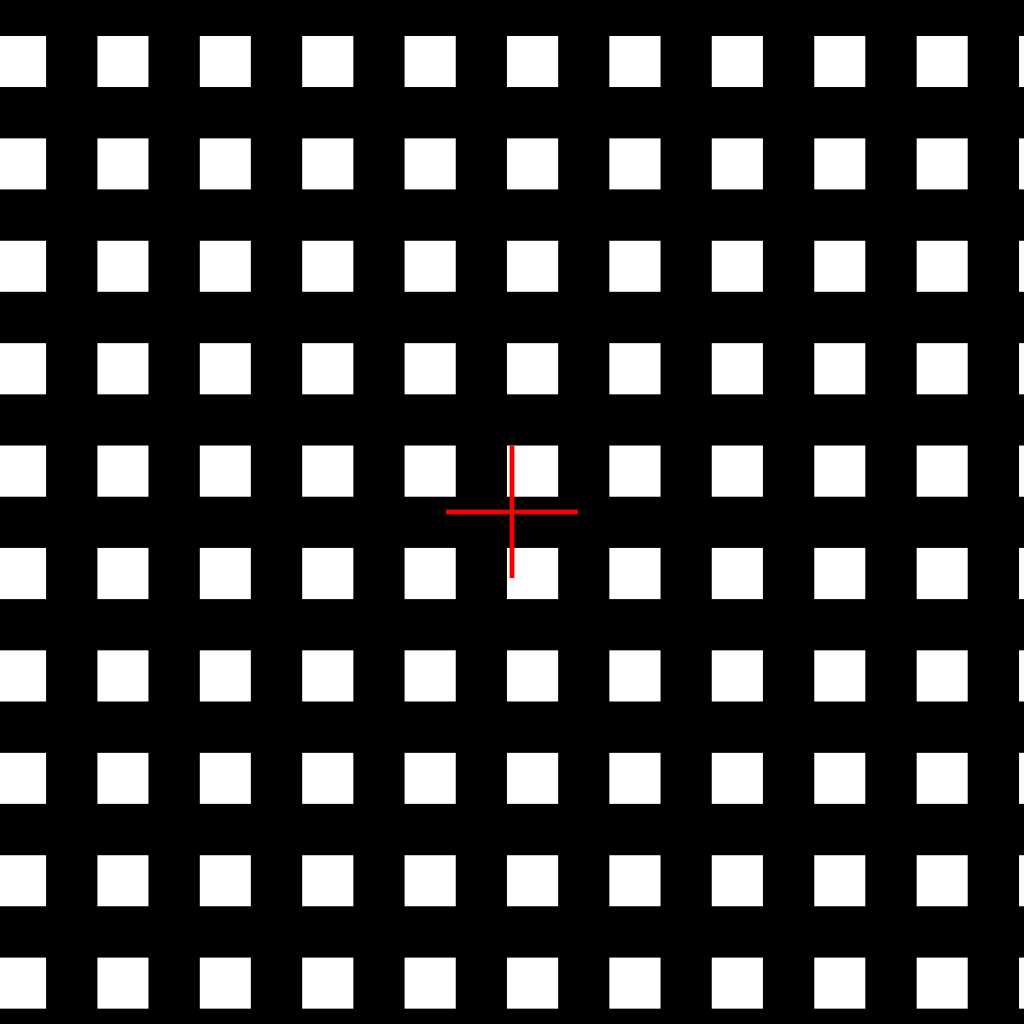}
\caption{}
\label{fig:p4mtilingshift}
\end{subfigure}
\caption{Centred (a) and generic (b) images of the same periodic tiling. The red cross indicates the centre of the image.}
\label{fig:p4mtiling}
\end{figure}

Let us first consider a synthetic image composed of square holes arranged on a square lattice. When working with periodic images, it is standard practice to centre the pattern, as illustrated in \autoref{fig:p4mtilingcent}. The centre of the image, marked by a red cross, corresponds to the common centre of the four surrounding white squares.In this particular example the symmetry properties of the image can be identified by visual inspection. The point group -- that is, the group of linear isometries -- is $\DD{4}$. A generating set consists of the fourfold rotation $\mb{r}_{4}$ together with a reflection across the horizontal axis, $\mb{h}$. When affine isometries are taken into account, the resulting wallpaper group is p4m.

It may happen that the image under study is not centred, as illustrated in \autoref{fig:p4mtilingshift}. This can result either from imperfections in the image acquisition process or from the absence of a well-defined centre, as is the case for quasiperiodic mesostructures. Such off-centred images are said to be in a generic position. Since the identification procedure must be independent of the image’s centring, the first step is therefore to examine the effect of such a shift.\par

\begin{figure}[H]
\centering
\begin{subfigure}[c]{0.4\linewidth}
\includegraphics[width=0.875\linewidth]{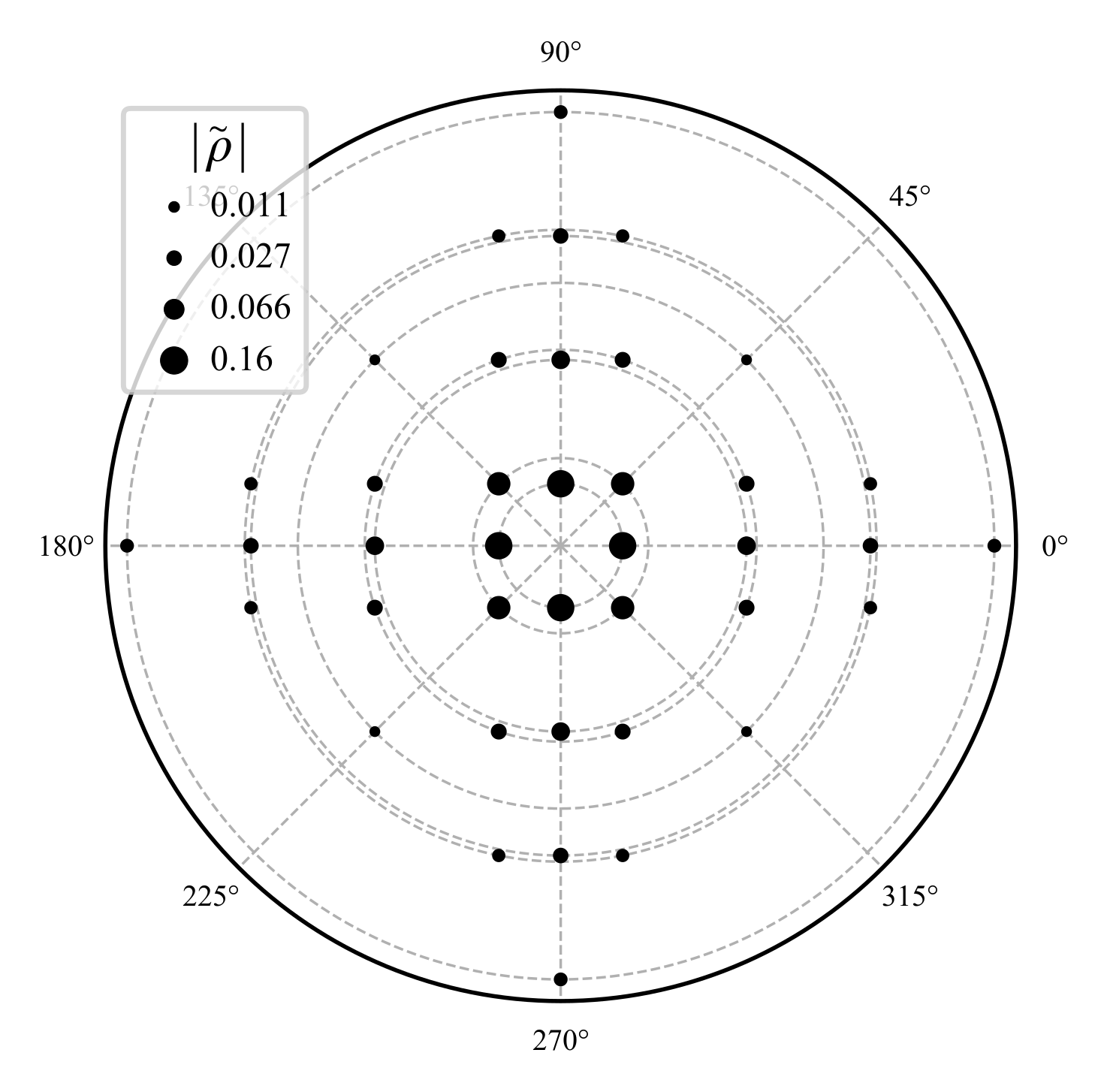}\\
\includegraphics[width=\linewidth]{diffractiondiagram_exact_pixel_p4msquare.png}
\caption{}
\label{fig:p4mfouriercent}
\end{subfigure}
\begin{subfigure}[c]{0.4\linewidth}
\includegraphics[width=0.875\linewidth]{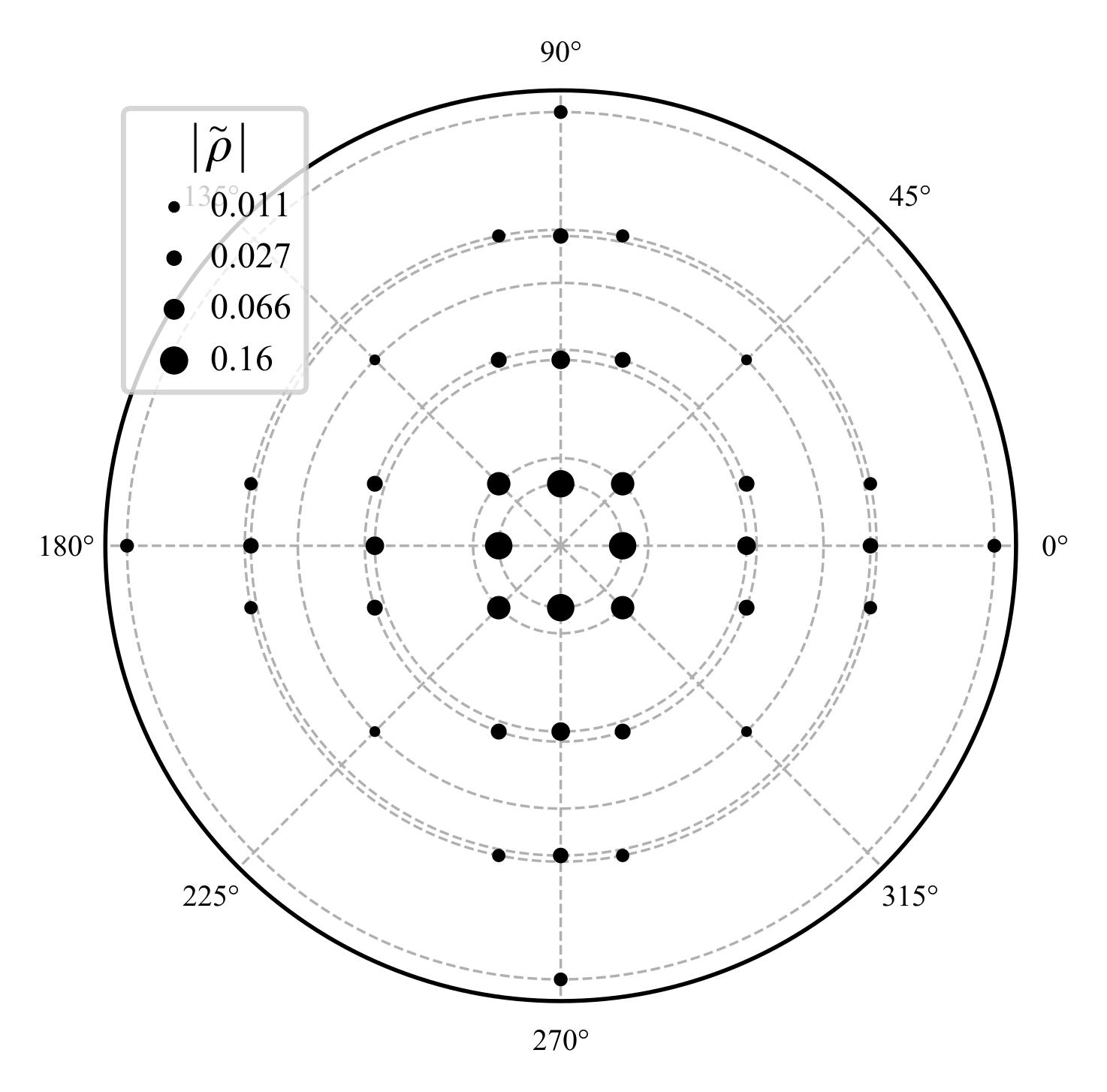}\\
\includegraphics[width=\linewidth]{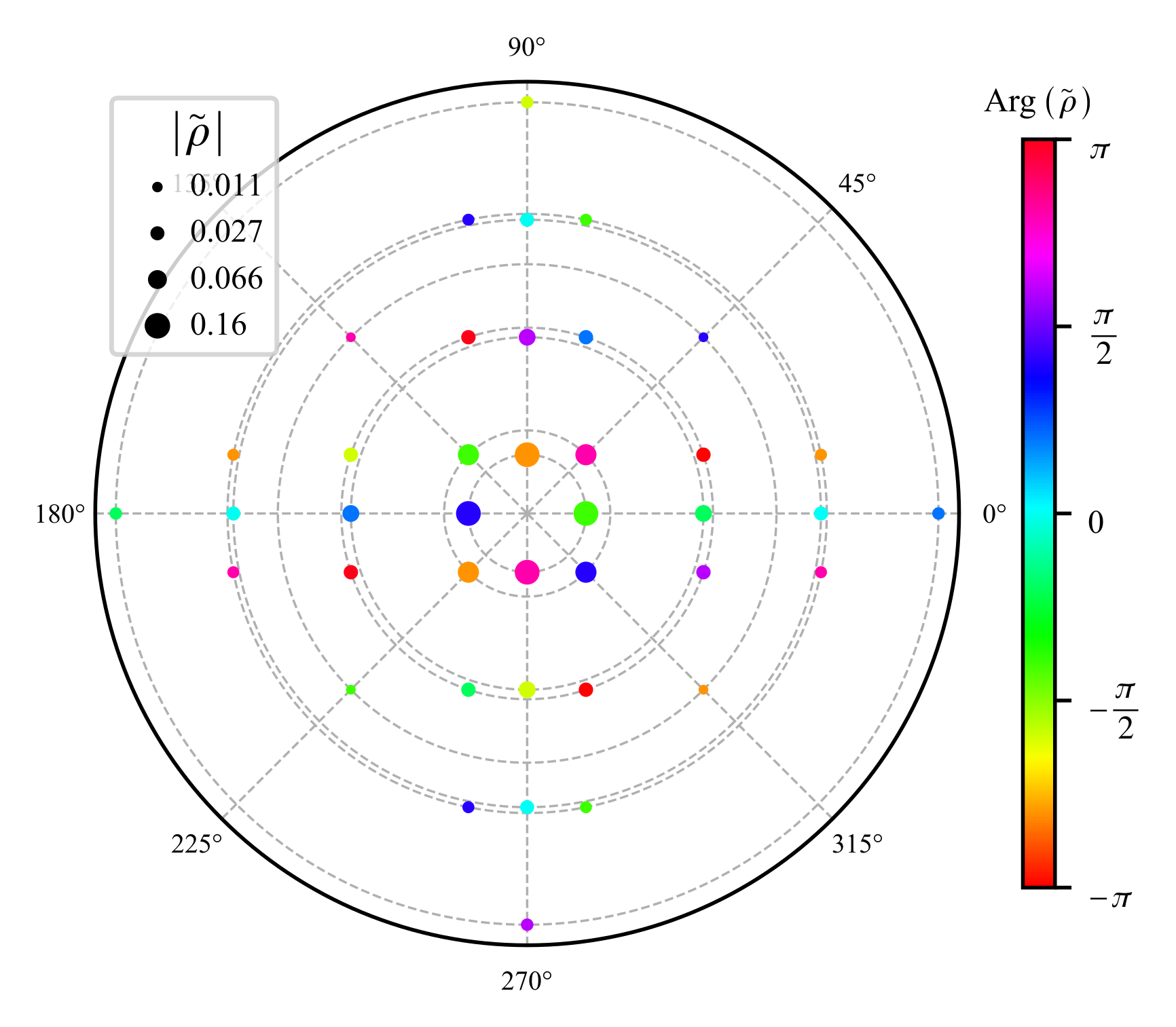}
\caption{}
\label{fig:p4mfouriershift}
\end{subfigure}
	\caption{Computed Fourier coefficients for the (a) centred and (b) generic images of the same periodic tiling shown in \autoref{fig:p4mtiling}. 
}
\label{fig:p4mfourier}
\end{figure}

The Fourier coefficients computed for the two images, each with a different centring, are presented in \autoref{fig:p4mfourier}. While the positions of the peaks and their amplitudes are essentially identical, their phases differ. These differences stem from the off-centred image, which introduces a phase shift in the complex arguments of the Fourier coefficients.

Even though the symmetry is obvious in the present example, the point group of the off-centred image is determined using the procedure introduced in \autoref{sec:pointgroup}. The two generators of the holohedry point group $\DD{4}$ -- the reflection $\vp{h}$ and the fourfold rotation $\mb{r}_{4}$ -- are tested. For the image in generic position, the maximum deviation values $\symerror$ are $4.9\times10^{-3}$ for $\mb{r}_{4}$ and $10^{-5}$ for $\vp{h}$, thereby confirming $\DD{4}$ as its point group.
These results demonstrate the robustness of the procedure with respect to the generic centring of periodic patterns. In the following, and for the sake of simplicity, all calculations for periodic materials are carried out on properly centred images.

\subsubsection{Comparison of square lattice heterogeneous materials}
\label{sec:square}
\begin{figure}[H]
\centering
\begin{subfigure}[c]{0.3\linewidth}
\includegraphics[width=\linewidth]{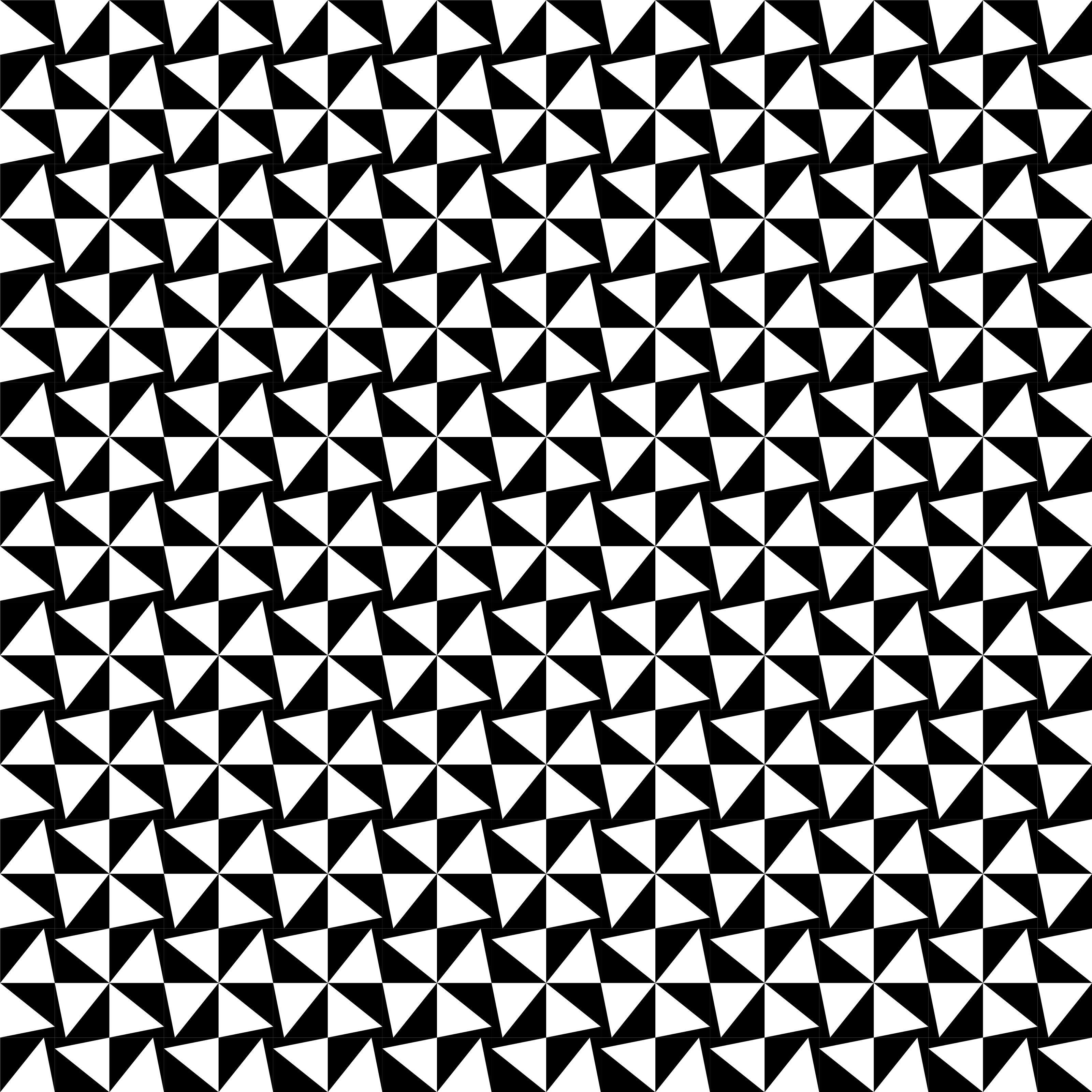}
\caption{p4}
\label{fig:p4}
\end{subfigure}
\hspace{1em}
\begin{subfigure}[c]{0.3\linewidth}
\includegraphics[width=\linewidth]{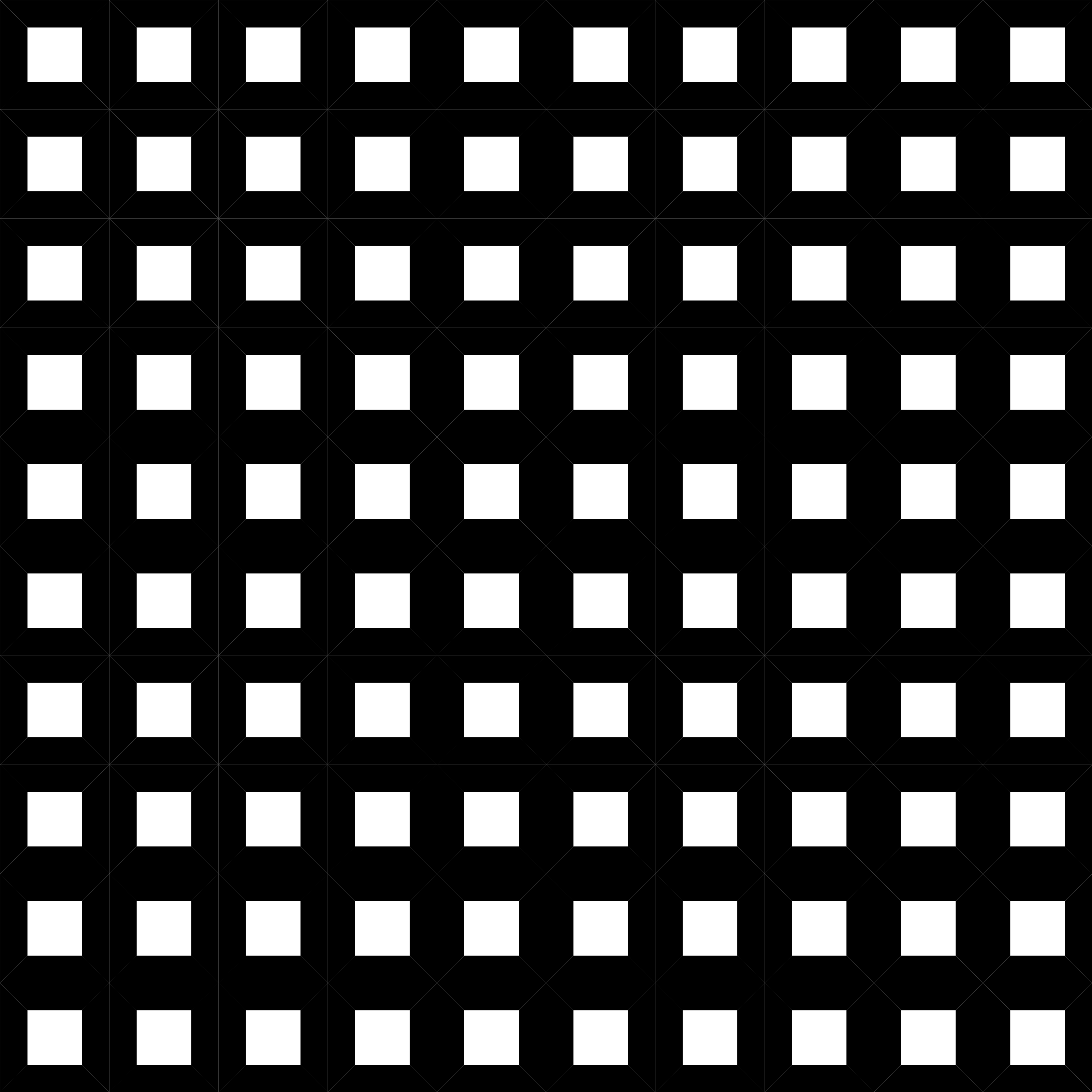}
\caption{p4m}
\label{fig:p4m}
\end{subfigure}
\hspace{1em}
\begin{subfigure}[c]{0.3\linewidth}
\includegraphics[width=\linewidth]{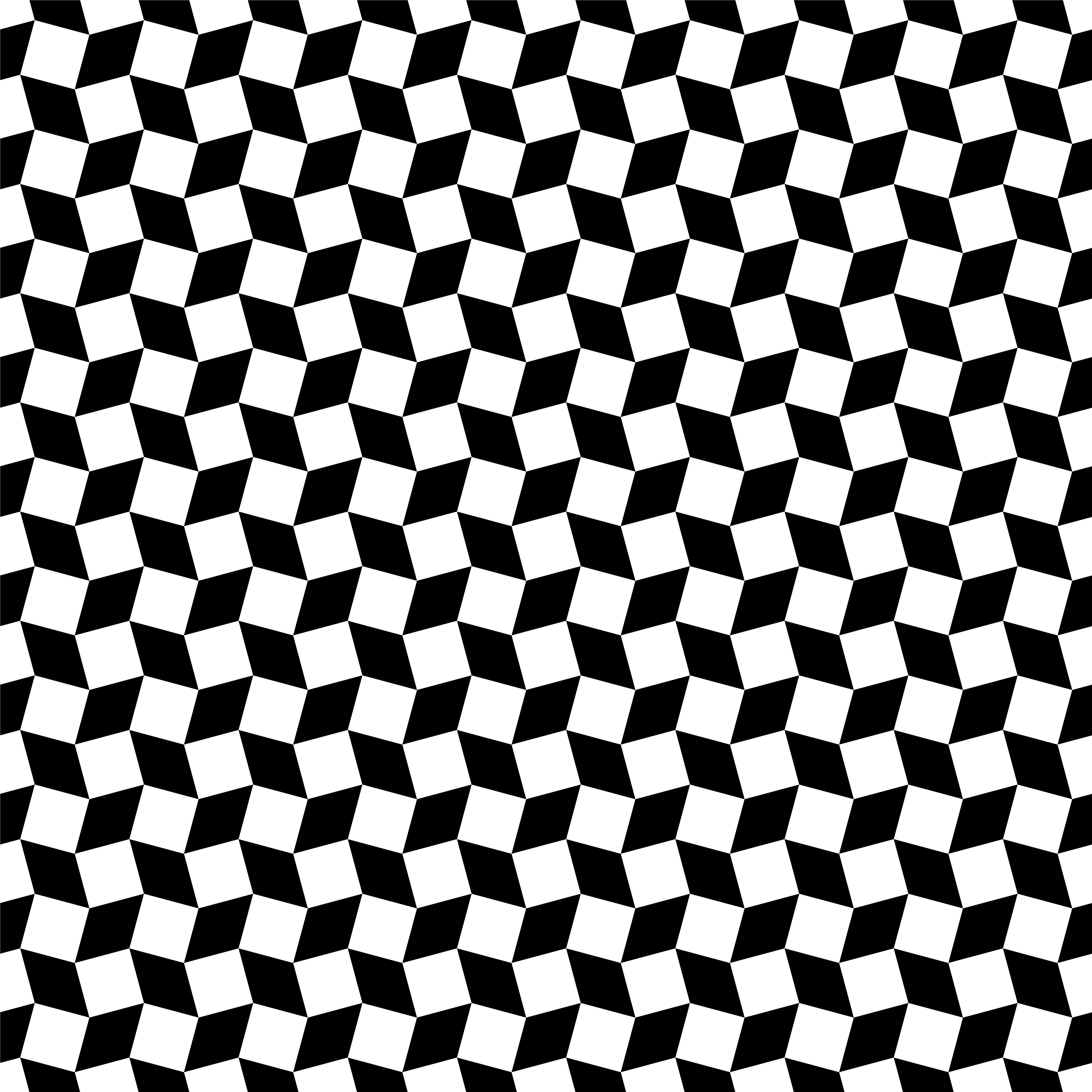}
\caption{p4g}
\label{fig:p4g}
\end{subfigure}
\caption{The three studied tilings with a square lattice.}
\label{fig:p4micro}
\end{figure}

In this section, we analyse the images presented in \autoref{fig:p4micro} of three heterogeneous materials that share the same square invariant lattice group but belong to different (wallpaper) space groups.

The symmetry properties of these mesostructures are detailed in the table below:

\begin{table}[H]
\centering
\begin{tabular}{c c c c}
\hline
\textbf{Label} & \textbf{Space Group} & \textbf{Properties} & \textbf{Point Group} \\
\hline
(a) & p4  & Symmorphic, chiral        & $ \ZZZ_{4}$ \\
(b) & p4m & Symmorphic, achiral                & $\DD{4}$ \\
(c) & p4g & Non-symmorphic, achiral            & $\DD{4}$ \\
\hline
\end{tabular}
\caption{Summary of the three wallpaper groups considered}
\end{table}

\begin{figure}[H]
	\centering
		\begin{tikzpicture}
			\node (a1) at (0,0) {\includegraphics[width=0.38\textwidth]{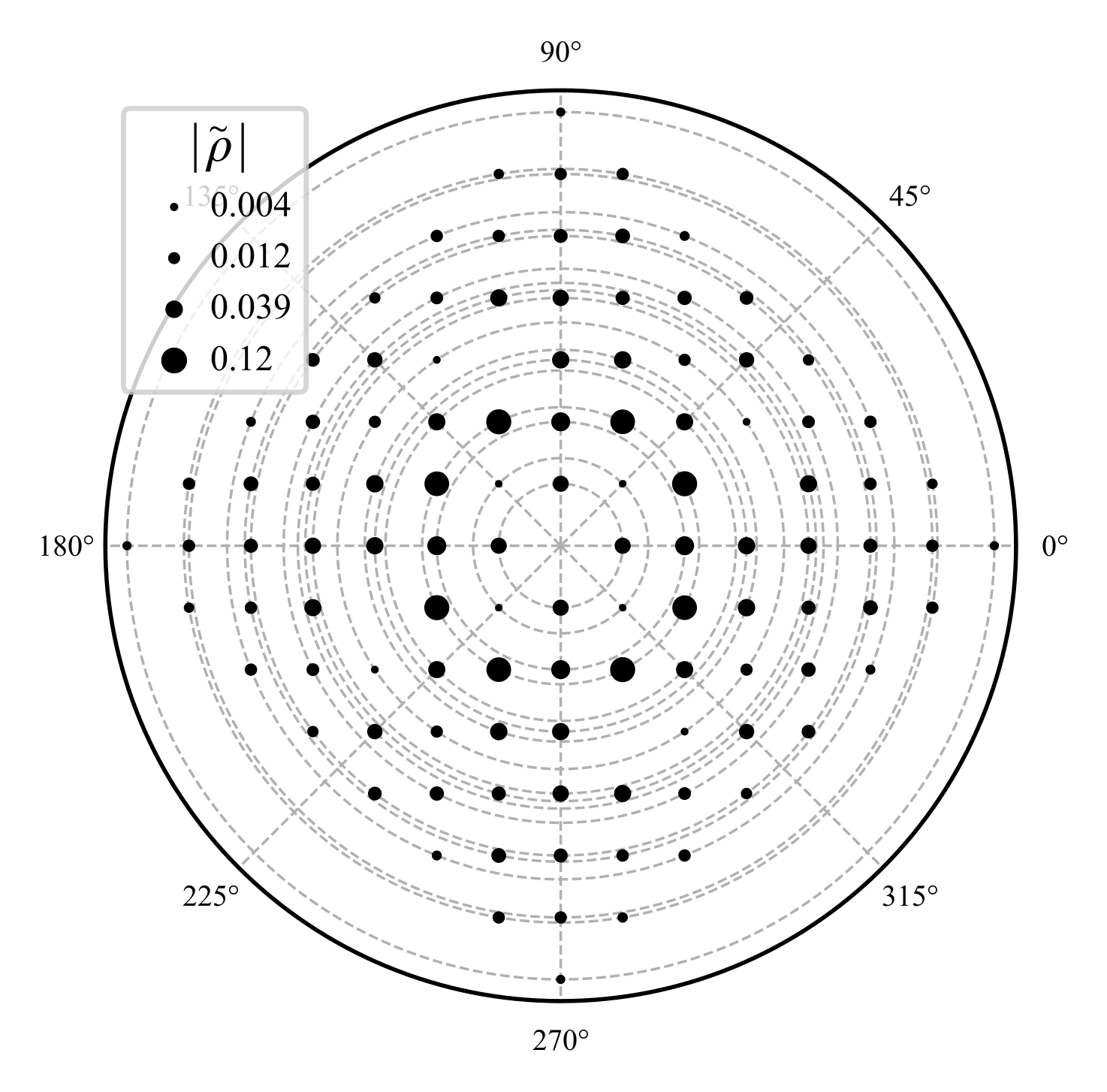}};
			\node[anchor=north] (b1) at (a1.south) {\includegraphics[width=0.38\textwidth]{diffractiondiagram_exact_pixel_p4msquareamplitudes.png}};
			\node[anchor=north] (c1) at (b1.south) {\includegraphics[width=0.38\textwidth]{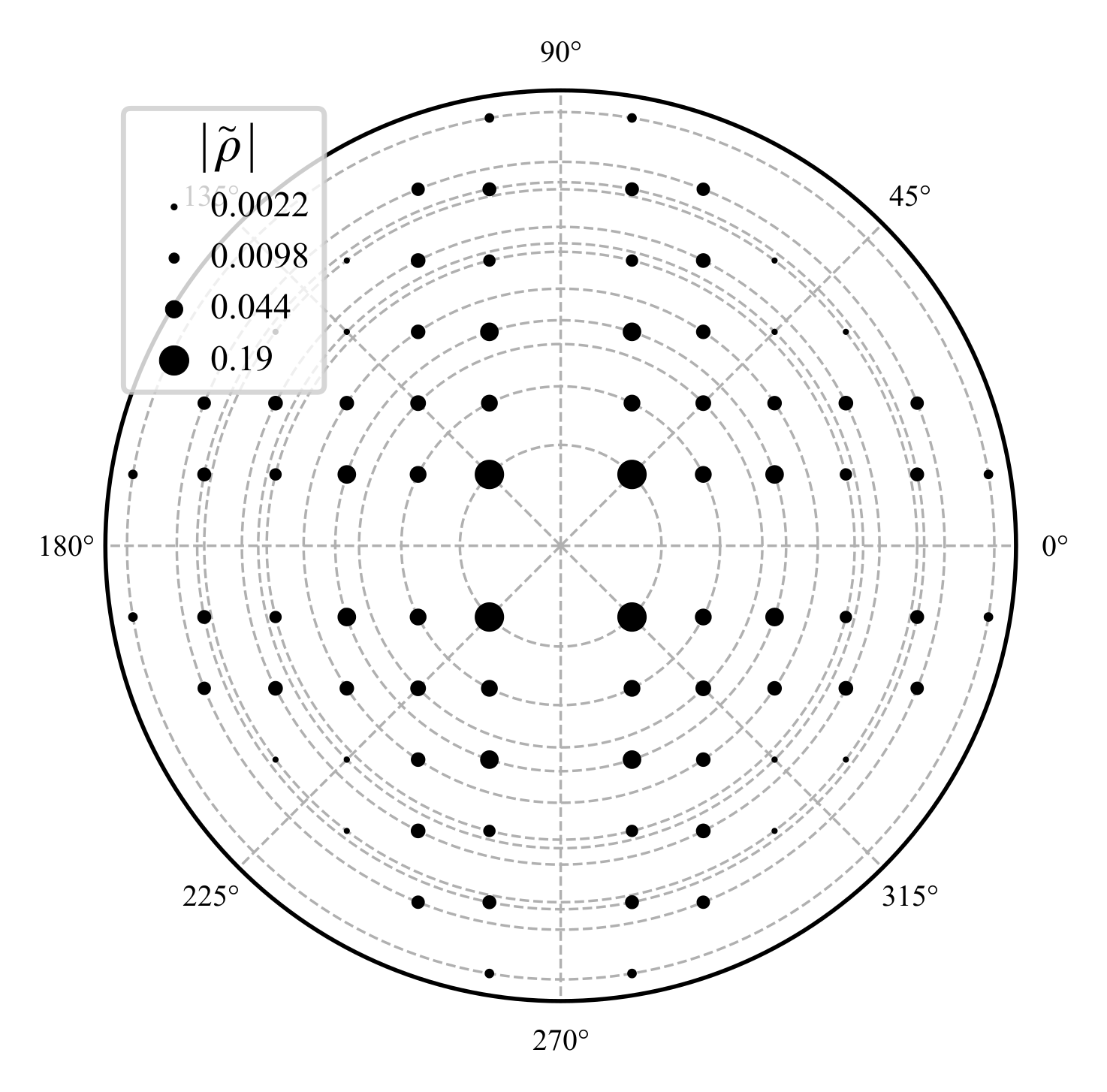}};

			\node[anchor=west] (a2) at (a1.east) {\includegraphics[width=0.41\textwidth]{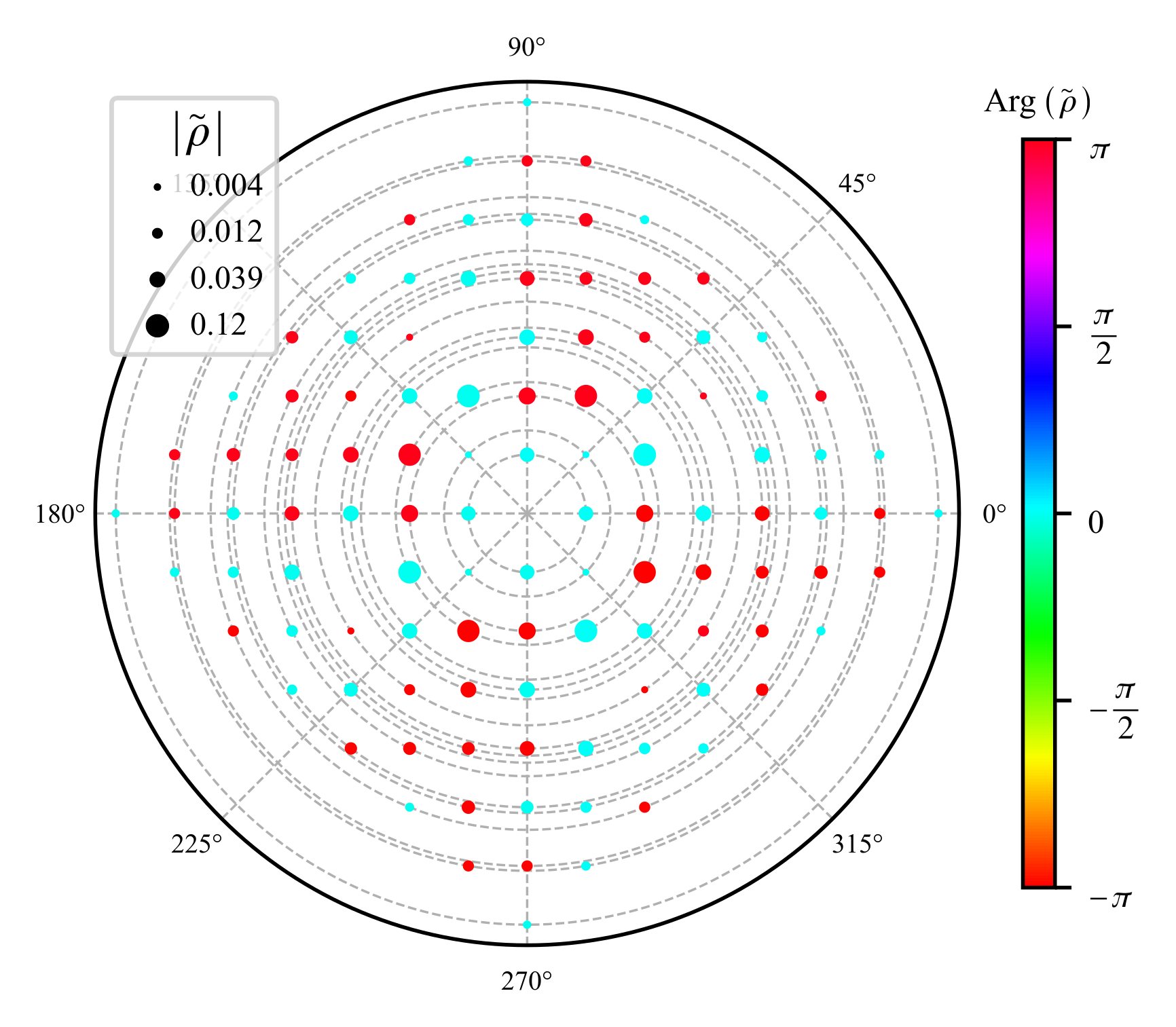}};
			\node[anchor=west] (b2) at (b1.east) {\includegraphics[width=0.41\textwidth]{diffractiondiagram_exact_pixel_p4msquare.png}};
			\node[anchor=west] (c2) at (c1.east) {\includegraphics[width=0.41\textwidth]{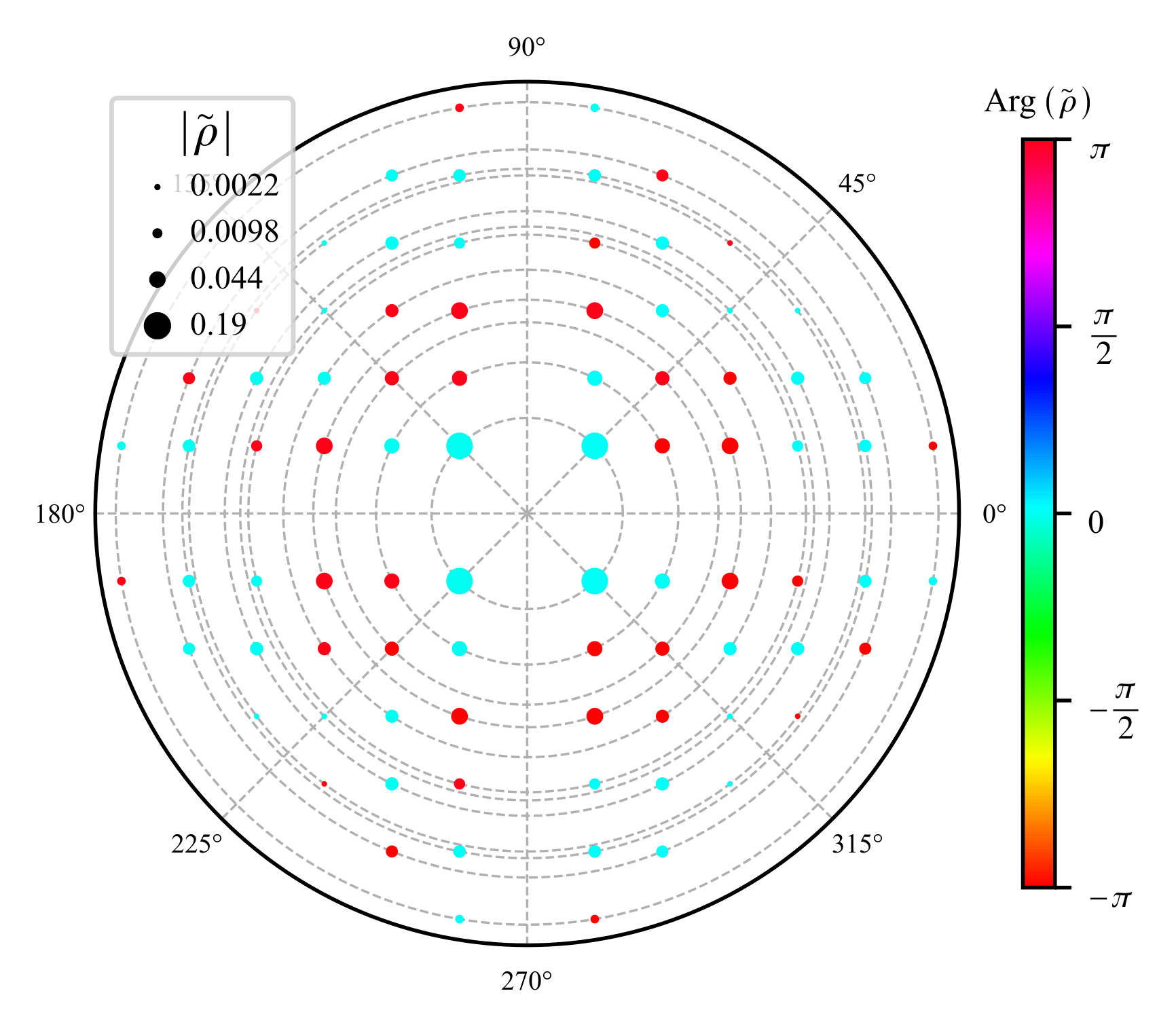}};
				\node[anchor=east,rotate=90,xshift=0.4cm] (a0) at (a1.west) {p4};
				\node[anchor=east,rotate=90,xshift=0.4cm] (b0) at (b1.west) {p4m};
				\node[anchor=east,rotate=90,xshift=0.4cm] (c0) at (c1.west) {p4g};
		\end{tikzpicture}
	\caption{The computed amplitudes (left) and phases (right)
	of the Fourier coefficients of the three tilings selected for the study.
	}
	\label{fig:all4amp}
\end{figure}

The Fourier coefficients of these images are presented in \autoref{fig:all4amp}. A natural question is to determine the consequences of these different symmetry properties in Fourier space.\par

First, the holohedry group $\DD{4}$ is revealed by the positions of the peaks in the various figures. By the positions of the peaks, we mean considering neither their phases nor their amplitudes, but only their locations.

The chirality of image \autoref{fig:p4}, corresponding to $p4$-invariant mesostructure, is directly reflected in the amplitudes of its Fourier coefficients. Reflection symmetry breaking does not manifest in the low-order circles of the amplitude and phase Fourier-coefficient diagrams. It becomes detectable starting at the 7th circle in the amplitude diagram and at the 4th circle in the phase diagram. As a consequence, it is important for the user to select a large enough set $\mathcal{M}^{\select}$ in order to detect these features.\par 

\begin{figure}[H]
    \centering
    \begin{subcaptionblock}{0.48\linewidth}
        \includegraphics[width=\linewidth]{diffractiondiagram_exact_pixel_p4gsquare.png}
        \caption{$\ZZZ_4$ centring}
    \end{subcaptionblock}
    \begin{subcaptionblock}{0.48\linewidth}
        \includegraphics[width=\linewidth]{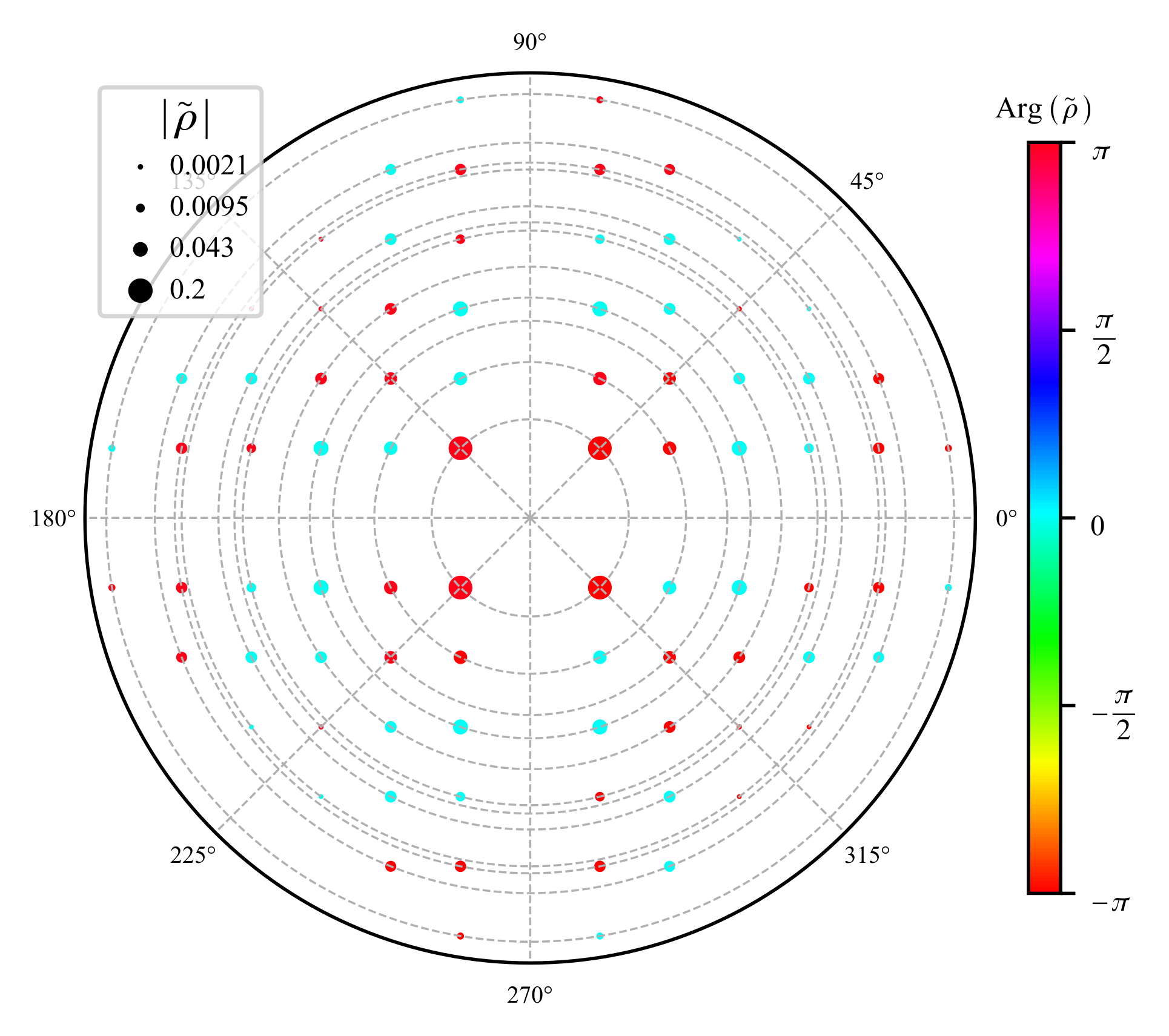}
        \caption{$\DD{2}$ centring}
        \label{fig:p4g-alternative-center-b}
    \end{subcaptionblock}
    \caption{\todoR{
    Fourier coefficients of p4g-symmetric tiling shown in~\autoref{fig:all4amp} for different centrings: (a) Centred on the red primitive unitcell of~\autoref{fig:p4m_s} (b) Centred on the yellow primitive unitcell of~\autoref{fig:p4g_s}.
    }}
    \label{fig:p4g-alternative-center}
\end{figure}

For the two mesostructures corresponding to the p4m and p4g groups, the amplitudes of their Fourier coefficients exhibit mirror symmetry for all observed points. However, while the phases of the p4m mesostructure also display mirror symmetry, the phases of the p4g Fourier coefficients lose this symmetry starting from the second dashed circle away from the origin. This difference arises from their symmorphic character: the space group of the first image is symmorphic, whereas that of the second is not.\\
\todoR{More precisely, due to the non-symmorphic nature of the  p4g space group, there exists no unit cell in which the full point group is realized. As a result, no choice of origin or centring yields Fourier coefficients invariant under the full point group $\DD{4}$. The centring chosen for the image is compatible with the $\ZZZ_4$
symmetry, as evidenced by the phase diagram. Adopting an alternative centring compatible with the $\DD{2}$ symmetry (yellow unit cell in \autoref{fig:p4g-alternative-center-b}) leads to \autoref{fig:p4g-alternative-center-b}, where the corresponding symmetries of the Fourier coefficients become apparent with diagonal mirrors. This result shows how the phases of the Fourier coefficients are sensitive to different centrings in real space.
}

As a consequence, the point group $\DD{4}$ of the (centred) image \autoref{fig:p4m} can be identified by direct inspection, whereas its detection in image \autoref{fig:p4g} is more subtle and requires a numerical verification of the gauge-linearity condition for its phase functions.

\begin{figure}[H]
  \centering
	\begin{subfigure}[c]{0.45\linewidth}
	\includegraphics[width=\linewidth]{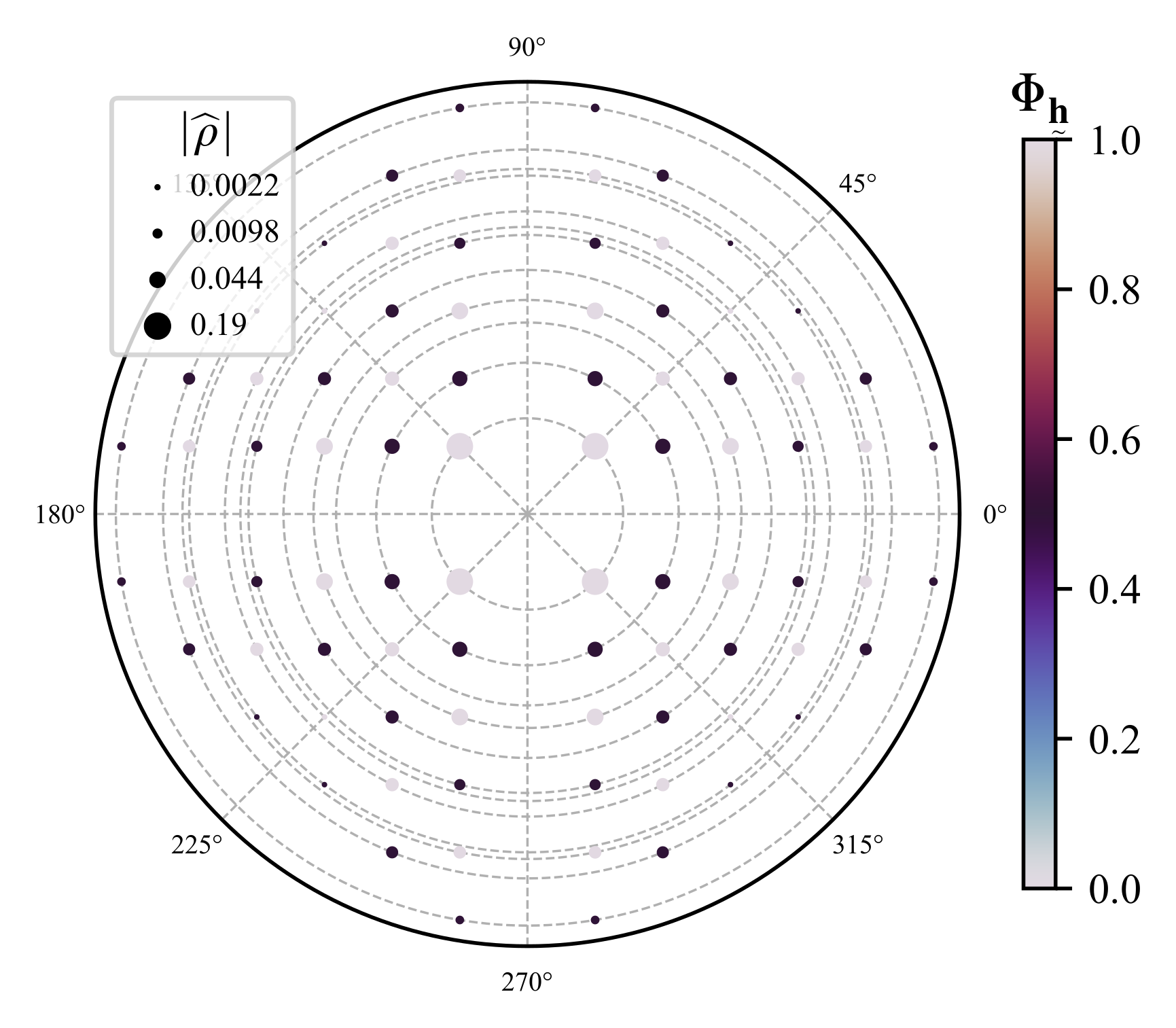}
	\caption{$\Phi_{\vp{h}}$}
	\label{fig:p4ggauges2}
	\end{subfigure}
	\hspace{2em}
	\begin{subfigure}[c]{0.45\linewidth}
	\includegraphics[width=\linewidth]{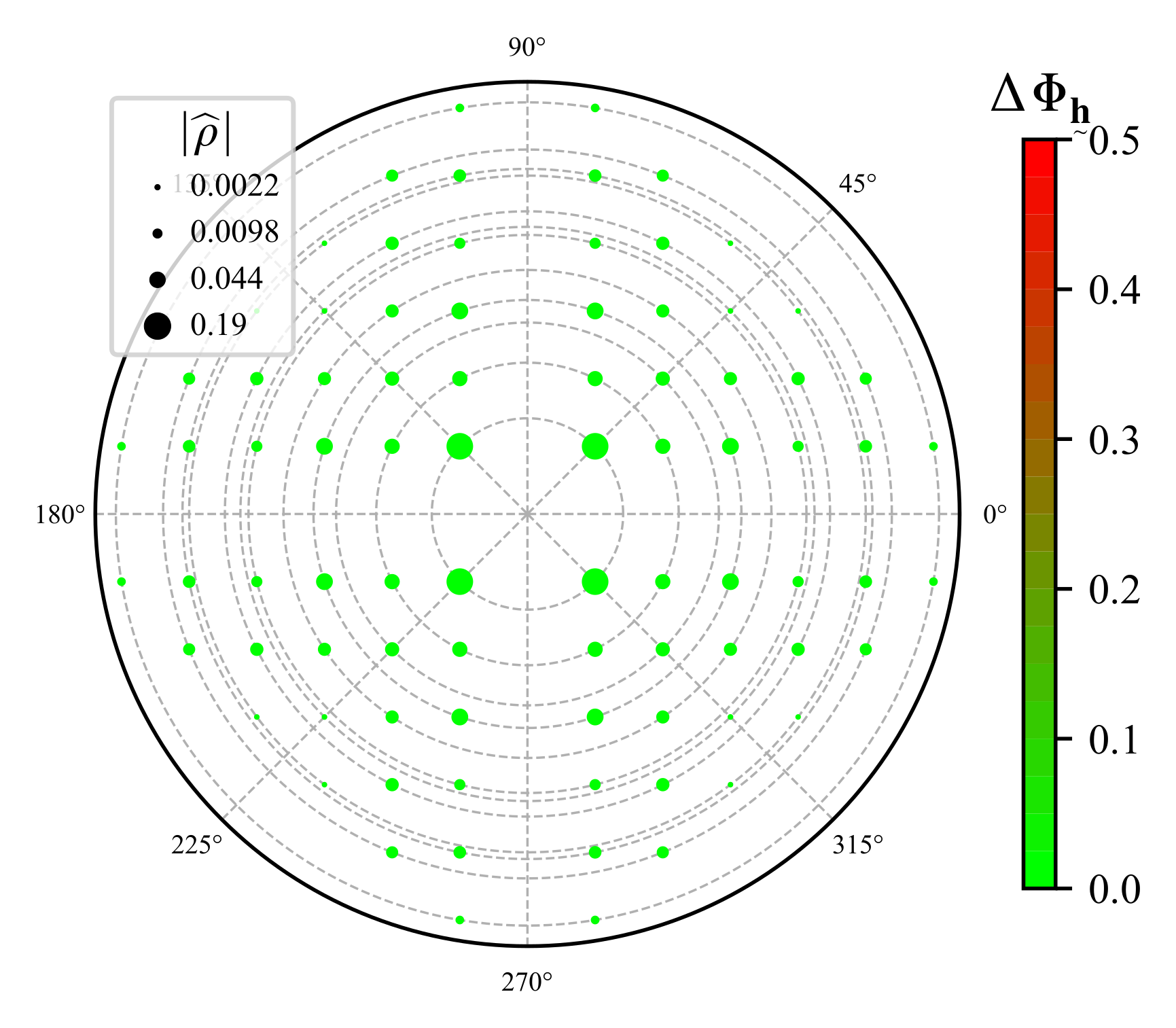}
	\caption{$\Delta\Phi_{\vp{h}}$}
	\label{fig:p4ggaugeerrors}
	\end{subfigure}
	\caption{Phase function of horizontal mirror acting on the p4g tiling: (a) Values of phase function $\Phi_{\vp{h}}$, (b) Deviations from gauge-linearity $\Delta\Phi_{\vp{h}}$.}\label{fig:p4ggauges}
	\label{fig:p4ggauges}
\end{figure}

\todoR{With this choice of centering, the rotational symmetry of \autoref{fig:p4g} becomes apparent in the Fourier coefficients. Consequently, it suffices to test gauge linearity with respect to the generator of the mirror symmetry.} The phase function $\Phi_{\vp{h}}$ and its gauge-linearity deviations $\Delta\Phi_{\vp{h}}$, corresponding to the horizontal mirror $\vp{h}$, are presented in \autoref{fig:p4ggauges}\footnote{While it may be intuitive to assume that the uniform translation in physical space associated to the glide reflection corresponds to a uniform shift of the complex argument in Fourier space, this is not the case. Indeed, since the associated translation is precisely half as long as the period of the tiling, only every other Fourier coefficient experiences a shift of $\pi$ in its argument, as shown in Figure~\ref{fig:p4ggauges2}.}.
The vanishing value of $\Delta\Phi_{\vp{h}}$ for all observed peaks confirms that the point group of \autoref{fig:p4g} is indeed $\DD{4}$.\\
This final example highlights that, in certain specific cases, it is necessary to go beyond visual inspection and quantitatively evaluate deviations from gauge linearity in order to determine the point group symmetry of a given image. It should be noted that the non-symmorphic character of the tiling can be tested algorithmically; however, for the sake of conciseness, the details of this verification are provided in the \ref{sec:appendix-symmorph}.

It is worth noting that, in the cases considered here, many features could be directly read from the Fourier coefficients because the images were non-generically centred on symmetry elements. For images in a generic position, an additional phase term would have prevented such direct interpretation, and the use of the algorithmic procedure would have been required for all images. However, the non-generic analysis is still valuable, as it reveals in these simple situations how the presence or absence of various physical symmetry elements is reflected in the Fourier coefficients.

For the first two images, the Fourier coefficients in~\autoref{fig:all4amp} are invariant with respect to their respective point groups. Accordingly, the phase functions associated to those transformations are equal to zero (modulo one) meaning that associated space groups  are symmorphic, as explained by~\autoref{eq:ConSym}. 
For images in a generic position, an additional phase term would have prevented such direct interpretation.

The situation is different for the Fourier coefficients of the last image. 
As already stated, no \todoR{horizontal }mirror symmetry is directly observable on the Fourier coefficients. This is expected, since for a non-symmorphic image there is no non-generic centring point at which the full point group can be realised. Consequently, several non-generic centring can be proposed, none of which would yield the full point group. A different non-generic centring of the image\todoR{, shown in~\autoref{fig:p4g-alternative-center},} yields invariance in Fourier space with respect to $\DD{2}$, a subgroup of $\DD{4}$. 
Consequently, in this case, the methodology introduced in Section~\ref{sec:detect-symm} becomes necessary.

\subsection{Quasiperiodic heterogeneous materials}
In this final subsection, the procedure is applied to images of various quasiperiodic mesostructures. We begin by examining the symmetry properties\footnote{\todoR{In this section, unless explicitly stated otherwise, the notion of symmetry is to be understood in the sense of indistinguishability.}} of the well-known Penrose tiling, using it as a detailed example to illustrate each step of the method. 
Subsequently, the procedure is applied to two other classes of tilings: the Ammann-Beenker tiling and the Fibonacci-Square tiling.

\subsubsection{Penrose tiling: $\DD{10}$- or $\DD{5}$-symmetric ?}
\label{sec:penrose}
\begin{figure}[H]
	\centering
	\begin{subcaptionblock}[t]{0.38\linewidth}
		\centering
		\vskip 0pt
        \includegraphics[width=\linewidth]{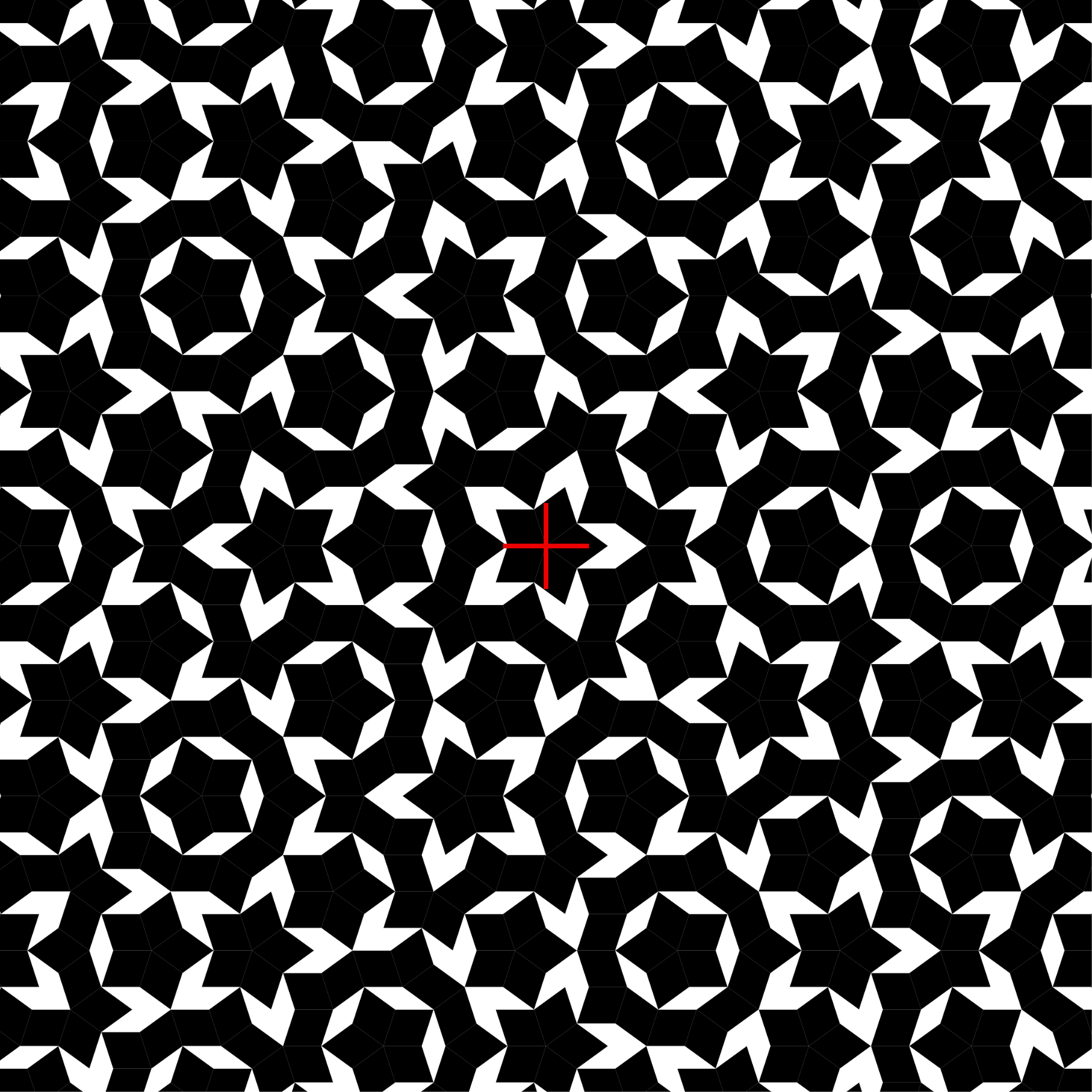};
		\caption{}
    \label{fig:penrose_tiling-a}
	\end{subcaptionblock}
	\begin{subcaptionblock}[t]{0.48\linewidth}
		\centering
		\vskip 0pt
		\includegraphics[width=0.8\linewidth]{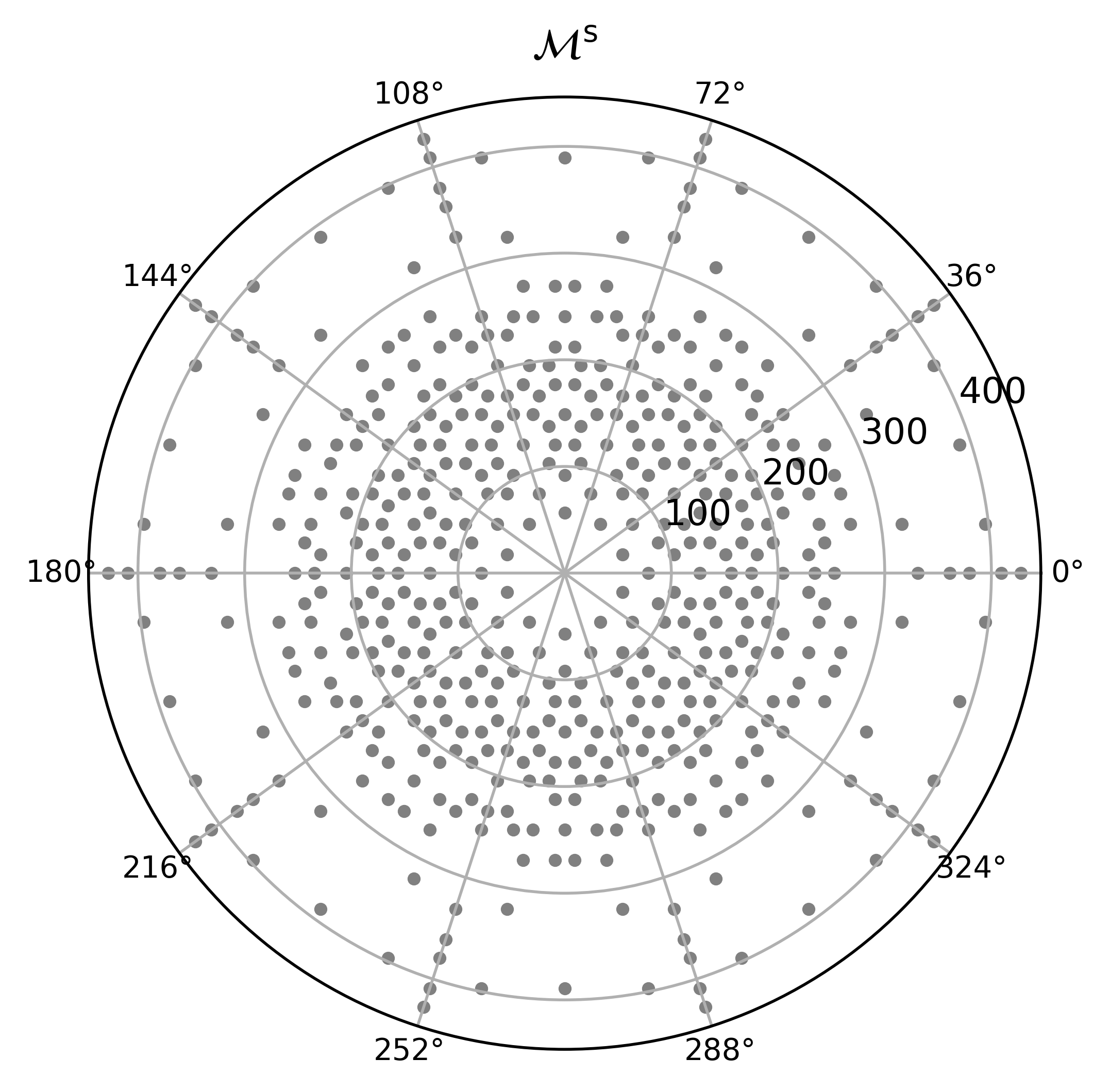}
		\caption{}
		\label{fig:penrose_tiling-b}
	\end{subcaptionblock}
	\caption{Classical Penrose tiling: (a) Unique centre of strong $\DD{5}$-symmetry by superposability (red) (b) Support of diffraction diagram $\mathcal{M}^{\select}$ with $\DD{10}$ as holohedry.}
	\label{fig:penrose_tiling}
\end{figure}

The first case we examine is the Penrose tiling, which, as noted in the introduction, is perhaps the most iconic example of a quasiperiodic tiling. Like other quasiperiodic structures, it features a unique point in the infinite tiling around which a dihedral group appears in the strong sense of superposability, as illustrated in \autoref{fig:penrose_tiling-a}. For this reason, it is sometimes regarded within the solid-mechanics community as possessing $\DD{5}$ symmetry \citep{IMEDIEGWU2023111922,rosa_stiff_2024}. This point could be chosen as centring point of the image. However, contrary to the periodic case, this point being unique, one cannot assume it will be contained in the studied image.\\

The support of the Penrose tiling diffraction diagram $\mathcal{M}^{\select}$, shown in \autoref{fig:penrose_tiling-b}, exhibits a higher degree of symmetry than $\DD{5}$, clearly indicating $\DD{10}$ as its holohedral point group. It follows that the point group of the image is either $\DD{10}$, or one of its subgroups. Following the methodology developed in this paper, we begin by testing the generators of $\DD{10}$, which, for reference, are taken to be $\mb{r}_{10}$ and $\vp{h}$.

\begin{figure}[H]
	\centering
	\begin{subcaptionblock}{0.47\linewidth}
		\includegraphics[width=\linewidth]{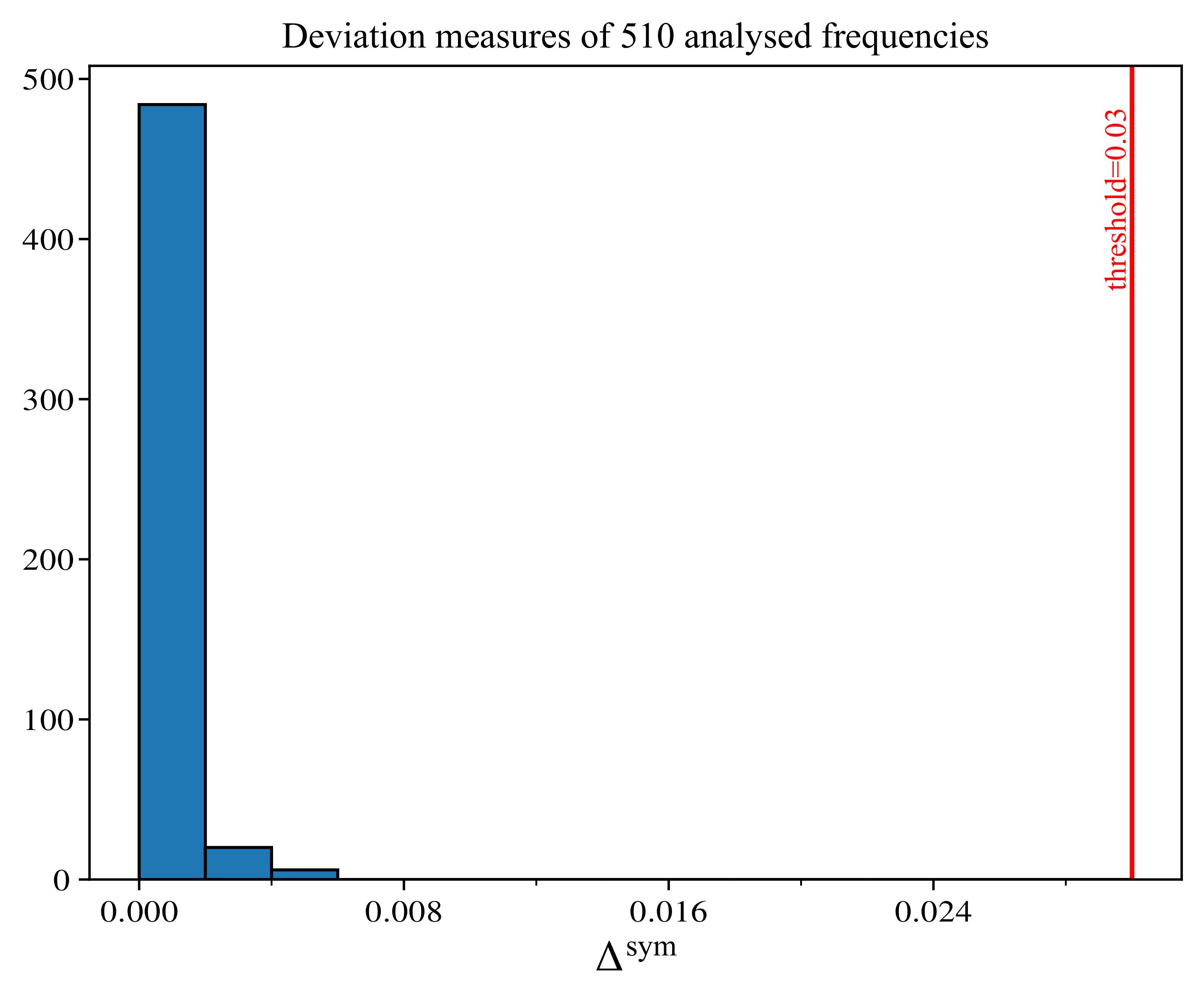}
		\caption{}
	\end{subcaptionblock}
	\begin{subcaptionblock}{0.47\linewidth}
		\includegraphics[width=\linewidth]{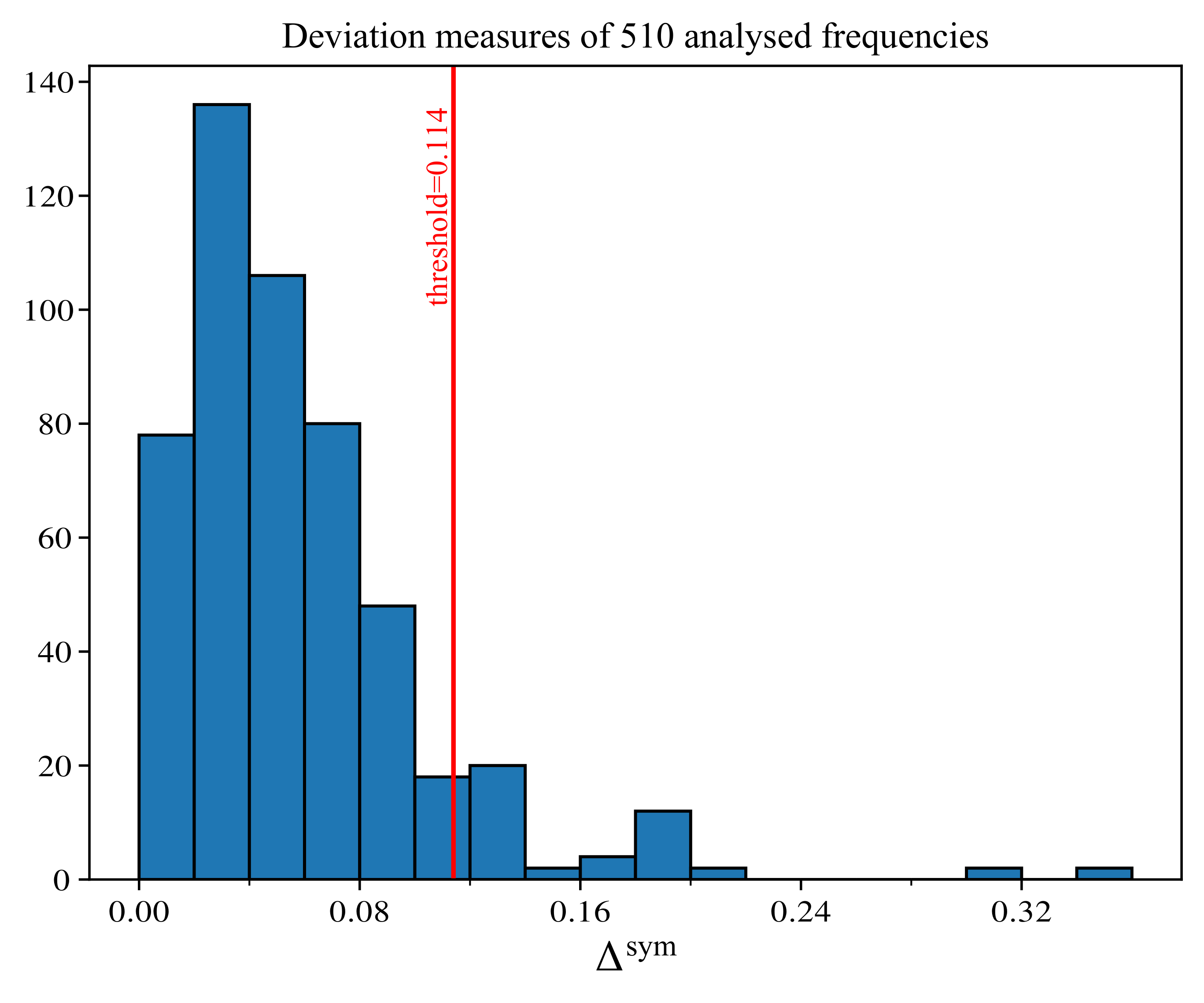}
		\caption{}
	\end{subcaptionblock}
	\caption{Deviation measures $\symerror$ for generators of $\DD{10}$ evaluated on a fragment of the canonical Penrose tiling: (a) Horizontal mirror (b) Generating rotation of $\frac{2\pi}{10}$. The vertical red lines correspond to the maximum deviations obtained from a degraded image of the p4m heterogeneous material introduced in \autoref{fig:p4mtiling}.}
	\label{fig:canonical_penrose_deviations}
\end{figure}

The deviations $\symerror$ of the generators of $\DD{10}$ from the conditions of indistinguishability are shown in~\autoref{fig:canonical_penrose_deviations}. In contrast to the periodic case investigated in the previous subsection, the deviations $\symerror$ are significantly larger and can no longer be regarded as approximately equal to $0$. As a consequence, it has been chosen to represent the results as histograms rather than the previously exposed polar plots.
The differing levels of accuracy can be explained by the distinct definitions of the Fourier coefficients in the two situations. In the periodic case, the Fourier coefficients are defined on a finite unit cell (cf.~\autoref{eq:fourcoeff-periodic}) and can therefore be extracted with high precision from a finite-sized image. In the quasiperiodic case, the Fourier coefficients are defined only in the infinite-domain limit (cf.~\autoref{eq:fourcoeff}). As a result, their computation from a finite-sized image implies some degree of approximation.\\

To assess whether the detected $\symerror$ should be interpreted as numerical artifacts or as a genuine lack of symmetry, we compare them with an artificially degraded image of the p4m lattice studied in subsection~\ref{sec:square}. This degraded image has side lengths that are not integer multiples of the unit cell, and, in addition, $2\%$ of its pixels are randomly inverted from black to white and vice versa. Since the true point group is known to be $\DD{4}$, all resulting deviations must originate from these deliberate numerical perturbations. In this case, the largest deviations $\symerror$ obtained are $0.03$ for the horizontal mirror and $0.114$ for the generating rotation.
Since most of the values of $\symerror$ shown in~\autoref{fig:canonical_penrose_deviations} lie below these levels, indicated by vertical red lines, this confirms that the group $\DD{10}$ is the point group of the generalised Penrose tiling in the weak sense of indistinguishability.\\

\begin{figure}[H]
	\centering
		\includegraphics[width=0.38\linewidth]{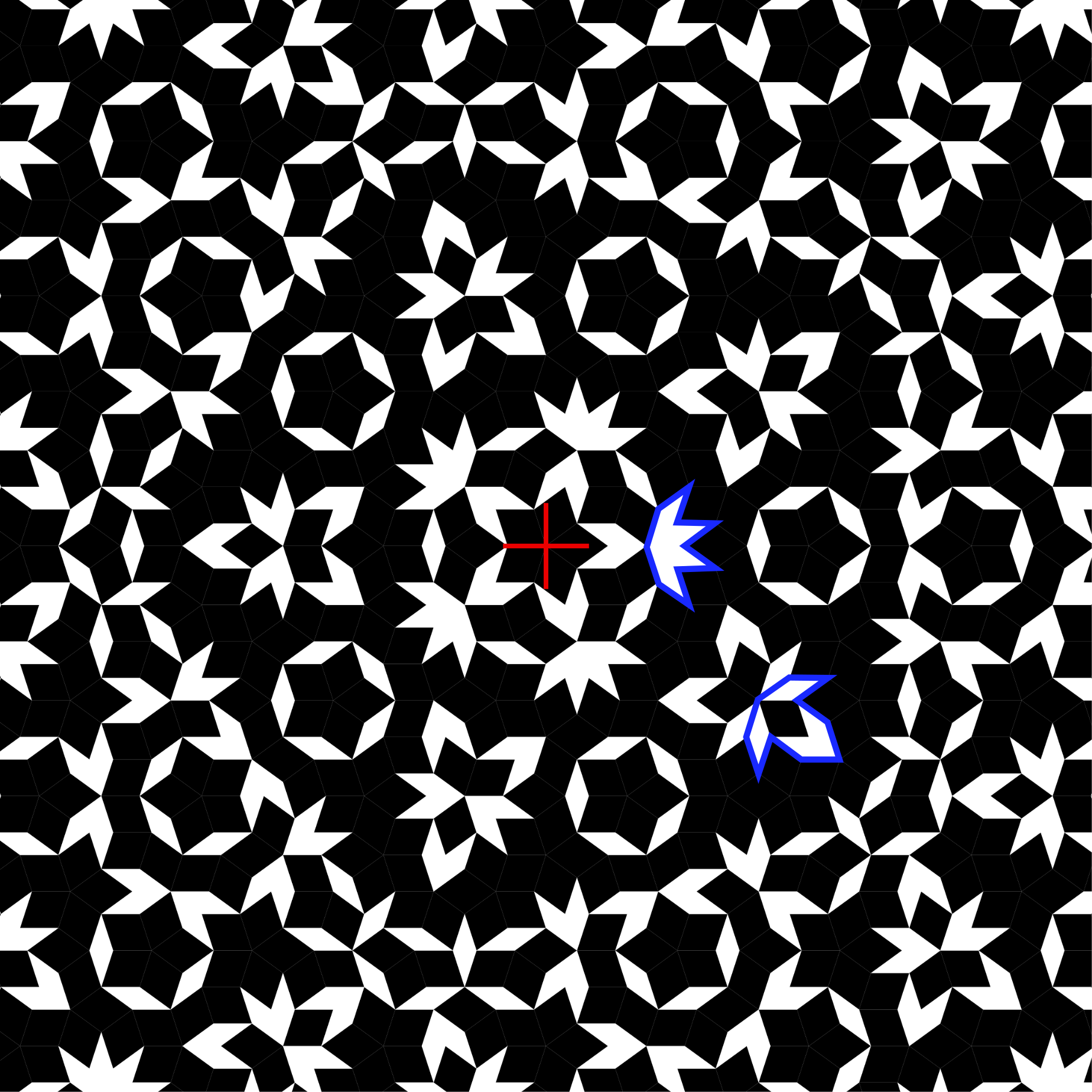}
	\caption{Fragment of generalised Penrose tiling. The unique centre of strong $\DD{5}$-symmetry by superposability is marked by the red cross. Patches of tiles that do not appear in the canonical penrose tiling are marked in blue.}
	\label{fig:penrosed5}
\end{figure}

It should be emphasized that, although this material is $\DD{10}$-indistinguishable, there exists no point in the infinite Penrose tiling at which the group $\DD{10}$ manifests in the strong sense of superposability. 
Aside from the canonical Penrose tiling, there exist variants that are truly $\DD{5}$-indistinguishable. The method is now applied to one of these variants to assess its true $\DD{5}$-indistinguishability. The studied example of such a tiling, corresponding to the parameter value $\gamma = 0.7$ in the parametrisation presented in~\citep{Ishihara:bw0492}, is shown in~\autoref{fig:penrosed5}. 
This tiling is difficult to distinguish from the canonical Penrose tiling by eye. There is, however, a subtle clue: this generalised version contains configurations of tiles that are forbidden in the canonical tiling, marked blue in~\autoref{fig:penrosed5}.

\begin{figure}[H]
	\centering
	\begin{subcaptionblock}{0.48\linewidth}
		\includegraphics[width=\linewidth]{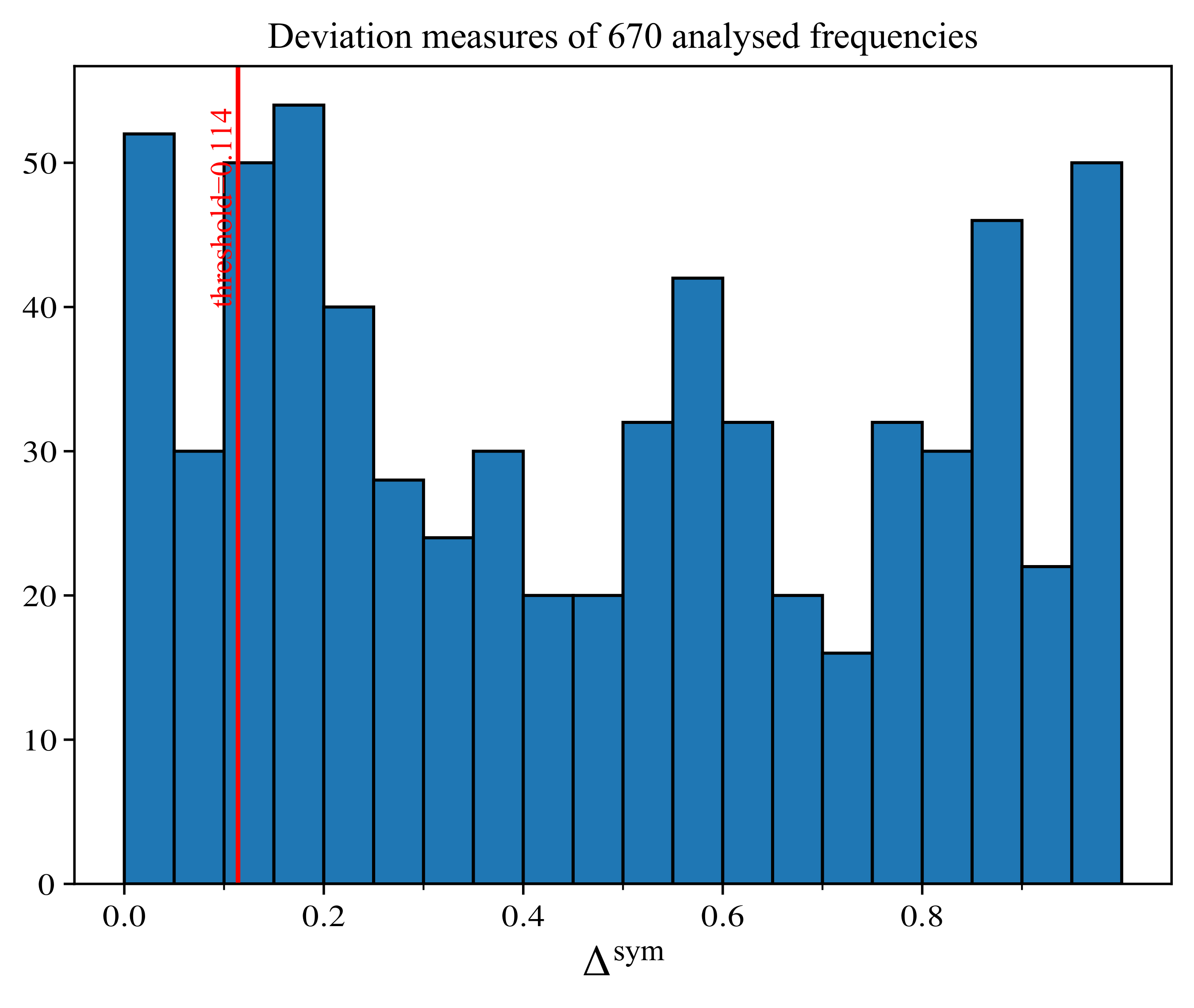}
		\caption{$\mb{r}_{10}$}
        \label{fig:penrosed5_rot_10_histogram}
	\end{subcaptionblock}
	\begin{subcaptionblock}{0.48\linewidth}
		\includegraphics[width=\linewidth]{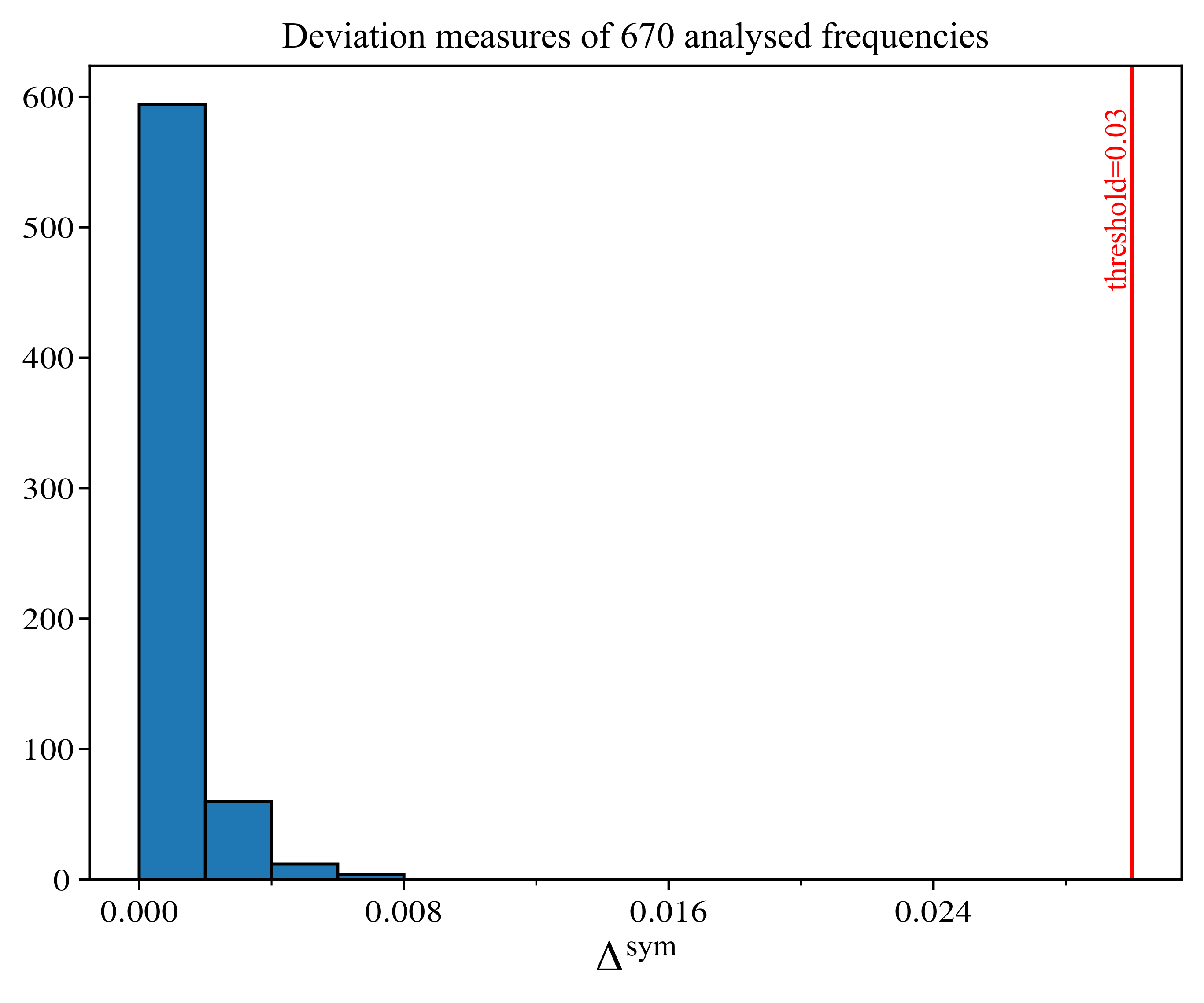}
		\caption{$\mb{r}_{5}$}
        \label{fig:penrosed5_rot_5_histogram}
	\end{subcaptionblock}
	\caption{Deviations $\symerror$ from the conditions of indistinguishability calculated for (a) the rotation $\mb{r}_{10}$ and (b) the rotation $\mb{r}_{5}$ on the generalised Penrose tiling shown in~\autoref{fig:penrosed5}. The vertical red line corresponds to the maximum deviations obtained from a degraded image of the p4m heterogeneous material introduced in \autoref{fig:p4mtiling}.}
	\label{fig:penrosed5_rot_histogram}
\end{figure}

The Fourier coefficients (not presented here for the sake of conciseness) reveal that this tiling possesses the same $\DD{10}$ holohedry as the canonical Penrose tiling. However, the magnitude of the deviations $\symerror$ associated with the rotation $\mb{r}_{10}$ differs significantly from those of the canonical Penrose tiling, as presented in Figure~\ref{fig:penrosed5_rot_10_histogram}. Only a small fraction of the analysed spatial frequencies exhibit deviations $\symerror$ below the acceptable level of $0.114$. Consequently, $\mb{r}_{10}$ cannot be regarded as a symmetry operation of the generalised Penrose tiling, thereby excluding $\DD{10}$ as the point group of this lattice. If the gauge linearity condition is now tested with respect to $\mb{r}_{5}$, one obtains the histogram in \autoref{fig:penrosed5_rot_5_histogram}, corresponding to the deviation measure $\symerror$. It is evident that the vast majority of the values lie below the threshold level. Performing the same analysis for the mirror generator (not presented here) confirms that $\DD{5}$ is, as expected, the point group of the generalised Penrose tiling in the sense of indistinguishability.

Before concluding this subsection, it is worth noting that, since the point groups of both tilings are not of the form $\DD{2^m}$, the tilings must necessarily be symmorphic.\\

This subsection studied the Penrose tiling, confirming its $\DD{10}$ indistinguishability. It underlined a difficulty for concluding on the indistinguishability of quasi-periodic tilings due to the discrepancy between the infinite-domain definition of the Fourier coefficients in the quasiperiodic case and the requirement of working with finite-sized images. Two thresholds have then been defined for the rotation and the mirror generators based on an analysis of a degraded image of a periodic tiling. These thresholds have then been applied to a generalized Penrose tiling in order to assess their validity on the case where the studied generator is not a weak symmetry of the tiling.

\subsubsection{Ammann-Beenker and Fibonacci-squares tiling}
\begin{figure}[H]
    \centering
		\begin{subcaptionblock}{0.38\linewidth}
			\includegraphics[width=\linewidth]{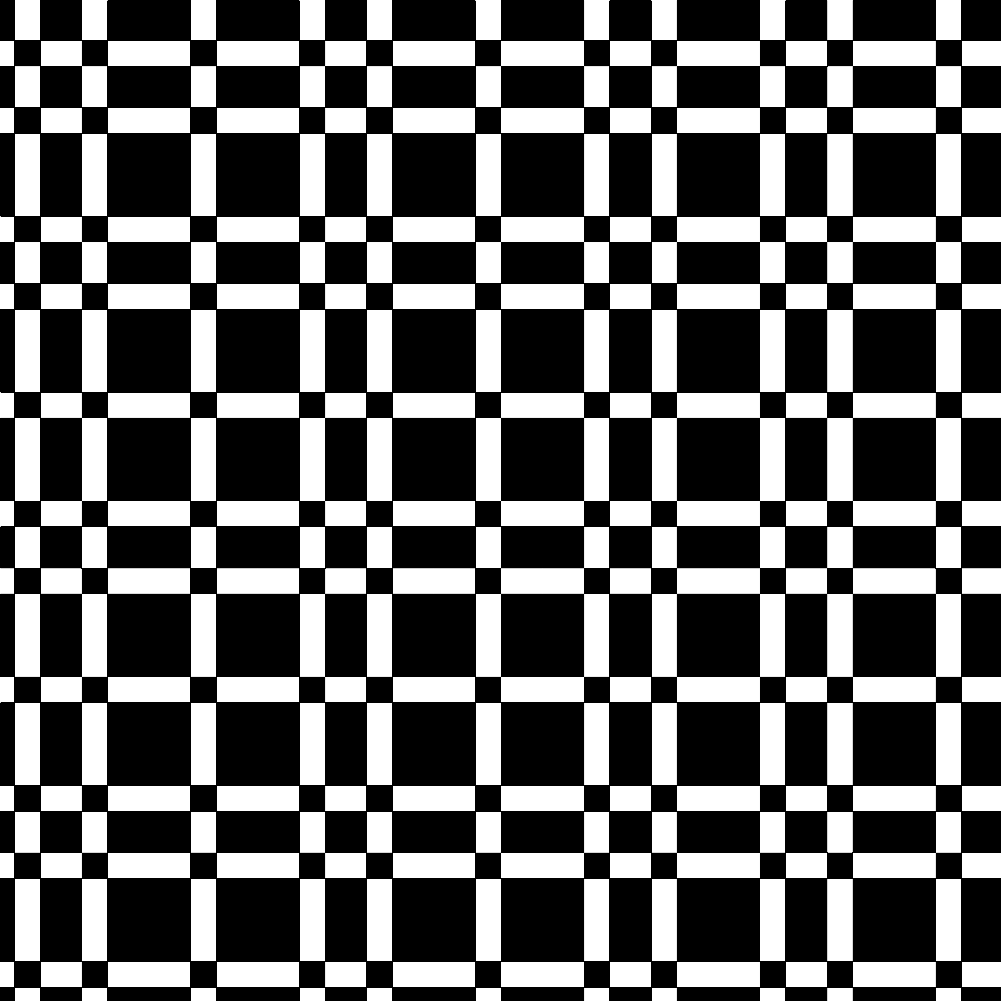}
			\caption{}
		\end{subcaptionblock}
        \hspace{0.5cm}
		\begin{subcaptionblock}{0.38\linewidth}
			\includegraphics[width=\linewidth]{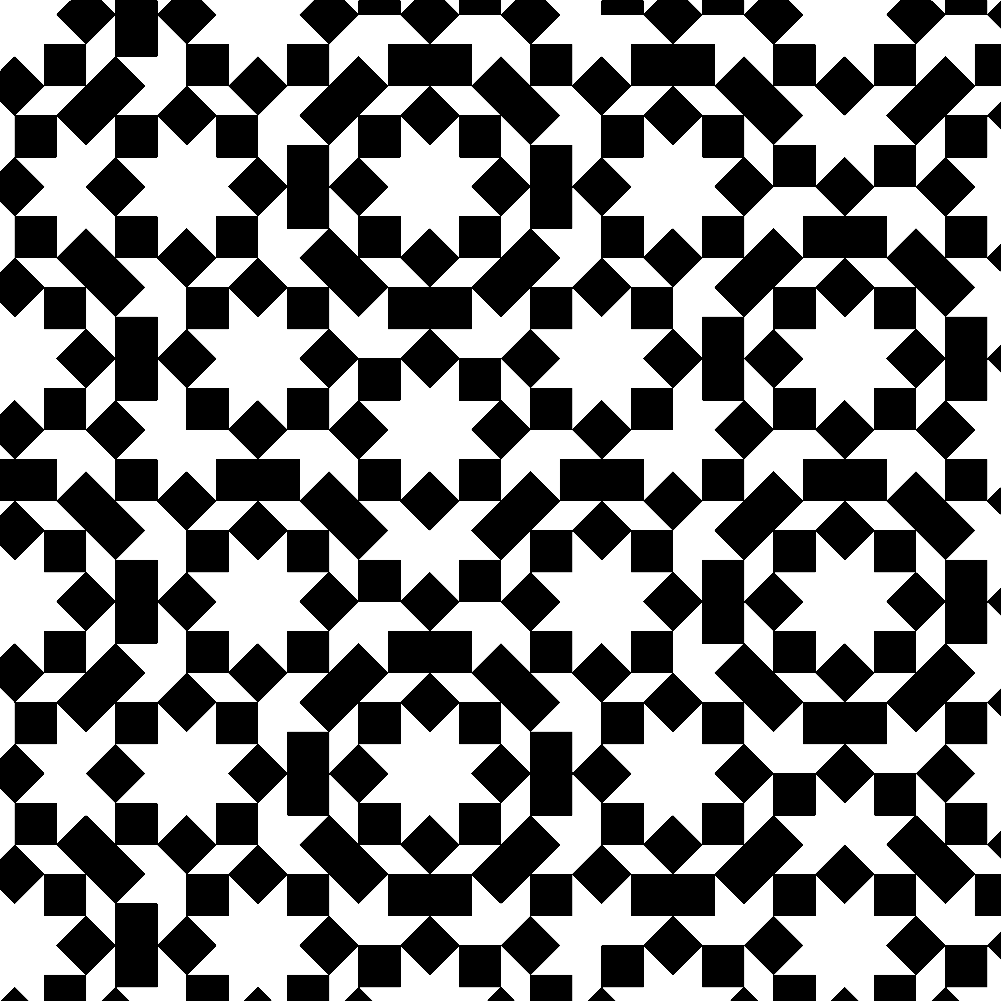}
			\caption{}
		\end{subcaptionblock}
		\caption{Generic fragment of the two chosen additional quasiperiodic tilings: (a) Fibonacci-squares tilings (b) Ammann-Beenker. A generic patch is selected that does not include their respective unique centre of superposability symmetry.}
		\label{fig:qptilings}
\end{figure}

\begin{figure}[H]
	\centering
		\begin{tikzpicture}
			\node (a1) at (0,0) {\includegraphics[width=0.43\textwidth]{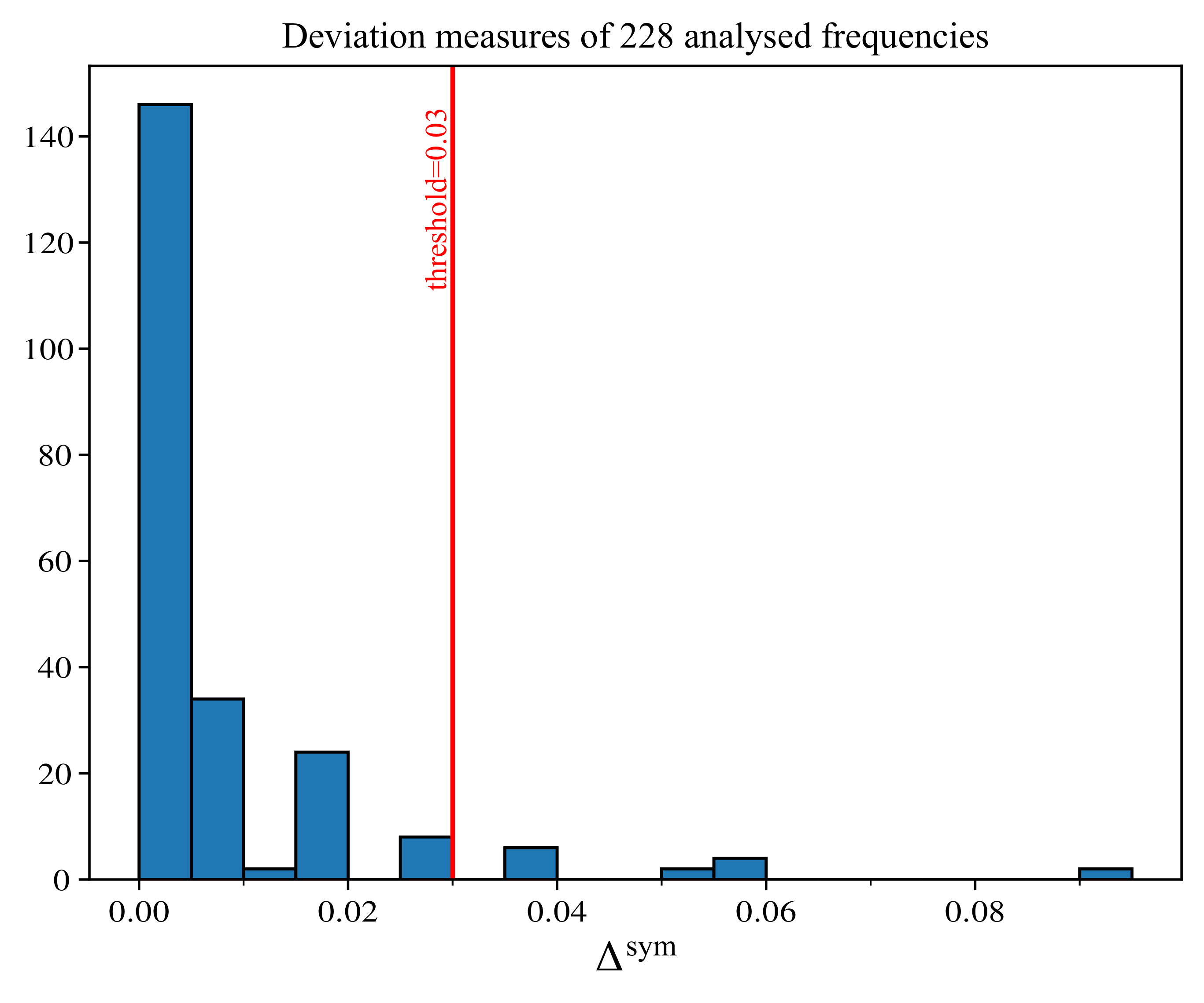}};
			\node[anchor=north] (b1) at (a1.south) {\includegraphics[width=0.43\textwidth]{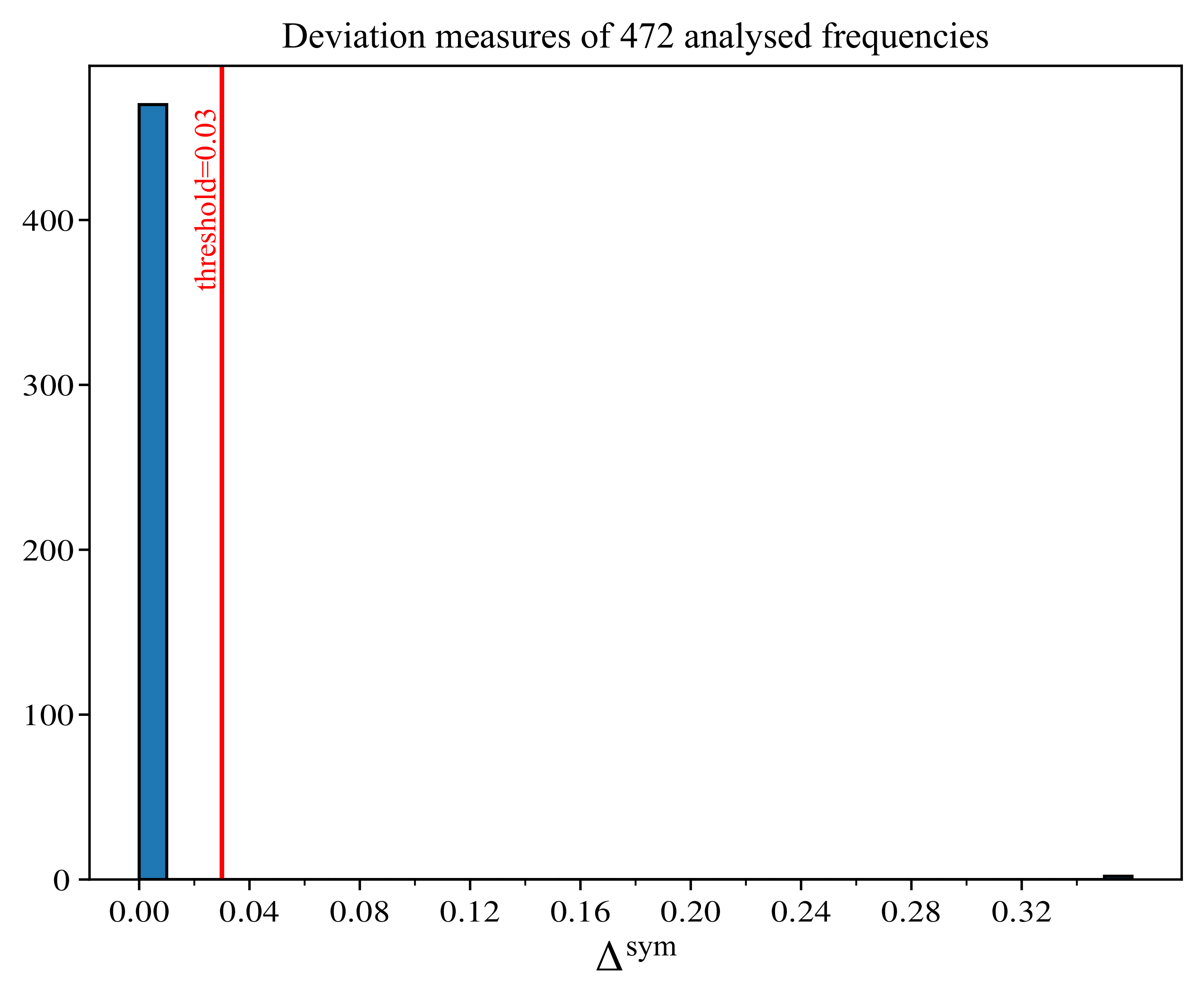}};

			\node[anchor=west] (a2) at (a1.east) {\includegraphics[width=0.43\textwidth]{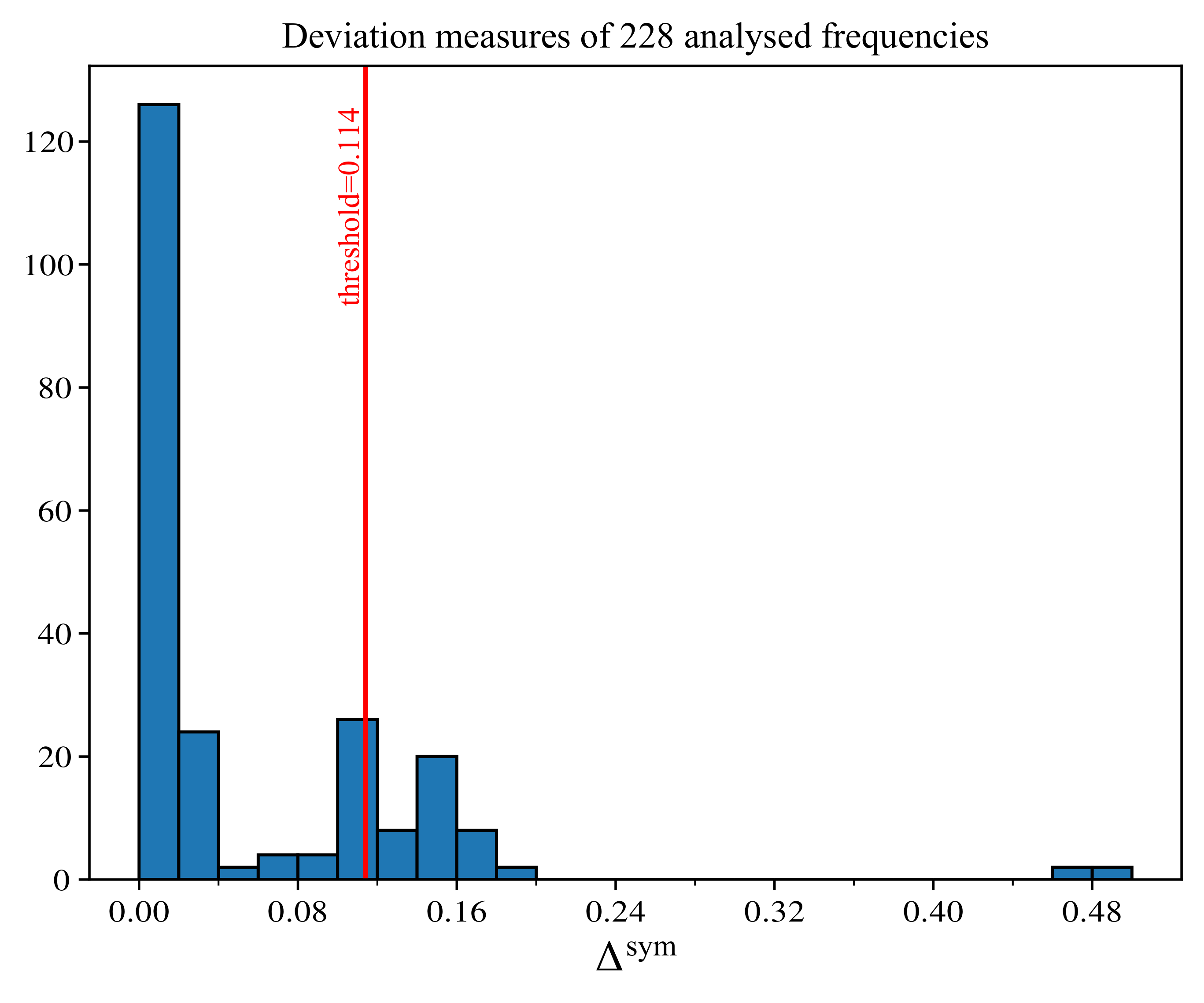}};
			\node[anchor=west] (b2) at (b1.east) {\includegraphics[width=0.43\textwidth]{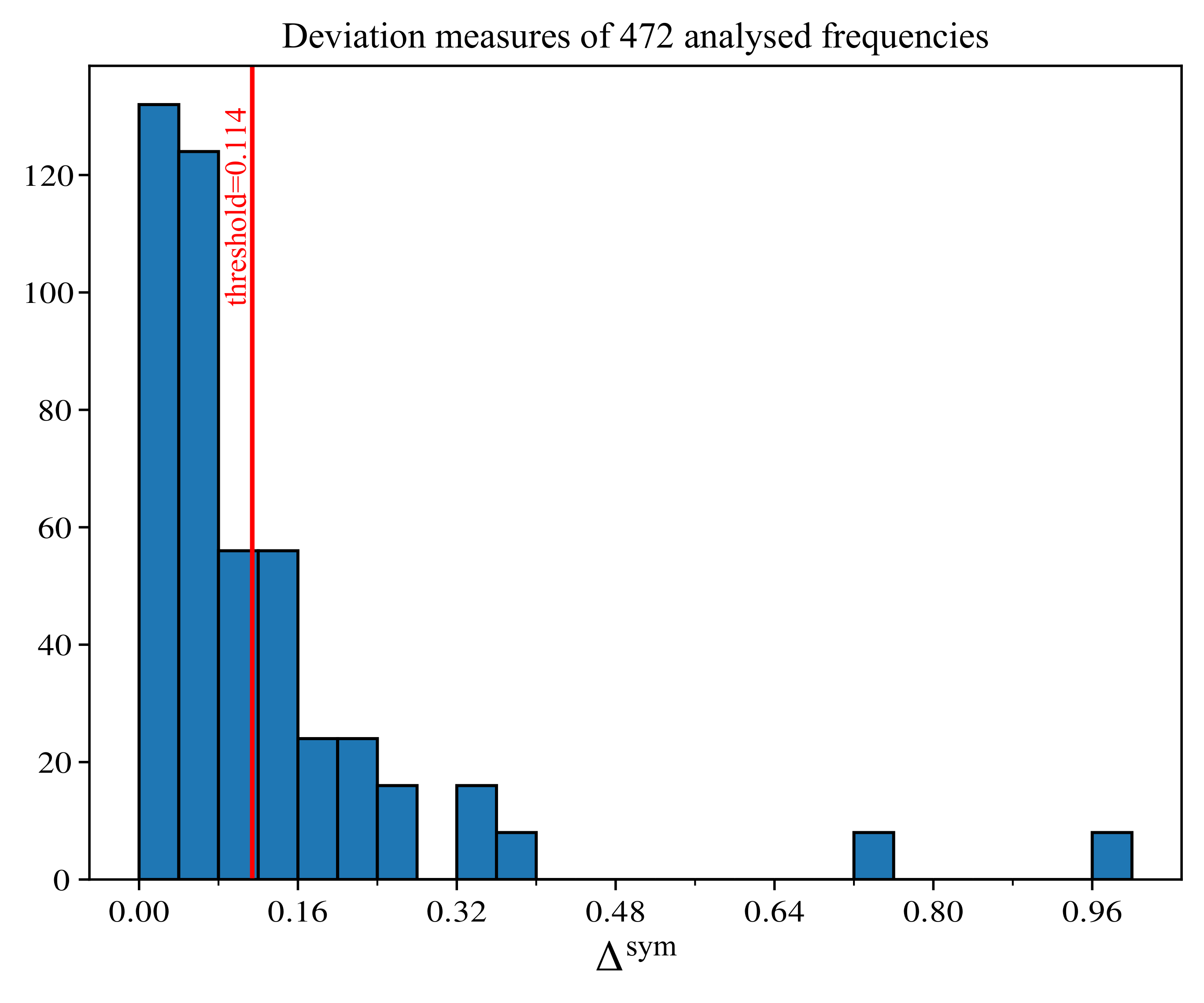}};

			\node[anchor=east,rotate=90,xshift=2.2cm,yshift=0.3cm] (a0) at (a1.west) {Fibonacci-squares, $\DD{4}$};
			\node[anchor=east,rotate=90,xshift=2.2cm,yshift=0.3cm] (b0) at (b1.west) {Ammann-Beenker, $\DD{8}$};
			\node[anchor=south] at (a1.north) {\textbf{Generating mirror}};
			\node[anchor=south] at (a2.north) {\textbf{Generating rotation}};
		\end{tikzpicture}
	\caption{Detected deviations $\symerror$ from indistinguishability for the generators of the point groups of the Ammann-Beenker and Fibonacci-squares tilings. The left column shows the horizontal mirror and the right column the generating rotation. The vertical red lines correspond to the maximum deviations obtained from a degraded image of the p4m heterogeneous material introduced in \autoref{fig:p4mtiling}. }
	\label{fig:qp_histograms}
\end{figure}

In this final subsection, we investigate the symmetry properties of two additional quasiperiodic tilings: the Fibonacci-squares tiling~\citep{LIFSHITZ2002186} and the Ammann-Beenker tiling~\citep{Beenker1982AlgebraicTO}. Fragments of these two tilings are shown in~\autoref{fig:qptilings}.
\todoR{The holohedral symmetry groups of the Fibonacci square and Ammann-Beenker tilings are identified as the dihedral groups $\DD{4}$ and $\DD{8}$, respectively}. The deviations $\symerror$ from indistinguishability of the generators of these groups are presented on
\autoref{fig:qp_histograms}. A clear majority of the detected deviations $\symerror$ are below the acceptance threshold for the Fibonacci-squares tiling as well as the horizontal mirror for the Ammann-Beenker tiling. The generating rotation $\DD{8}$ on the Ammann-Beenker tiling shows slightly higher deviations $\symerror$. However, those deviations are judged as acceptable compared to~\autoref{fig:penrosed5_rot_10_histogram}, which shows a transformation that is clearly not a weak symmetry. \todoR{Accordingly, the indistinguishability point group of the Fibonacci square tiling is $\DD{4}$, while that of the Ammann–Beenker tiling is $\DD{8}$.}\\
Both tilings have a point group compatible with non-symmorphism. However, a gauge function can be found in both cases using the methodology described in section~\ref{sec:detect-symm} that cancels the phase functions of both the point group generators. This proves the symmorphism of these two tilings. The detailed calculations are omitted here for the sake of conciseness.


%% file: conclusion.tex
\section{Conclusion}

This article introduces the symmetry criterion of indistinguishability for heterogeneous materials. This criterion generalises the conventional criterion of superposability to the quasiperiodic case. It originates from the works of \cite{rabson_1991} and \cite{mermin_1992} in condensed matter physics, where it describes the symmetries of (quasi)crystals on the molecular scale.\\
The criterion of indistinguishability in Fourier space consists of two conditions. Indistinguishable Fourier coefficients have (i) equal amplitudes and (ii) complex phases differing by a gauge function that respects the condition of gauge-linearity.\\
We propose a method that first extracts a set of Fourier coefficients from an image of a heterogeneous material and then deduces from them symmetry properties of the material.
First, the point group is determined by means of a deviation measure that indicates the distance from the conditions of indistinguishability\footnote{The quantitative nature of this measure is in itself an improvement of the conventional criterion of superposability, as it allows to evaluate the extend to which a given transformation constitutes a symmetry operation.}.
Second, the symmorphic character of the studied material is evaluated by determining the existence of a gauge function that nullifies all of the phase functions that are induced by the elements of its point group.\\
Finally, the practicability of the proposed method is demonstrated by applying it to artificially generated images of both two-dimensional periodic and quasiperiodic materials.
\todoR{As a result, we show that the symmetry group, understood in terms of indistinguishability, and hence in the sense of homogenisation, of the classical Penrose tiling is $\DD{10}$ rather than $\DD{5}$, as is sometimes stated in the literature.}\\

The proposed method currently requires for the user to manually identify the gauge function that determines symmorphism. Automating this step is now a work in progress. Moreover, one possible extension of this method would be to provide an automatic detection of the fundamental frequencies which are now manually identified by the user. Finally, when handling materials that do not exhibit a generalised lattice of peaks in Fourier space, such as the Thue-Morse tiling~\citep{Gazeau_2008}, the ability to extrapolate the values of a gauge function from the fundamental frequencies is lost and a redefinition of the deviation measure $\Delta \Phi_{\mb{Q}}$ derived from the complex phases is required. \todoR{More generally, it would be of interest to extend our algorithm to more general classes of tilings than those considered here, including aperiodic tilings (Thue-Morse, Pinwheel), bicrystalline moiré patterns \cite{gratias_2023}, among others. This constitutes a study in its own right, potentially involving further numerical developments, and will be addressed in future work. Another direction for further development lies in the possibility of inverting the proposed method to guide the design of tailored architectured materials.}\\

The proposed method may be applied beyond the scope of two-dimensional synthetic images presented in this article.
Possible use cases include experimentally obtained images of real materials or numerically obtained images from numerical simulations. However, care should be taken when selecting these images in order to have enough material inside the chosen square image to analyse in order to generate accurate enough Fourier coefficients and to be able to appropriately identify the peaks. Using experimental or numerical images, this method could be implemented into the bifurcation detection procedure described in  \citep{poncelet_2023} in order to extend their procedure to quasiperiodic materials. Furthermore, the theoretical framework of indistinguishability is  suitable for treating three-dimensional heterogeneous materials \citep{rabson_1991}. The additional complexity of the three-dimensional case stems primarily from the greater set of possible point groups, which exceeds the dihedral and cyclic groups of two dimensions~\citep{miller_symmetry_1972}.

%% file: appendix.tex
\section{Detection of symmorphism for the periodic square images}
\label{sec:appendix-symmorph}

\begin{figure}[H]
	\centering
	\includegraphics[width=0.45\linewidth]{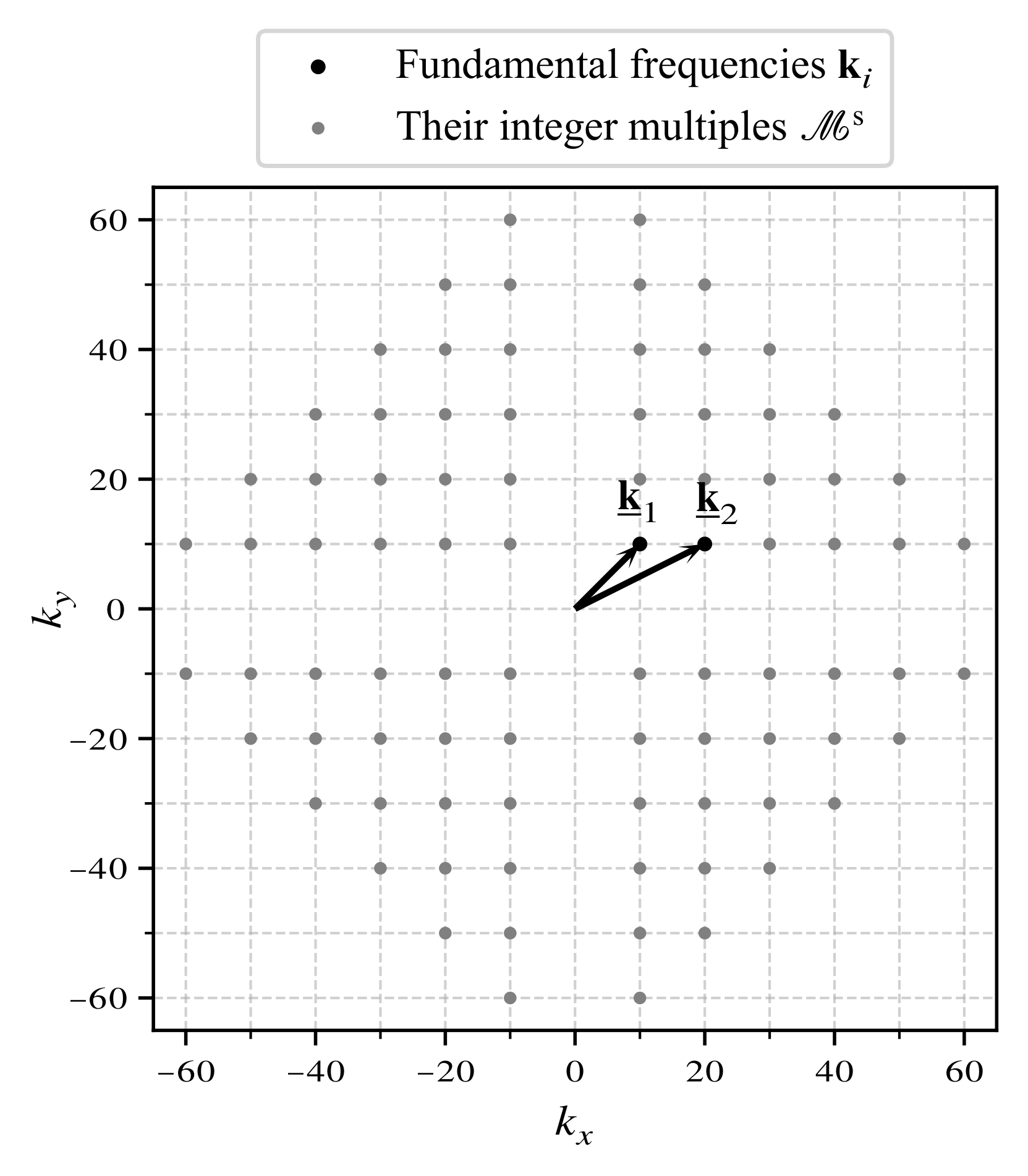}
	\caption{Selected spatial frequencies of the p4g symmetric heterogeneous material in Fourier space together with the fundamental frequencies $\vk_1$ and $\vk_2$.}
	\label{fig:p4g-frequencies}
\end{figure}

To this end, the fundamental frequencies are selected to be $\left[\vk_1\right]=
\begin{bmatrix}
	10\\
	10
\end{bmatrix}
$
and
$\left[\vk_2\right]=
\begin{bmatrix}
	20\\
	10
\end{bmatrix}
$, shown in \autoref{fig:p4g-frequencies}.
The seemingly more intuitive choice for the square lattice of $\left[\vk_1\right]=
\begin{bmatrix}
	10\\
	0
\end{bmatrix}
$
and
$\left[\vk_2\right]=
\begin{bmatrix}
	0\\
	10
\end{bmatrix}
$ is not used because, even though these points belong to the reciprocal lattice, they do not carry a detectable Fourier coefficient.\\


The phase functions ${\Phi}_{\mb{r}}$ and ${\Phi}_{\mb{h}}$ associated to the generators of the point group $\DD{4}$ being gauge functions, they are completely determined by their values on the fundamental frequencies $\vk_i$.
These values are given by
$\left[{\Phi}_{\mb{r}}\right]=
\begin{bmatrix}
    \Phi_{\mb{r}}(\vk_1)\\
 	\Phi_{\mb{r}_4}(\vk_{2})
 \end{bmatrix} =
\begin{bmatrix}
	0\\
	0
\end{bmatrix}
$ and
$\left[{\Phi}_{\mb{h}}\right]=
\begin{bmatrix}
   \Phi_{\mb{h}}(\vk_1)\\
 	\Phi_{\mb{h}_4}(\vk_{2})
 \end{bmatrix} =
\begin{bmatrix}
	0\\
	0.5
\end{bmatrix}
$, where the notation $[]$ is introduced to denote the column vector of values of a gauge functions on the fundamental frequencies.\\
As ${\Phi}_{\mb{r}}$ is already equal to zero, the same holds for the gauge function $\chi_{\mb{r}}$ introduced in~\autoref{eq:chi_r_def}.
Consequently, only the second gauge function $\chi_{\vp{h}}$ introduced in~\autoref{eq:chi_h_def}, which serves to nullify ${\Phi}_{\mb{h}}$, remains to be determined.\\
The horizontal mirror $\mb{h}$ is expressed with respect to the fundamental frequencies $\vk_i$ as
\ben
\left[\mb{h}\right]=
\begin{bNiceMatrix}
	10 & 20 \\
	10 & 10 \\
	\CodeAfter
	\UnderBrace[shorten]{2-1}{2-1}{\vspace{1cm}\vk_1} 
  \UnderBrace[shorten]{2-2}{2-2}{\vspace{1cm}\vk_2} 
\end{bNiceMatrix}^{-1}
\begin{bmatrix}
	1 & 0 \\
	0 & -1 \\
\end{bmatrix}
\begin{bNiceMatrix}
	10 & 20 \\
	10 & 10 \\
	\CodeAfter
	\UnderBrace[shorten]{2-1}{2-1}{\vspace{1cm}\vk_1} 
  \UnderBrace[shorten]{2-2}{2-2}{\vspace{1cm}\vk_2} 
\end{bNiceMatrix}
=
\begin{bNiceMatrix}
	-3 & -4 \\
	2 & 3 \\
\end{bNiceMatrix}
.
\een
 \\
Consequently, its phase function ${\Phi}_{\mb{h}}$ transforms under application of ${\chi}_{\vp{h}}$ following \autoref{eq:gaugtrans} as
\ben
\left[{\Phi}_{\mb{h}}'\right]\equiv\left[{\Phi}_{\mb{h}}\right]+
\cdot
\left(
\begin{bmatrix}
	-3 & -4\\
	2 & 3\\
\end{bmatrix}
-
\begin{bmatrix}
	1 & 0\\
	0 & 1\\
\end{bmatrix}
\right)
^\mathrm{T}
\left[{\chi}_{\vp{h}}\right]
=
\left[{\Phi}_{\mb{h}}\right]+
\begin{bmatrix}
	-4 & -4\\
	2 & 2
\end{bmatrix}
^\mathrm{T}
\cdot
\left[{\chi}_{\vp{h}}\right]
.
\een
As ${\chi}_{\mb{h}}$ is limited to values of either $0$ or $0.5$ (see \autoref{sec:detect-symm}), the second summand in the above equation is limited to integer values and thereby unable to induce a change in ${\Phi}_{\mb{h}}$. Accordingly, there can be no gauge function that nullifies both ${\Phi}_{\mb{r}_4}$ and ${\Phi}_{\mb{h}}$ and the p4g heterogeneous material is determined to be non-symmorphic.